\documentclass[10pt]{article}

\usepackage{bm}
\usepackage{latexsym}
\usepackage{dcolumn}
\usepackage{amsmath,amsfonts,amssymb}
\usepackage{graphicx,epsfig}
\usepackage{color}
\usepackage[active]{srcltx}

\def\g{\sqrt{-g}\,}
\def\gu#1#2{{g^{#1#2}}}
\def\gl#1#2{{g_{#1#2}}}
\def\pdsl#1#2{(\partial#1/\partial#2)}        
           
\def\pdlr#1#2{\left(\frac{\partial#1}{ \partial#2}\right)}

\def\eq#1{{Eq.~(\ref{#1})}}
\def\beq{\begin{equation}}
\def\eeq{\end{equation}}
\def\bea{\begin{eqnarray}}
\def\eea{\end{eqnarray}}
\def\benu{\begin{enumerate}}
\def\eenu{\end{enumerate}}
\def\nn{\nonumber}

\def\l{\left}
\def\r{\right}

\def\DM{\mathrm{d}}
\def\hatxi{\widehat{\xi}}
\def\hatn{\widehat{n}}

\def\pdsl#1#2{(\partial#1/\partial#2)}        
           
\def\pdlr#1#2{\left(\frac{\partial#1}{ \partial#2}\right)}

\def\gu#1#2{{g^{#1#2}}}
\def\gl#1#2{{g_{#1#2}}}
\def\g{\sqrt{-g}\,}

\def\rc#1#2#3#4{{R^{#1}_{\phantom{#1}#2#3#4}}}

\def\hatxi{\widehat{\xi}}
\def\hatn{\widehat{n}}
\def\tn{\textsl n}
\def\tslt{\textsl t}
\def\rnn{R^{\tn {A}}_{~ \tn { A}}}
\def\rtt{R^{\tslt { A}}_{~ \tslt { A}}}
\def\rnna{R^{\tn { A}}_{~ \tn { B}}}
\def\rtta{R^{\tslt { A}}_{~ \tslt { B}}}

\def \k{\overset{(0)}{  {K}^{\alpha}_{\beta} }}
\def \kp{\overset{(1)}{  {K}^{\alpha}_{\beta} }}

\def\pdsl#1#2{(\partial#1/\partial#2)}        
           
\def\pdlr#1#2{\left(\frac{\partial#1}{ \partial#2}\right)}
\def\g{\sqrt{-g}\,}
\def\mr{\mathcal{R}}
\def\rc#1#2#3#4{{R^{#1}_{\phantom{#1}#2#3#4}}}

\def\gu#1#2{{g^{#1#2}}}
\def\gl#1#2{{g_{#1#2}}}

\def\va{Virasoro algebra}

\def\mbt{\bm} 

\newcommand{\myeq}[2]{\begin{equation} \begin{split} \label{#1} #2 \end{split} \end{equation}}

\def\DM{\mathrm{d}}

\newcommand{\ph}[1]{\phantom{#1}}
\newcommand{\D}{\ensuremath{\nabla}}

\newcommand{\mdof}{microscopic degrees of freedom}

\newcommand{\bh}{black hole}

\newcommand{\LL}{Lanczos-Lovelock }

\def\sec#1{{Sec.~\ref{#1}}}

\title{\LL models of gravity}
 \author{T. Padmanabhan\footnote{paddy@iucaa.ernet.in}, \\
IUCAA,
 Post Bag 4, Ganeshkhind, Pune - 411 007, India 
 \\
 \\
Dawood Kothawala\footnote{dawood@physics.iitm.ac.in}\\
Department of Physics, IIT Madras, Chennai - 600 036, India.}

\begin{document}

\maketitle

\begin{abstract}
\LL\ models of gravity represent a natural and elegant generalization of Einstein's theory of gravity to higher dimensions. They  are characterized by the fact that the field equations only contain up to second derivatives of the metric  even though the action functional can be a quadratic or higher degree polynomial in the curvature tensor. Because these models share several key properties of Einstein's theory they serve as a useful set of candidate models  for testing the emergent paradigm for gravity. This review highlights several geometrical and thermodynamical aspects of \LL\ models which have attracted recent attention.
\end{abstract} 

\tableofcontents


%
%

\section{Introduction}

The principle of equivalence, along with the principle of general covariance, suggests that gravity has a geometrical interpretation in terms of the spacetime structure. These principles are adequate to determine the \textit{kinematics} of gravity (viz., how curvature of spacetime influences the dynamics of spacetime) in an elegant manner.  Much of the beauty attributed to general theory of relativity arises from this geometrical interpretation and the kinematics alone.

To complete the picture, one also needs to determine the \textit{dynamics}, viz.,  how matter determines the spacetime structure. Based on the structure of other physical theories, it appears reasonable to expect that this should arise through a second order differential equation relating the metric tensor to suitably defined source built from matter variables.  Unfortunately, we do not have any equally elegant principle to determine these field equations of gravity. 

It is therefore reasonable to study the most general system of field equations which are second order in the independent variables, which are the components of the metric tensor. Under very reasonable assumptions, we are then led to a class of models first discussed by Lanczos \cite{ll1} and Lovelock \cite{ll2}. These field equations follow from an action functional which is a \textit{polynomial} in the curvature tensor and reduces --- uniquely --- to Einstein's  general relativity when $D=4$. In higher dimensions ($D>4$) the \LL\ models differ from Einstein's theory and have a richer structure.

The purpose of this review is to describe several features of the \LL\ models. Given the fact that they are more general than Einstein's theory and have been the subject of intensive study in the recent years, it is not possible to cover all the varied aspects of these models. In order to retain  focus in the review, we will concentrate on the aspects of \LL\ models closely related to the emergent paradigm for gravity \cite{rop,TPreviews,TPstructuralasp}. It turns out that several features related to the thermodynamics of horizons carry over from Einstein's general relativity to \LL\ models. Since the emergent paradigm of gravity uses the thermodynamic features of horizons as a key input, this allows a natural extension of this paradigm to all \LL\ models.  The discovery that several features of the emergent paradigm transcends Einstein's theory of gravity strongly suggests that the relationship between gravitational dynamics and horizon thermodynamics is more general than was first recognized. In fact, this has been a strong motivation to study \LL\ models of gravity in the recent years. Such studies have also unravelled several peculiar  geometrical features of \LL\ models which deserve attention on their own. We will describe these features as well, as we go along.

 Unless otherwise specified, we work with a $D$ dimensional spacetime with signature $(-,+,+, \ldots)$.  Latin alphabets $i, j, \ldots$ run over spacetime indices, whereas Greek symbols $\mu, \nu, \ldots$ run over indices representing a $(D-1)$ dimensional hypersurface. This would  be usually a spacelike hypersurface, except in Sections \ref{sec:4d4}, \ref{sec:4d5} where it is timelike. Finally, we use upper case Latin letters $A, B, \ldots$ for indices representing a $(D-2)$ dimensional spacelike hypersurface. The convention for Riemann tensor we follow is: $R^a_{\phantom{a}bcd}=2\partial_{[c} \Gamma^a_{\phantom{a} d]b} + 2 \Gamma^a_{\phantom{a} m[c} \Gamma^m_{\phantom{m} d]b}$. 
We use the symbol `$\equiv$' to denote the fact that the equation defines a variable or a notation.

This review is structured in terms of five parts consisting of Sections 2 to 5. Section 2 introduces \LL\ models and their major structural and geometrical features. Section 3 begins by introducing the concept of horizon entropy in theories more general than Einstein gravity and uses it to study several aspects of horizon entropy in these models. Sections 4 and 5 describe the connetion between \LL\ model and the emergent gravity paradigm.

\section{\LL\ models: Structural Aspects}

As emphasized above, one distinguishing feature of \LL\ models is that, although the action functional of the theory could be an arbitrary higher order polynomial in the curvature tensor, it leads to equations of motion which do not contain derivatives of the metric tensor of order higher than two. Originally, this class of theories were arrived at from a somewhat indirect route by Lovelock \cite{ll2}. 
 He  wrote down the field equations in the  form 
$E_{ab} = (1/2) T_{ab}$ where $T_{ab}$ is the energy momentum tensor of matter fields and demanded that $E_{ab}$ satisfies the following requirements: (i) $E_{ab}$ is
symmetric in $a$ and $b$; (ii) it is a function of metric and its first two derivatives; (iii) it is divergence-less. Having determined the most general form of $E_{ab}$ satisfying these criteria, he obtained a Lagrangian from which one can obtain these field equations. (Note that we have not demanded $E_{ab}$ to be \textit{linear} in the second derivatives of the metric. If such a restriction is  \textit{also} added, one obtains Einstein's theory with a cosmological constant.) 
It is, however, possible to obtain \LL\ models in a more streamlined and transparent manner along different lines (see e.g., Chapter 15 of Ref.~\cite{gravitation}) which is what we will pursue here.

\subsection{Lagrangians from curvature and metric: general results} \label{sec:2a1}

 Let us  consider a {\it general} Lagrangian built out of Riemann tensor and the metric, but not containing derivatives of the Riemann tensor, and ask the question: What is the condition on such a Lagrangian so as to ensure that equations of motion should contain no derivatives of the metric tensor higher than second order? The answer, as it turns out, uniquely picks out the \LL\ Lagrangians. 

We will now describe this result highlighting the main conceptual points, many of which are of interest in their own right. We start with the action functional in $D-$dimensional spacetime:
\begin{equation}
A= \int_\mathcal{V} d^Dx\sqrt{-g}\, L 
\label{genaction1}
\end{equation}
where $L$ is a scalar built from the metric and the curvature tensor.
We can think of the Lagrangian $L$ as a functional of any one of ($R_{abcd}, \rc abcd, R^{ab}_{cd} ...$) in combination with either $\gu ab$ or $\gl ab$. The variation of the Lagrangian with respect to curvature and metric will bring in tensors defined by the derivative of Lagrangian with respect to different kinds of curvature components and the metric, like, for e.g., 
\begin{equation}
 P^{abcd} \equiv \pdlr{L}{R_{abcd}}_{g_{ij}}; \qquad P^{ab}\equiv \pdlr{L}{g_{ab}}_{R_{ijkl}}
 \label{tpone}
\end{equation}
It is obvious from the above definition that $P^{ab}$ is symmetric and $P^{abcd}$ has the algebraic properties of the curvature tensor:
\begin{equation}
 P^{abcd} = -P^{bacd} = -P^{abdc};\quad  P^{abcd} = P^{cdab}; \quad  P^{a[bcd]} =0
 \label{tpsymm}
\end{equation} 
Curiously enough, the \textit{mere fact that $L$ is a scalar imposes several further relations} among these derivatives which play a crucial role in deriving the \LL\  field equations \cite{tp-aspects}. The most important ones among these relations can be expressed in terms of the tensor
\begin{equation}
\mathcal{R}^{ab} \equiv P^{aijk} \rc bijk                                         
\end{equation} 
which plays the role analogous to Ricci tensor of Einstein's theory. It can be shown (see \cite{tp-aspects}; Appendix \ref{app:aspects-of-eom} summarises these results with proof) that the following relations hold: To begin with, $P^{ab}$ and $\mr^{ab}$ are closely related to each other:
\begin{equation}
 \pdlr{L}{g_{ab}}_{R_{ijkl}} = - 2 \rc bijk \pdlr{L}{R_{aijk}}_{g_{ab}}; \quad \mathrm{i.e.,}\ P^{ab}=-2 \mr^{ab}
 \label{key1}
\end{equation} 
It immediately follows that $\mr^{ab}$ is symmetric; this is by no means an obvious result and does \textit{not} follow from the algebraic properties of curvature tensor. Using \eq{key1}, one can obtain several other relations connecting the partial derivatives when  
 $(g^{ab}, R_{abcd})$ or $(g^{ab}, \rc abcd)$ or $(g^{ab},R^{ab}_{cd})$ are used 
as  independent variables 
to describe the system. 
 If we use the pair $(g^{ab}, R_{abcd})$ as independent variables, we have 
\begin{equation}
 \pdlr{L}{g^{im}}_{R_{abcd}} = - \pdlr{L}{g_{im}}_{R_{abcd}} = 2\mr_{im}
 \label{derdown}
\end{equation} 
On the other hand, if we use the pair $(g^{ab}, \rc abcd)$ as independent variables, one can show that
\begin{equation}
 \pdlr{L}{g^{im}}_{\rc abcd} = \mr_{im}
 \label{deronethree}
\end{equation} 
and, more interestingly, if we use the pair $(g^{ab},R^{ab}_{cd})$ as independent variables, we get
\begin{equation}
 \pdlr{L}{g^{im}}_{R^{ab}_{cd}} = 0
 \label{dertwotwo}
\end{equation} 
Finally, it can be proved that the tensor $\nabla_m \nabla_n P^{amnb}$ is symmetric in $a, b$.

An \textit{intuitive} way of understanding these results is as follows: To construct    scalar \textit{polynomials} of arbitrary degree in the curvature tensor,  using only the curvature tensor $R_{abcd}$ and metric $g^{ij}$ (with index placements as chosen), we need two factors of the metric tensors to contract out the four indices of each curvature tensor in each term. 
So  \eq{derdown} is clearly true if $L$ is a polynomial in $R_{abcd}$.
Any arbitrary scalar function of curvature and metric can be represented as a  (possibly infinite) series expansion in powers of the curvature tensor. Because \eq{derdown} is linear in $L$, if it holds for a  polynomial of arbitrary degree in curvature tensor, then it should hold for arbitrary scalar functions which possess a series expansion. This (intuitive) argument  also  gives the result in \eq{dertwotwo} which states that in scalar polynomials constructed from $R^{ab}_{cd}$ the metric cannot appear and all the contractions must be with Kronecker deltas! This is obvious because, if we use $g^{ab}$ to contract any two lower indices, that will leave two upper indices hanging loose which cannot be contracted out because we do not have $g_{ab}$ available as an independent variable. Once we have \eq{dertwotwo} the other results can be obtained from it quite easily. Of course, the argument based on polynomials does not capture the full generality of our results, which hold for arbitrary Lagrangians that are not necessarily polynomials.

Having obtained these useful identities, 
let us now proceed with the variation of the action treating the Lagrangian as a function of $R_{abcd} $ and $\gu ab$ with the index placements as shown. This leads to the result:
\begin{equation}
\delta A = \delta \int_\mathcal{V} d^Dx\sqrt{-g}\, L = \int_\mathcal{V} d^Dx \, \sqrt{-g} \, E_{ab} \delta g^{ab} + 
\int_\mathcal{V} d^Dx \, \sqrt{-g} \, \nabla_j \delta v^j
\label{deltaa}
\end{equation}
where
\begin{eqnarray}
E_{ab} &\equiv& \frac{1}{\sqrt{-g}}\left( \frac{\partial \sqrt{-g}L}{\partial g^{ab}}\right)_{R_{abcd}} 
- \mr_{ab}
-2\nabla^m\nabla^n P_{amnb}\nonumber\\  
&=&  \mr_{ab} - \frac{1}{2} g_{ab} L -  2 \nabla^m \nabla^n P_{amnb}
\label{eab}
\end{eqnarray} 
and
\begin{equation}
\delta v^j \equiv [2P^{ibjd}(\nabla_b\delta g_{di})-2\delta g_{di}(\nabla_cP^{ijcd})]
\label{genvc}
\end{equation}
In arriving at the second equality in \eq{eab} we have used \eq{derdown}. Note that each of the three terms in $E_{ab}$ in \eq{eab} is individually symmetric in $a,b$.

In order to obtain sensible field equations from $\delta A = 0$, we need to ignore the boundary term involving $\delta v^i$. From \eq{genvc} it is clear that both the variation of the metric and its first derivative  normal to the boundary contributes at the boundary. In other words, we do \textit{not} get a sensible variational principle if we only assume $\delta \gl ab =0$ at the boundary. (This problem, of course, is well known even in Einstein's theory.) To obtain a well defined action principle, we have to add some boundary terms to our action such that the variations of the normal derivatives of the metric are cancelled out. We will say more about it in \sec{sec:2a3} but will assume, for the moment, that the boundary term involving $\delta v^i$ can be ignored or cancelled. In that case, taking the total action to be $A_{tot}=A+A_{matter}$ and defining the energy momentum tensor for matter from the variation of $A_{matter}$ with respect to the metric in the usual manner, our field equations will reduce to $E_{ab} = (1/2) T_{ab}$ where $T_{ab}$ is the matter energy momentum tensor. 

\subsection{\LL action functional}

Since $P^{abcd}$ involves second derivatives of the metric, the term $\nabla^m\nabla^n P_{amnb}$ in \eq{eab} will contain up to fourth derivatives of the metric. Hence, to satisfy our requirement that equations of motion should not have derivatives higher that second, we need to impose the condition:
\begin{equation}
\nabla_a P^{abcd}=0 
\label{eq:divP-condition}
\end{equation}
(It turns out that the seemingly more general condition, $\nabla^m\nabla^n P_{amnb}=0$ does not lead to anything different.) 
Due to symmetries of $P_{abcd}$, this makes $P^{abcd}$ divergence-free in all indices. In this case, the equations of motion become
\begin{equation}
\mr_{ab} - \frac{1}{2} g_{ab} L
=\left(\frac{\partial \sqrt{-g}L}{\partial g^{ab}}\right)_{\rc abcd} 
= \frac{1}{2} T_{ab}
\label{ordder1}
\end{equation} 
In arriving at the first equality we have used \eq{deronethree}.
Clearly, the field equations do not involve more than second derivatives
of the metric but  are, in general, nonlinear functions of the second derivatives (the only exception being the Einstein-Hilbert Lagrangian). Note that the structure of $E_{ab}$ is very similar to that in Einstein's theory with $\mr_{ab}$ playing the role analogous to Ricci tensor. 

Our original task of finding an action functional, leading to equations of motion which contain no derivatives  higher than the second order, now reduces to finding all scalar functions of curvature and metric such that \eq{eq:divP-condition} is satisfied. We shall now describe how one such class of Lagrangians can be constructed but will not bother to  prove that our construction is actually unique --- for which we refer the reader to original literature \cite{ll1,ll2}.

For this purpose, let us consider a general class of Lagrangians of polynomial form in curvature: 
\begin{eqnarray}
L = \sum \limits_{m=0}^{M} L_m\l[(\mathrm{Riem})^m \r]
\end{eqnarray}
where  $L_m$ contains $m$ factors of the Riemann tensor, chosen with index placements $R^{ab}_{cd}$. Our condition in  \eq{dertwotwo}
then implies that the Lagrangian is independent of the metric tensor and all the indices of the curvature tensor $R^{ab}_{cd}$ must be contracted out using Kronecker deltas. All we have to do is to find the correct manner of contracting the indices so as to satisfy \eq{eq:divP-condition}. Ignoring a constant term corresponding to $m=0$, the next simplest case corresponds to $m=1$ which should be a Lagrangian linear in $R^{ab}_{cd}$. The only way of contracting the indices of $R^{ab}_{cd}$  using Kronecker deltas leads to 
\begin{equation}
L_{EH}=\delta^{13}_{24}R^{24}_{13}; \qquad \delta^{13}_{24}=\delta^1_2\delta^3_4-\delta^3_2\delta^1_4
\end{equation} 
where $\delta^{13}_{24}$ is the determinant tensor involving products of two Kronecker deltas and we are using an obvious notation as regards contraction of indices where the numeral $n$ actually stands for an index $a_n$ etc. This clearly leads to Einstein's theory with $L_{EH}\propto R$.

The next order Lagrangian (for $m=2$) will be quadratic in $R^{ab}_{cd}$. This implies that $P^{abcd}$  (defined by \eq{tpone}) will be linear in the curvature tensor. The corresponding Lagrangian can be found by obtaining  the most general fourth rank tensor built from metric and curvature which is (i) linear in curvature and (ii) possesses symmetries of curvature tensor and (iii) divergence-free in all indices. The choice is unique and given by
 \begin{equation}
P^{abcd}=R^{abcd} -  G^{ac}g^{bd}+ G^{bc}g^{ad} + R^{ad}g^{bc} - R^{bd}g^{ac} 
\label{ping}
\end{equation} 
(In four dimensions, this tensor is essentially the double-dual of Riemann.) 
Using this we can find the $m=2$ Lagrangian to be 
\begin{equation}
L=\frac{1}{2}\left(g_{ia}g^{bj}g^{ck}g^{dl}-4g_{ia}g^{bd}g^{ck}g^{jl}
+\delta^c_a\delta^k_ig^{bd}g^{jl}\right)R^i_{\phantom{i}jkl}R^a_{\phantom{a}bcd}
=\frac{1}{2}\left[R^{abcd}R_{abcd}-4R^{ab}R_{ab}+R^2\right]
\end{equation}  
This is the first non-trivial example of a Lagrangian leading to second order equations of motion from a Lagrangian which is nonlinear in the curvature tensor. The Lagrangian above is called Gauss-Bonnet Lagrangian, and a more explicit construction of it is given in Appendix \ref{app:gblag}. The Gauss-Bonnet Lagrangian can also be written in the form:
\begin{equation}
L_{GB}=\delta^{1357}_{2468}R^{24}_{13}R^{68}_{57}
\end{equation} 
where $\delta^{1357}_{2468}$ is again the determinant tensor involving four factors of Kronecker deltas.

The examples of Einstein-Hilbert and Gauss-Bonnet actions
suggest an 
obvious generalization for higher orders that indeed leads to the Lanczos-Lovelock Lagrangian, which --- in  
$D-$dimensions  --- is given by the sum:
\begin{eqnarray}
L &=& \sum_m c_m L_m
\label{eqnx} \\
L_m^{(D)} &=& \frac{1}{16 \pi} \frac{1}{2^m} \delta^{a_1 b_1 \ldots a_m b_m}_{c_1 d_1 \ldots c_m d_m} R^{c_1 d_1}_{~ a_1 b_1} \cdots R^{c_m d_m}_{~ a_m b_m}
\label{eq:ll-action}
\end{eqnarray}
where the tensor appearing in the right hand side  of \eq{eq:ll-action}    is a completely antisymmetric determinant tensor defined  as:
\begin{eqnarray}
\delta^{i a_1 b_1 \ldots a_m b_m}_{j c_1 d_1 \ldots c_m d_m} = {\mathrm{det}} \left[ \begin{array}{c|ccc}
\delta^i_j & \delta^i_{c_1} & \cdots & \delta^i_{d_m} 
\\
\hline
\\
\delta^{a_1}_j &  & & 
\\
\vdots & & \delta^{a_1 b_1 \ldots a_m b_m}_{c_1 d_1 \ldots c_m d_m} &
\\
\delta^{b_m}_j & & &  \vphantom{\bigg{|}} \end{array} 
\right] \;
\end{eqnarray}
for $m \geq 0$.
 The lowest order terms, $m=0,1$ correspond to cosmological constant and the Einstein-Hilbert action respectively, as can be easily seen by expanding the alternating determinant. For $m=1$, $L_1=(16 \pi)^{-1} R$, and the factor of $16 \pi$ in the definition of $L_m$ essentially changes the right hand side of equations of motion from the conventional $8 \pi T_{ab}$ to $(1/2) T_{ab}$. (Getting an {\it explicit} expression for $L_m$'s remains a complicated task since it requires evaluating the determinant of a $2m \times 2m$ matrix; fortunately, we shall never require explicit form of $L_m$'s for the results which we wish to cover in this review!)

The $c_m$'s in \eq{eqnx} are arbitrary coupling constants which are a priori unconstrained. Some restrictions  on $c_m$'s might arise while considering specific solutions in \LL\ theory (such as black holes), or these might be  related to certain parameters of a fundamental (quantum) theory when we consider low energy effective action that contain such \LL\ terms. This second situation arises in string theory  where $c_2$, for e.g., is related to the string scale, $c_2 \propto l_s^2$. For most of our discussion in this review, we shall assume $c_m$'s to be arbitrary.

To complete our argument we now need to prove that, for this Lagrangian, \eq{eq:divP-condition} is indeed satisfied. To do this, note that, being homogeneous in $R_{abcd}$, the $L_m$'s can be written as 
\begin{eqnarray}
L_m &=& \frac{1}{m} \frac{\partial L_m}{\partial R_{abcd}} R_{abcd} = \l( 1/m \r) P^{abcd}_{(m)} R_{abcd}
\end{eqnarray}
 It is obvious that this tensor is same as the tensor $P_{abcd}$ defined earlier, restricted to a particular value of $m$.  The condition in \eq{eq:divP-condition} is therefore ensured if 
\begin{eqnarray}
\nabla_a P^{abcd}_{(m)} = 0 
\end{eqnarray}
This is easily proved using the standard Bianchi identity for the curvature tensor:
\begin{eqnarray}
\nabla_a P^{ab}_{cd\ (m)} &\propto& \nabla_a \delta^{a b \star \star \ldots \bullet \bullet}_{c d \star \star \ldots \bullet \bullet} 
\underbrace{ R^{\star \star}_{\star \star} \cdots R^{\bullet \bullet}_{\bullet \bullet} }_{(m-1) \rm{factors}}
\\
&\propto& \delta^{a b \star \star \ldots \bullet \bullet}_{c d \star \star \ldots \bullet \bullet} 
\nabla_{[a} R^{\star \star}_{\star \star]} \cdots R^{\bullet \bullet}_{\bullet \bullet} 
\\
&=& 0
\end{eqnarray}
where in the second step we have used complete antisymmetry of determinant tensor, and the final equality follows from the Bianchi identity $\nabla_{[a} R^{ij}_{b c]} = 0$. We have also introduced a convenient symbolic notation of using similar symbols (such as $\star$, $\bullet$ etc.) to denote contracted indices. This helps us display relevant expressions (especially in later sections) in a form which conveys same information in a much more transparent manner.
Therefore, the \LL\ Lagrangians certainly would lead to desired equations of motion (not containing higher derivatives of the Riemann tensor). That they are unique in doing so requires some more work, but the demonstration for Gauss-Bonnet given in Appendix \ref{app:gblag} should at least make it plausible.

Just as in the above result, almost all of the unique features of the \LL\ models arise due to (i) the combinatorial properties of the determinant tensor and (ii) the fact that the $L_m$'s are homogeneous polynomials of order $m$ in the Riemann tensor. 
In particular, the \LL\ Lagrangians are \textit{finite} order polynomials; i.e., although the sum in \eq{eqnx} is defined for all $m \in \mathbb{Z}$, only a finite number of  $L_m$'s are non-zero in a given spacetime dimension $D$. This follows essentially from complete antisymmetry of the determinant tensor and can be seen as follows. 
For even $D$, the determinant tensor vanishes if $m > D/2$, whereas for odd $D$, it vanishes if $m>(D-1)/2$. Therefore, for a given $D$, there is a critical value $m=m_c$ such that the action is non-zero only for $L_m$ 's with $m \leq m_c$, where $m_c = [(D-1)/2]$ with $[x]$ denoting the smallest integer not smaller than $x$. The dimension $D=2m$ is called the \textit{critical dimension} for a given $L^D_m$, since $L^{D<2m}_{m}$ vanishes identically.

 The structure of the \LL\ action in $D=2m$ is interesting in its own right, and highlights certain topological significance of the \LL\ actions. This will be discussed in detail in \sec{sec:afincd}. In particular, the critical dimension for Einstein-Hilbert action ($m=1$) is $D=2$ in which $G_{ab}$ identically vanishes; similarly the critical dimension for the Gauss-Bonnet Lagrangian ($m=2$) is $D=4$ and it  does not contribute to the equations of motion in $D=4$. We will say more about critical dimensions in \sec{sec:afincd}.
 
It is possible to write down an alternate form of the \LL\ action, which will  be shown (in \sec{sec:2a4}) to be   related to the  `holographic' structure of these actions. This alternate form is a direct generalization of  the so called $\Gamma\Gamma$ (or Dirac) Lagrangian for Einstein gravity.
We will sketch the basic structure symbolically, and will describe the details while discussing holographic structure of \LL\ actions in \sec{sec:afincd}. Consider any Lagrangian which can be written as $L=Q_{abcd} R^{abcd}$ such that (i) $\nabla_a Q^{abcd}=0$ and (ii) $Q_{abcd}$ has all the algebraic symmetries of Riemann tensor. We know from the above discussion that \LL\ models belong to this class. We can then (symbolically) decompose $L$ as follows
\begin{eqnarray}
L \sim Q R &\sim& Q \l( \partial \Gamma + \Gamma \Gamma \r)
\nn \\
&\sim& Q \Gamma \Gamma + \partial \l( Q \Gamma \r) + \Gamma \nabla Q
\nn \\
&\sim& Q \Gamma \Gamma + \partial \l( Q \Gamma \r)
\label{eq:x2-QGG-form}
 \end{eqnarray}
One immediately sees, in the final expression above, (a generalization of) the structure of the  $\Gamma\Gamma$ Lagrangian familiar from Einstein theory. The surface term in the above split is  important  in studying the thermodynamical properties of horizons in \LL\ models, as we shall show in a later section.  

Before concluding this section we will also comment on the Palatini form of the variational principle for the general Lagrangian. In Einstein theory, the conventional metric variation of the action yield equations of motion which are equivalent to those obtained upon treating metric and connection as independent quantities and varying $g^{ab}$ and $\Gamma^i_{jk}$ independently. The latter variational principle (the so called Palatini variation) actually forces the connection to be metric compatible. For the general Lagrangian $L[g^{ab},R_{abcd}]$, it is easy to show that, when the connection and metric are varied independently, we obtain
\bea
\delta(L_m \g) = \nabla_i \l( 2 \g \; g^{al} P^{ij}_{kl} \delta \hat \Gamma^{k}_{ja} \r) +\g \;  
[\mr_{ab}-\frac{1}{2}g_{ab}L] 
\delta g^{ab} - \nabla_i \l( 2 P^{ij}_{kl} g^{al} \g \r) \delta \hat \Gamma^{k}_{ja} 
\label{pala}
\eea 
where $\hat \Gamma^k_{\phantom{k}ja}$ is an independent connection. If we ignore the total divergence term and set the coefficients of $\delta g^{ab}$
and $\delta \hat \Gamma^{k}_{\phantom{k}ja}$ independently to zero, then we get  the conditions 
\begin{equation}
\mr_{ab}-\frac{1}{2}g_{ab}L = \frac{1}{2}T_{ab}; \qquad
\nabla_i \l( 2 P^{i(j}_{kl} g^{a)l} \g \r)
= 0    
\label{twocond}                                                
 \end{equation} 
It is difficult to make sense of the second equation in general and obtain the correct field equations. But if we restrict ourselves to Lagrangians which satisfy the condition $\nabla_i P^{ij}_{kl}=0$, then the vanishing of the coefficient of $\delta \hat \Gamma^{k}_{\phantom{k}ja}$ requires $\nabla_i \l( g^{al} \g \r)=0$. This in turn can be shown to lead to $\nabla_i \g = 0 = \nabla_i g_{ab}$, leading to the standard relation between metric and connection. The vanishing of the coefficient of $\delta g^{ab}$ now leads to the correct equations of motion. Thus for the \LL\ Lagrangians, the Palatini form of the variation leads to the same field equations, just as in Einstein's theory.

The converse is also true \cite{qemms}. If we assume that  $\nabla_i \g = 0 = \nabla_i g_{ab}$ (which makes connection and metric compatible), then the second condition in \eq{twocond} leads to $\nabla_iP^{ij}_{kl}=0$. That is, 
the compatibility of metric and connection (which is sometimes taken as  a manifestation of the equivalence principle) for the generic Lagrangian of the form $L[g_{ab}, R_{abcd}]$ picks out \LL\ Lagrangians uniquely. (Same conclusion was also reached in \cite{bjb}. See also \cite{nkd-n1}.)

\subsection{Explicit form of the equations of motion}

The equations of motion for the \LL\ model,  given by \eq{ordder1}, can be explicitly computed by first determining the form   of $P^{abcd}$ from the Lagrangian in \eq{eq:ll-action}. This gives, for a generic \LL\ action $L=\sum_m c_m L_m$, the equations:

\begin{eqnarray}
E^a_{b} &=& \sum_{m} {c_m E^{a}_{b(m)}}=\frac{1}{2} \; T^a_{b}
\nonumber
\end{eqnarray}
where
\begin{eqnarray}
E^i_{j(m)} &=& \frac{1}{16 \pi} \frac{m}{2^m} \delta^{a_1 b_1 \ldots a_m b_m}_{j_{~} d_1 \ldots c_m d_m} R^{i d_1}_{a_1 b_1} \cdots R^{c_m d_m}_{a_m b_m} 
- \frac{1}{2} \delta^i_{j} L_m
\nonumber \\
&=&  - \frac{1}{2} \frac{1}{16 \pi} \frac{1}{2^m} \delta^{i a_1 b_1 \ldots a_m b_m}_{j c_1 d_1 \ldots c_m d_m} R^{c_1 d_1}_{a_1 b_1} \cdots 
R^{c_m d_m}_{a_m b_m}
\label{eom-lovelock}
\end{eqnarray}
The equivalence of the two expressions  in \eq{eom-lovelock} can be  established through the identity;
\begin{eqnarray}
\l[ \delta^{i}_{j} \delta^{a_1 b_1 \ldots a_m b_m}_{c_1 d_1 \ldots c_m d_m} - 2 m \; \delta^{i}_{c_1} \delta^{a_1 b_1 \ldots a_m b_m}_{j d_1 \ldots c_m d_m} \r] R^{c_1 d_1}_{a_1 b_1} \cdots R^{c_m d_m}_{a_m b_m} 
= \delta^{i a_1 b_1 \ldots a_m b_m}_{j c_1 d_1 \ldots c_m d_m} R^{c_1 d_1}_{a_1 b_1} \cdots R^{c_m d_m}_{a_m b_m}
\label{eq:alter-tensor}
\end{eqnarray}
which, in turn, is  easily verified by noting that the alternating tensor in the right hand side of Eq.\;(\ref{eq:alter-tensor}) can be written as a determinant:
\begin{eqnarray}
\delta^{i a_1 b_1 \ldots a_m b_m}_{j c_1 d_1 \ldots c_m d_m} = {\mathrm{det}} \left[ \begin{array}{c|ccc}
\delta^i_j & \delta^i_{c_1} & \cdots & \delta^i_{d_m} 
\\
\hline
\\
\delta^{a_1}_j &  & & 
\\
\vdots & & \delta^{a_1 b_1 \ldots a_m b_m}_{c_1 d_1 \ldots c_m d_m} &
\\
\delta^{b_m}_j & & &  \vphantom{\bigg{|}} \end{array} 
\right] \;
\end{eqnarray}
The first term on the left-hand-side of Eq.\;(\ref{eq:alter-tensor}) therefore comes from the multiplication of $\delta^i_j$ with the lower right block of the above matrix. The remaining terms in the determinant can be grouped as 
\begin{eqnarray}
&-& \l[ \delta^i_{c_k} \delta^{a_1 b_1 \;\; \cdots \;\; a_m b_m}_{j c_1 d_1 \cdots d_{k-1} d_{k} \cdots c_m d_m} - \delta^i_{d_k} \delta^{a_1 b_1 \;\; \cdots \;\; a_m b_m}_{j c_1 d_1 \cdots c_{k} c_{k+1} \cdots c_m d_m} \r]
\label{eq:alter-tensor-1}
\end{eqnarray}
where $1 \le k \le m$. Since this whole determinant is multiplied by the product of curvature tensors, only the piece antisymmetric in the pair $\{c_k, d_k\}$ will be picked up, producing a factor of $2$ for each pair $\{c_k, d_k\}$. Further, each of the $m$ such pairs contribute the same amount due to the symmetries of the alternating tensor and the curvature tensor. This gives another factor of $m$, so that the contribution of the remaining terms in the above determinant becomes equal to $2 m$ times the contribution of any particular term, say, the term corresponding to the pair $\{c_1, d_1\}$. Noting the overall minus sign in (\ref{eq:alter-tensor-1}), we obtain the second term on the left-hand-side of Eq.\;(\ref{eq:alter-tensor}), thereby proving the desired equality.

There are three more results which are easily obtained from the explicit form of the equations of motion:
\begin{enumerate}
 \item 
The field equations resulting from a $m$-th order \LL\ Lagrangian, which involves $E^a_b$  in \eq{eom-lovelock}, become vacuous when $D=2m$, the  critical dimension for a given \LL\ term. This is immediately obvious from the second form of \eq{eom-lovelock}, on recalling that the determinant tensor is completely anti-symmetric, and hence the number of values that the $2m+1$ indices of the tensor can take (which of course is the same as the number of spacetime dimensions $D$), must be greater than or equal to $2m+1$. For $D \le 2m$ the $E^a_b$ vanishes identically.

\item Since  $E_{ab}= \mr_{ab} - \frac{1}{2} g_{ab} L$
 arises  from the variation of a generally covariant scalar with
respect to $g^{ab}$, it follows that $\nabla_a E^a_b =0$ which is the analogue of standard Bianchi identity. (The proof proceeds exactly as in the case of standard Bianchi identity; see e.g., page 255 of Ref. \cite{gravitation}).
This in turn implies that
\begin{equation}
\nabla_a \left(P_b^{\phantom{b}kij} \rc akij\right) = \frac{1}{2} \partial_b L
\label{LLbianchi} 
\end{equation} 
One consequence of this result is the following: Suppose we have a theory in which the field equations given by are the projection of the \LL\ field equations on to null vectors. That is, suppose the field equations of the theory are given by:
 \begin{equation}
2 P_b^{\phantom{b}kij} \rc akij n^b n_a = 
T^a_b n^b n_a
\label{nullLL}
\end{equation} 
for \textit{all} null vectors $n_a$ with $T^a_b$ being a covariantly conserved symmetric tensor. This implies the equations
\begin{equation}
2 E^a_b = \left[ 2P_{b}^{\ph{b}ijk}R^{a}_{\ph{a}ijk}-\delta^a_bL\right]=
  T{}_b^a +\Lambda\delta^a_b   
\label{entunc71a}
\end{equation}
where $\Lambda $ is a \textit{constant}. To see this, note that \eq{nullLL}
implies that 
$2P_{b}^{\ph{b}ijk}R^{a}_{\ph{a}ijk} - T{}_b^a = -\lambda \delta{}_b^a$
where $\lambda(x) $ is some scalar.
Rewriting this equation as $2E^a_b + (L+\lambda) \delta^a_b = T^a_b$,
 taking the divergence and using $\nabla_a E^a_b = 0 =  \nabla_a T^a_b$
 we get $\partial_b (L+\lambda) =0$. Therefore, $\lambda = - L + $ constant.
In other words, a theory with null projection of the equations of motion is equivalent to the original theory with a cosmological constant added to it. 
\item
It can be shown that \cite{nkd} 
 for each \LL\ term of order $m$, one can construct a fourth rank tensor $\mathcal R^{(m)}_{abcd}$ from $P^{(m)}_{abcd}$ and $R^{(m)}_{abcd}$, with all the algebraic symmetries of the Riemann tensor and having the property that, the trace of it's Bianchi derivative $\nabla_{[a} \mathcal{R}^{i(m)}_{|j|bc]}$ vanishes, yielding a divergence-free tensor which is the analogue of Einstein tensor for \LL models. Moreover, it can be argued \cite{nkd} 
that \LL\ Lagrangians are the only ones which admit such a description in terms of a Bianchi derivative. It is also worth mentioning that, in \cite{nkd-n2}, it has been proved, using the above mentioned generalized Riemann tensor for pure \LL\ actions, that for any pure \LL\ term of order $m$, the vacuum is trivial with respect to $\mathcal{R}^{i(m)}_{jkl}$ in all odd dimensions $D=2m+1$; i.e., it corresponds to $\mathcal{R}^{i(m)}_{jkl}=0$, which one might call as \textit{Lovelock-flat}. This generalizes the well known fact in Einstein gravity ($m=1$), that the vacuum solutions in $D=3$ are Riemann flat.

The additional characterizations of \LL\ models like the one mentioned above (as well as the one arising from validity of the Palatini variational principle) are not connected (at least in any obvious way) to the original requirement of quasi-linearity of equations of motion from which these models were first obtained.  While it is possible that all such ``uniqueness" results for \LL\ theories are eventually connected to (and derivable from) the topological significance of the \LL\ action in the critical dimension, we do not have at present in the literature a proof of the same. Such a proof might have important implications for several of the results we shall discuss in this review, which hinge on very special algebraic properties of certain tensors encountered in the \LL\ theory.

\item
It also follows from \eq{ordder1} that, for the $m$th  Lanczos-Lovelock Lagrangian, $L_{(m)}$, the trace of the equations of motion is proportional to the Lagrangian:
\begin{equation}
E\equiv g^{ab}E_{ab}=\frac{g^{ab}}{\g}\frac{\partial \sqrt{-g}L_{(m)}}{\partial g^{ab}}=-[(D/2)-m]L_{(m)}
\label{treom}
\end{equation} 
This \textit{off-shell} relation (that is, 
 a relation valid even field equations are not imposed)
 is easy to prove from the fact that we need to introduce $m$ factors of $g^{ab}$  to proceed from $R^a_{\phantom{a}bcd}$ to $R^{ab}_{cd}$ and that $\sqrt{-g}$ is homogeneous function of $g^{ab}$ of degree $-D/2$.
Using this  we get $L_{(m)}=-2E/(D-2m)=-T/(D-2m)$ which allows us to rewrite the field equations of the $m$-th order \LL\ theory in the form:
 \begin{equation}
2P_{b\ (m)}^{\ph{b}ijk}R^{a}_{\ph{a}ijk}=
  T{}_b^a -\delta^a_b\frac{1}{(D-2m)}T   
\label{entunc71z}
\end{equation}
This is similar in structure to 
 $R^a_b=8\pi( T{}_b^a -\delta^a_b(1/2)T)$, to which it reduces to in $D=4$ for the $m=1$ case. Note that this result does \textit{not} hold for the theory obtained by adding together \LL\ terms of different order.
 
\end{enumerate}

\subsection{Diffeomorphism invariance and the Noether current}

In any generally covariant theory, the infinitesimal coordinate transformations $x^a \to x^a + \xi^a(x)$ lead
to the conservation of a current (called the Noether current) that can be obtained as follows:
The variation of the gravitational Lagrangian
resulting from arbitrary variations of $\delta g^{ab}$ can be expressed as
form,
\begin{equation}
\delta(L\sqrt{-g }) =\sqrt{-g }\left( E_{ab} \delta g^{ab} + \nabla_{a}\delta v^a\right). \label{variationL} 
\end{equation}
where the $\delta v^a$ term leads to the boundary contribution.
When the variations in $\delta g^{ab}$  arise due to the diffeomorphism $x^a \rightarrow x^a + \xi^a$ we have, $\delta (L\sqrt{-g} ) = -\sqrt{-g} \nabla_a (L \xi^a)$, with $\delta g^{ab} = (\nabla^a \xi^b + \nabla^b \xi^a)$. Substituting these in \eq{variationL} and using the (generalized) Bianchi identity $
\nabla_a E^{ab} = 0 
$, we obtain the off-shell
conservation law $\nabla_a J^a = 0$, for the current,
\begin{equation}
J^a \equiv \left(2E^{ab} \xi_b+L\xi^a + \delta_{\xi}v^a  \right) 
\label{current}
\end{equation}
where $\delta_{\xi}v^a$ represents the boundary term which arises for the specific variation
of the metric in the form $ \delta g^{ab} = ( \nabla^a \xi^b + \nabla^b \xi^a$). 
It is also convenient to introduce the antisymmetric tensor (called the Noether potential) $J^{ab}$ by $J^a \equiv \nabla_b J^{ab}$.
Using the known expression in \eq{genvc} for $\delta_{\xi}v^a $ in \eq{current}, it is possible to write an explicit expression for the current $J^a$ for any diffeomorphism invariant theory. For the general
class of theories we are considering, the $J^{ab}$ and $J^a$ can be expressed  in the form
\begin{equation}
J^{ab} = 2 P^{abcd} \nabla_c \xi_d - 4 \xi_d \left(\nabla_c P^{abcd}\right)
\label{noedef}
\end{equation} 
\begin{equation}
J^a = -2 \nabla_b \left (P^{adbc} + P^{acbd} \right ) \nabla_c \xi_d + 2 P^{abcd} \nabla_b \nabla_c \xi_d - 4 \xi_d \nabla_b \nabla_c P^{abcd} 
\end{equation} 
where $P_{abcd}\equiv (\partial L/\partial R^{abcd})$. The Lagrangians we are considering here depend, upto this point, on metric and Riemann tensor, but {\it not} on higher derivatives of the curvature tensor. However, the expressions in this section hold even in the latter case, the only difference being that the tensor $P_{abcd}$ will now be defined with additional terms involving terms such as $\partial L/\partial \l(\nabla R\r)$ etc. Algebraically, $P_{abcd}$ simply gives the ``equation of motion" with $R_{abcd}$ treated as the dynamical variable.

In the case of \LL\ models, $\nabla_aP^{abcd}=0$ and  one can obtain from the \eq{genvc} the expression for the boundary term $\delta v^i$ to be: 
\begin{equation}
\delta v^i= -2 P_{k}^{\phantom{k}aji} \; \delta_{\bm \xi} \Gamma^{k}_{\phantom{k}aj}
\end{equation}
where the Lie derivative of the connection (which is a generally covariant tensor) is defined through:
\begin{equation}
\mathcal{L}_{\bm \xi} \Gamma^c_{\phantom{c}ab} = \nabla_a \nabla_b \xi^c - R^c_{\phantom{c} a b m} \xi^m                                                                                                          \end{equation} 
This leads to the  expressions for Noether potential and current given by
\begin{equation}
J^{ab} = 2 P^{abcd} \nabla_c \xi_d; \qquad
J^a =  2 P^{abcd} \nabla_b \nabla_c \xi_d =
\left(2\mr^{ab} \xi_b + \delta_{\xi}v^a\right)
\label{eq:JabLL}
\end{equation}
In this case, we can also write
the current as:
\begin{equation}
 J^i = 2\mr^{ib} \xi_b + \delta_{\xi}v^i=2\mr^{ab} \xi_b -2 P_{k}^{\phantom{k}aji} \; \delta_{\bm \xi} \Gamma^{k}_{\phantom{k}aj}
\label{jgamma}
\end{equation}  
This result shows that,  near any event $\mathcal{P}$ where  $\xi^a$ behaves like an (approximate) Killing vector and satisfies the conditions
$
 \mathcal{L}_{\bm \xi} g_{ab} = \nabla_{( a} \xi_{b)} = 0
$
and
$
 \mathcal{L}_{\bm \xi} \Gamma^c_{\phantom{c}ab}= 0
$
(which a true Killing vector will satisfy everywhere), we get
\begin{equation}
 J^a \equiv 2\mr^{ab} \xi_b. 
 \label{current1}
\end{equation}

Incidentally, the expression for the Noether current can also be obtained from the Palatini variation in \eq{pala}. Assuming that the metric is compatible with the connection and considering variations arising from diffeomorphisms for which $\delta(L_m \g) = - \g \ \nabla_i \l(L \xi^i\r)$, we obtain an expression for the conserved current as:
\bea
\nabla_i \l(
2 E^i_j \xi^j + L \xi^i + 2 g^{al} P^{ij}_{\phantom{ij}kl} \ \delta_{\bm \xi} \Gamma^{k}_{\phantom{k}ja}
\r) = 0
\eea
where $\Gamma^a_{bc}$ is now the metric compatible connection. This yields the same expression as in \eq{jgamma}.

\subsection{Action functional in the critical dimension} \label{sec:afincd}

As we said before, the field equations resulting from a $m$-th order \LL\ Lagrangian become trivial when $D=2m$, known as the  critical dimension for a given \LL\ term.  The \LL\ action itself also vanishes identically for $D<2m$. But for $D=2m$, the action itself does not vanish. This raises the question: What is the structure of \LL\ action itself in the critical dimension? The vacuous nature of field equations requires that the \textit{variation} of the action will be a pure surface term in the critical dimension. But it is not obvious whether the \textit{action itself} is a pure surface term. We discuss these and related issues in this section, beginning by stressing  why such issues are interesting and non-trivial.

Let us begin with $D=2$, which is the critical dimension for $m=1$ \LL\ theory, the Lagrangian of which is simply $R$ leading to  Einstein-Hilbert action. It is trivial to show, using the language of differential forms, that the integrand of the Einstein-Hilbert  action $R \sqrt{-g} $  in $D=2$ can be written as an exterior derivative of a quantity which can be expressed in terms of the \textit{tetrads} (see \eq{R=dOmega} below). But what we are interested in is whether or not  $R \sqrt{-g} $ can be expressed as $\partial_j R^j$ for a doublet of functions ($R^0, R^1$) built from the \textit{metric and its derivatives} without using tetrads in a \textit{general} coordinate system. That is,  the line element is:
\begin{equation}
 ds^2 = A (t,x) dt^2 +2 C(t,x) dx dt + B(t,x)dx^2,
\end{equation} 
(in which we have not imposed any gauge or coordinate conditions) we want to find an explicit expression for  ($R^0, R^1$) in terms of  $A, B, C$ and their derivatives. This is again trivial to accomplish in  the conformally flat gauge with $C=0,A=B$ in which $R\sqrt{-g}=\partial_j\partial^j \ln A$. This has the required form but it will \textit{not} survive if we transform to an arbitrary coordinate system. 

Since we already know  the answer in terms  of differential forms, one might think that there is no room for  controversy about obtaining an explicit expression for $R^j$. However, a paper by Deser and Jackiw \cite{Deser:1995ne} seems to suggest that this is impossible, which led Kiriushcheva and Kuzmin \cite{Kiriushcheva:2006nn} to  conclude that there are fundamental differences between the vierbein formalism and the standard formalism   and, because of those differences, the Einstein-Hilbert Lagrangian being a total derivative in the vierbein formalism does not imply that it is a total derivative in the standard formalism.  
Obviously,   this issue deserves clarification: on one hand, several papers claim (at least implicitly) that the action itself is a total derivative for all \LL\ models in critical dimensions but without concrete proof; on the other hand, there are arguments,  summarized in the articles by Kiriushcheva and Kuzmin \cite{Kiriushcheva:2005kk,Kiriushcheva:2006nn}, that these claims are incorrect.  

These contradictory conclusions originate largely due to lack of an explicit expression for $R^j$, as a function of the metric and its derivatives, satisfying $L_m \sqrt{-g} = \partial_j R^j$.  We will first summarise  some recent results \cite{tp-yale}  establishing that $R^j$  can indeed be found in terms of the metric and its derivatives and giving an explicit expression for $R^j$ for the case of $D=2$. This $R^j$ is not unique since one can always add to it any set of functions $f^j(x)$ which satisfy $\partial_jf^j=0$. There is, in fact,  one natural $R^i$ associated with each unit normalized vector field in $D=2$ and any two of them differ by an $f^j$ with  $\partial_jf^j=0$.

We will also summarize similar results for $D=4$. The \LL\ term, for which  $D=4$ is the critical dimension, is the $m=2$ Gauss-Bonnet term. The question again is whether one can write $L_{\rm GB} \sqrt{-g} = \partial_j R^j$. An expression for $R^j$ was given in \cite{Cherubini:2003nj} based on an expression originally given in \cite{precherubini}. Unfortunately, this expression was obtained through an invalid identification of frame indices with spacetime indices and  turns out to be incorrect. We will describe how correct expressions can be obtained and  outline a proof that any \LL\ Lagrangian in the  critical-dimension can be written as a total derivative of functions of the metric and its derivatives. 

Let us begin with $D=2$. In reference \cite{tp-yale},  three distinct methods for obtaining an explicit expression for $R^j$ in this case are given, of which we discuss the one based on Cartan formalism because it is easy to generalise to an arbitrary \LL\ term, $L_m$. For this purpose, we introduce an orthonormal basis $\omega^a = \omega^a_i dx^i$ (satisfying $g_{ij} = \omega^a_i \omega^b_j \eta_{ab}$) as a basis for the cotangent space. The spin connection, defined by the first Cartan structural equation (see, for e.g., Appendix A of \cite{tp-yale}) must be antisymmetric: $\Omega_{01} = -\Omega_{10}$.  The Einstein-Hilbert two-form can be written (up to an overall sign) as:
\myeq{R=dOmega}{
R \sqrt{-g} d^2x &=  \Theta^{ab} \wedge * (\omega_a \wedge \omega_b) \\
&= d \Omega_{01} d^2x,
}
where $\Theta^{ab} = \left( d \Omega^{ab} + \Omega^a_{~s} \wedge \Omega^{sb} \right)$ is the curvature two-form.  So, to find $R^j$ such that $R \sqrt{-g} = \partial_j R^j$  we only need to calculate explicitly the spin connection $\Omega_{01}$ and take its exterior derivative.  This gives \cite{tp-yale} the result:
\begin{equation} \begin{split} \label{FG}
R^0 &= \frac{1}{\sqrt{-g}} \left[ -\lambda \frac{g_{01}}{g_{00}}g_{00,1} + (1-\lambda)\frac{g_{01}}{g_{11}}g_{11,1} - 2(1-\lambda)g_{01,1} + g_{11,0} \right] \\
R^1 &= \frac{1}{\sqrt{-g}} \left[ -(1-\lambda) \frac{g_{01}}{g_{11}}g_{11,0} + \lambda \frac{g_{01}}{g_{00}}g_{00,0} - 2\lambda g_{01,0} + g_{00,1} \right],
\end{split} \end{equation}
which is an one-parameter family of solutions labelled by a constant $\lambda$.  This parameter can be thought of as describing the gauge freedom available in defining an orthonormal basis for our cotangent space. The $R^j$s for two different values of $\lambda$ are an example of non-uniqueness  mentioned earlier.  One can directly verify that  $f^j\equiv [R^j(\lambda_1)- R^j(\lambda_2)]$ satisfies $\partial_jf^j=0$, as it should. This gauge freedom is entirely unphysical, and one can fix it without  any extra condition on the spacetime or the metric.  The gauge-fixed Cartan formalism is equivalent to the standard formalism for General Relativity. 
An explicit expression for $R^j$  for the $m=2$ (the Gauss-Bonnet term) can also be found in Ref.~\cite{tp-yale}. 

We will now outline a general method for finding $R^j$'s for a generic \LL\ term, referring the reader to \cite{tp-yale} for technical details. There  exists a simple geometrical interpretation for the \LL\ Lagrangians being total derivatives of expressions involving tetrads. The Lagrangian ${L}_m$ is simply the Euler density of a $2m$-dimensional manifold (without a boundary), as can be seen, for example, using the results in the section on index theorems in ref. \cite{eguchi}.  In $2m$ dimensions, ${L}_m$ is a topologically invariant scalar density because it is a closed form and therefore locally exact \cite{Bidal}.  The Chern-Simons form $Q_m$ is defined to quantify this equality: evaluating it between two connections $\Omega$ and $\Omega'$, one obtains ${L}_m(\Omega) - {L}_m(\Omega') = dQ_m(\Omega,\Omega')$. 
In particular, if we restrict ourselves to a local coordinate patch, then we can introduce  a flat connection $\Omega'=0$ and can therefore write ${L}_m = dQ_m(\Omega,0)$.  Hence, the existence of the Chern-Simons form is  a valid proof that ${L}_m$ is an exact differential.  Moreover, this means that $R^j$ can be thought of as providing the local coordinate description of the Chern-Simons form. Once it is clear that we can express ${L}_m$ as an exact differential in the form language, we can proceed as in the two-dimensional case, and compute the set of (non-unique) functions $R^j$ such that $L_m \sqrt{-g}= \partial_j R^j$.  One simply has to define an orthonormal basis $\omega^a$, calculate the spin connections $\Omega^{ab}$, and use them in the expression already known in terms of differential forms. Therefore, $L_m \sqrt{-g}$ can  be written as a total derivative of a set of functions of the metric and its derivatives in any critical dimension.

We conclude this section by showing how $R^j$ takes a very simple form in any spacetime possessing a Killing vector. This is a direct consequence of the results obtained in ref.\cite{tp-sk-holo} and the explicit proof goes as follows:
Suppose the spacetime metric admits a timelike Killing vector $\xi^a$ which we can choose, without loss of generality, to be: $\xi^a = (1,0,\cdots,0)$.
Since $E_{ab}=0$, in the critical dimension, it follows from \eq{eab} in the case of \LL\ models with $\nabla_a P^{abcd}=0$  that 
\begin{equation}
L_m=2P^{0cde} R_{0cde}
\end{equation} 
 We now note that
\begin{equation}
L_m=2P^{0cdb}R_{0cdb} = P^{0cdb}R_{acdb}\xi^a = 2P^{0cdb}\nabla_c \nabla_d \xi_b 
= \frac{1}{\sqrt{-g}}\partial_c \left( 2\sqrt{-g}P^{0cdb}\nabla_d \xi_b \right) 
\end{equation} 
where we used the facts that $P^{abcd}$ is divergence-free  and has the same symmetries as the Riemann tensor.  It follows that $L_m\sqrt{-g}=\partial_c R^c$ 
with $R^c=2\sqrt{-g}P^{0cdb}\nabla_d \xi_b$. Using $\nabla_d \xi_b = \Gamma_{b0d}$,
we find that
\begin{equation} \label{KillingFG} R^j = -2 \sqrt{-g} P_a^{\phantom{a}b j 0} \Gamma_{0 b}^a 
\end{equation}
In this case $R^j$ is actually a component of the Noether potential corresponding to diffeomorphism invariance, giving a direct physical interpretation to the local version of the Chern-Simons form in spacetimes which have symmetries. (For such spacetimes, the Noether current also gives the horizon entropy provided a Killing horizon exists; we shall discuss this further in Sec.~\ref{sec:3b1}.) 

Obviously, similar results hold for spacetimes which have a spacelike Killing vector; if the Killing vector is taken to be, say, $\xi^j=\delta^j_1$ in a coordinate system, we can obtain a similar expression starting from the identity $E^1_1=0$. When the spacetime has more than one Killing vector, we  obtain, in general, more than one choice of $R^j$ which is again an example of the non-uniqueness mentioned earlier. 

\subsection{Boundary terms for the action principle} \label{sec:2a3}

Part of the interest in the \LL\ models arises from the fact  that they admit a 
standard initial value problem, because the equations of motion contain only up to second derivatives of the metric. At the level of the action, this means that one can hope to derive the equations of motion from a well defined variational principle by fixing the metric on given boundaries, without having to fix also the canonical momenta. For this to happen, one must be able to find suitable boundary terms which depend on  the  intrinsic geometry of the boundary, the variation of which can cancel the variation of normal derivatives of the metric on the boundary. Such a prescription is needed even for EH action, and the appropriate term is the well known York-Gibbons-Hawking term, given by the integral of the trace of the extrinsic curvature, $K=h^{\mu \nu} K_{\mu \nu}$, on the boundary \cite{yorkgibbonshawking}. The question arises as to whether one can construct a similar term for a generic \LL\ action. 
The answer is yes, and the boundary term itself is most easily obtained in the Hamiltonian formalism, which gives the canonical momenta the trace of which is proportional to the boundary term. For want of a better name, we shall call the \LL\ boundary terms as GYGH (Generalized York-Gibbons-Hawking). The first such analysis, to the best of our knowledge, seems to have been provided by Teitelboim and Zanelli \cite{teitelboim-zanelli}. 

Although we shall not discuss the details of the Hamiltonian formalism for \LL, it is relevant to mention one aspect of the same. For \LL\ action, the canonical momenta $\pi_{\mu \nu}$ can indeed be found as a function of extrinsic and intrinsic curvature of the boundaries, but the corresponding expression cannot be inverted in closed form to find $K_{\mu \nu}$ in terms of $\pi_{\mu \nu}$ (i.e., one cannot express ``velocities" in terms of ``conjugate momentum"). More importantly, $K_{\mu \nu}$ would generically be multiple valued functions of $\pi_{\mu \nu}$ for arbitrary \LL\ couplings (the $c_m$'s). Hence the Hamiltonian evolution is not unique, and the  generators of diffeomorphisms cannot be expressed in terms of canonical variables. We refer the interested reader to Ref.\cite{teitelboim-zanelli} and Ref.~\cite{deser-canonical} for further details on the Hamiltonian formulation of \LL\ models. Since we will be working with the Lagrangian formalism all along in this review, the expression for boundary terms in terms of $K_{\mu \nu}$ is sufficient for the specific topics we wish to cover in later parts of this review. Hence we shall focus on the same.

In what follows, we give  the boundary term for a timelike boundary. The corresponding expression for a spacelike boundary, which is the case dealt with in Ref.\cite{teitelboim-zanelli}, is similar except for a few (non-trivial) sign changes which one must be careful about. The timelike case turns out to be more relevant for study of quantum black holes since one usually describes the near-horizon physics by replacing the horizon with a {\it timelike} surface -- the stretched horizon -- in the spirit of the membrane paradigm. In fact, the boundary terms given in this section will be used later in Secs.~ \ref{sec:4d4}, \ref{sec:4d5} where we will obtain some important conceptual insights into horizon entropy  essentially by studying the near-horizon structure of these boundary terms. 

For the \LL\ models, the final form of the relevant boundary term can be written as \cite{misko}
\begin{eqnarray}
A_{\rm sur (m)} &=& \int \limits_{\partial M}  \DM^{(D-1)} x  \; \sqrt{-h} \; C_m
\nn \\
C_m &=& 2 m \int \limits_0^1 \DM s \; \delta^{i_1 i_2 i_3... i_{2m-1}}_{j_1 j_2 j_3...
j_{2m-1}} \; 
K^{j_1}_{i_1} \left( \frac{1}{2}R^{j_2 j_3}_{i_2 i_3}-s^2 K^{j_2}_{i_2}
K^{j_3}_{i_3}\right)
\ldots
\left(\frac{1}{2}R^{j_{2m-2} j_{2m-1}}_{i_{2m-2} i_{2m-1}}-s^2
K^{j_{2m-2}}_{i_{2m-2}} K^{j_{2m-1}}_{i_{2m-1}}\right)
\nn \\
\end{eqnarray}
The above form can be further simplified, and the complete action can be written as
\begin{eqnarray}
S = \sum_{m} \Biggl\{ \;
\int \limits_{M} \DM^D x \; \sqrt{-g} \; c_m L_m \; \pm \; 
\underbrace{
\frac{2}{D-2m} \int \limits_{\partial M} \DM^{(D-1)} x \; \sqrt{-h} \; h_{\mu \nu} t^{\mu \nu}_{(m)}
}_{A_{\rm sur (m)}}
\; \Biggl\}
\label{eq:surface-term-defn}
\end{eqnarray}
where
\begin{eqnarray}
8 \pi t^{\nu}_{(m) \mu} &=&  \frac{c_m m!}{2^{m+1}} \, \sum_{s=0}^{m-1} \widetilde{C}_s \, \widetilde{\pi}_{(s) \mu}^{\nu} 
\nonumber \\
\widetilde{\pi}^{\nu}_{(s) \mu} &=& \delta _{\lbrack
\mu \mu_{1}\cdots \mu_{2m-1}]}^{[\nu \nu_{1}\cdots \nu_{2m-1}]}\,{\mathcal R}_{\nu_{1}\nu_{2}}^{\mu_{1}\mu_{2}}\cdots {\mathcal R}_{\nu_{2s-1}\nu_{2s}}^{\mu_{2s-1}\mu_{2s}}%
\,K_{\nu_{2s+1}}^{\mu_{2s+1}}\cdots K_{\nu_{2m-1}}^{\mu_{2m-1}} 
\label{junctionconditionLLOLea}
\end{eqnarray}%
and the coefficients $\widetilde{C}_s$ are given by
\begin{equation}
\widetilde{C}_s = \frac{4^{m-s}}{s!\left( 2m-2s-1\right) !!}
\label{coefficient}
\end{equation}
As the notation already anticipates, the object $t^{\nu}_{(m) \mu}$ defined above can be shown to be the surface stress tensor in \LL\ theory; that is,
\begin{eqnarray}
t^{(m)}_{\mu \nu} = \frac{2}{\sqrt{|h|}} \frac{\delta A_{\rm sur (m)} }{\delta h_{\mu \nu}}
\end{eqnarray}
In the above expressions, $\mathcal R_{\mu \rho \sigma \nu}$ denotes the {\it projection} of full spacetime curvature tensor on the $(D-1)$ hypersurface $\partial M$. It is more convenient to express the stress tensor in terms of the {\it intrinsic} curvature $\hat R_{\mu \rho \sigma \nu}$ of the hypersurface (i.e., the one defined using the induced metric $h_{\mu \nu}$), through the Gauss-Codazzi relation 
\begin{eqnarray}
\hat R_{\mu \rho \sigma \nu} = \mathcal R_{\mu \rho \sigma \nu} - K_{\mu \sigma} K_{\rho \nu} + K_{\mu \nu} K_{\rho \sigma}
\end{eqnarray}
After some simple combinatorial manipulations, the surface stress energy tensor can be rewritten as
\begin{eqnarray}
8 \pi t^{\nu}_{(m) \mu} &=&  \frac{c _{m} m!}{2^{m+1}} \, \sum_{s=0}^{m-1} C_s \, \pi
_{(s) \mu}^{\nu} 
\nonumber \\
\pi^{\nu}_{(s) \mu} &=& \delta _{\lbrack
\mu \mu_{1}\cdots \mu_{2m-1}]}^{[\nu \nu_{1}\cdots \nu_{2m-1}]}\,{\hat R}_{\nu_{1}\nu_{2}}^{\mu_{1}\mu_{2}}\cdots {\hat R}_{\nu_{2s-1}\nu_{2s}}^{\mu_{2s-1}\mu_{2s}}%
\,K_{\nu_{2s+1}}^{\mu_{2s+1}}\cdots K_{\nu_{2m-1}}^{\mu_{2m-1}} 
\label{junctionconditionLL}
\end{eqnarray}%
with 
\begin{equation}
C_s=\sum_{q = s}^{m-1} \frac{ (-2)^{q-s} 4^{m-q} \; \binom{q}{s} }{q!\left( 2m-2q-1\right) !!}
\end{equation}
One can check that for $m=1$ and $m=2$, the above expression reduces to that of the surface stress tensor in Einstein \cite{mtw} and Gauss-Bonnet \cite{junctionGB} theories respectively. 

Although we have quoted the structure of the surface terms, we have not formally shown that these terms indeed make the variational principle well defined. In fact, this is a remarkable feature of the \LL\ actions and the reason why things work out again lies in the topological origin of the \LL\ actions. We refer the reader to \cite{eguchi, myers87} 
for a detailed description of this point. It is however worth giving a sketch of the main argument here.

As we mentioned in the previous section,  in the critical dimension $D=2m$, the \LL\ action $L^{(2m)}_m$ gives Euler character of a manifold $\cal M$ \textit{without} a boundary; in presence of a boundary $\partial \cal M$, one needs to add a specific boundary term, say $-Q^{(2m)}_m$, to obtain the correct Euler characteristic. Since Euler characteristic can not change under smooth variations of a metric with a fixed boundary, the variation of $\int_{\cal M} L^{(2m)}_m - \int_{\partial \cal M} Q^{(2m)}_m$  with respect to normal derivatives of the metric on the boundary must vanish; in other words, the corresponding variations in $L^{(2m)}_m$ and $Q^{(2m)}_m$ must cancel. This {\it cancellation} is independent of spacetime dimensions, and hence the above result  is valid even away from the critical dimension. So the boundary term we need is essentially the term which --- when added to the bulk term --- will make the sum equal to the Euler characteristic in the critical dimension. From this one can show that the form of the surface term  is given by the Chern-Simons form defined on $\partial \cal M$. (The above points are easily understood in Einstein gravity, where the standard York-Gibbons-Hawking term is precisely what is needed to make the full action Euler characteristic in $D=2$, which is the critical dimension for Einstein-Hilbert term corresponding to $m=1$.) This leads to the boundary terms quoted above.

We stress a few important characteristics of the above surface term, some of which will be relevant from the point of view of topics to be discussed in Sec.~\ref{sec:4d5}.

\begin{enumerate}

\item The surface term for a given \LL\ term of order $m$ has an apparent pole at the critical dimension $D=2m$. There is, however, no real pole since $ t^{\nu}_{(m) \mu}$ itself has a zero at $D=2m$ which precisely cancels the pole.  This will also become evident when we give explicit form for $t^{\nu}_{(m) \mu}$ near a static horizon in Sec.~\ref{sec:4d5}. In fact, what happens at $D=2m$ (which is the critical dimension) is straightforward to understand from the fact that, in $D=2m$ the corresponding \LL\ term is the Euler density of the manifold, and the above surface term must be included if the manifold has a boundary. 

\item For a given $m$, it is easy to see  that the boundary term is a polynomial containing {\it odd powers} of $K_{\mu \nu}$ of degree $1$ to $(2m-1)$. For e.g., for the Gauss-Bonnet term ($m=2$), the boundary term will be a linear combination of terms linear and cubic in $K_{\mu \nu}$. 

\item The notation for the boundary tensor anticipates the fact that the {\it jump} in $t_{\mu \nu}$ across a shell, say, can be 
related to stress-energy tensor of the material comprising the shell, to yield the (generalization of) Israel junction conditions for \LL\ models. This fact is useful in the study of  collapse and black hole formation in \LL\ models. Further, the same junction conditions can be used to set-up the membrane paradigm for black hole solutions in \LL\ theory, in the sense that one can replace the details of region hidden by the horizon by endowing the stretched horizon with the above stress tensor. We expect the membrane paradigm to work in this case just as it does in Einstein theory, due to the initial value problem being well defined in \LL\ models. We will discuss this further in section \ref{sec:4d5}.

\end{enumerate}

\subsection{Holographic structure of \LL\ action} \label{sec:2a4}

In a classical theory, action functional is just a device to obtain equations of motion and to encode the symmetries of the system. The actual value of the action functional is of no importance. In contrast, action functional plays a more crucial role in quantum theory: in the semiclassical limit, it gives the phase of the wave functional and in the full description it appears in the path integral. So the off-shell structure of the action functional is  important in quantum theory. In  \LL\ models we have already seen that the action has a peculiar structure and  requires a special treatment if it has to lead to a well-defined variational principle. By and large, this requires careful handling of the surface term in the action functional.

The existence of the surface term brings in another  aspect in the \LL\ models viz. that the  bulk and surface terms of the action are closely related to each other and duplicates the information to certain extent. This is reminiscent of the notion of holography, which represents --- in the most general sense --- a correspondence between bulk and boundary degrees of freedom.  For lack of better terminology we will call the \textit{correspondence between bulk and boundary terms in the action } (rather than in the  degrees of freedom) as ''holography of the action functional''. (In this review, we shall use the term `holography' as explained above and it should not be confused with the usage of this term in e.g., string theory.) In this section, we shall explore this feature and show that \LL\ actions possess a holographic structure which eventually  gets reflected in the  horizon thermodynamics in these models. 

The basic idea we wish to build upon was already expressed, symbolically, in the decomposition of the Lagrangian in \eq{eq:x2-QGG-form}. We will first describe a toy example of higher derivative action in ordinary mechanics to demonstrate the fact that such an action must have a very special structure if it has to yield equations of motion which are still second order in time derivatives \cite{tp-ayan}. It is then not a surprise that the generalization to gravitational actions picks out the \LL\ action uniquely as the one exhibiting holography.
 
\subsubsection{Higher derivative actions yielding second order equations of motion}

Consider  a dynamical variable  $q(t)$ in classical  mechanics described by a Lagrangian $L_q(q,\dot q)$.
 Varying the action obtained from integrating this Lagrangian
in the interval $(t_1,t_2)$ while keeping $q$ fixed at the endpoints, we obtain the Euler Lagrange equations for the system $(\partial L_q/\partial q)=dp/dt$, 
where $p(q,\dot q)\equiv (\partial L_q/\partial \dot q)$. (The subscript $q$ on $L_q$  denotes  the variable that is kept fixed at the end points.) The Lagrangian depends only up to first derivatives of the dynamical variable and the equations of motion are --- in general --- second degree in the time derivative. 

Interestingly, there exists a wide class of Lagrangians $L(\ddot q,\dot q,q)$ which depend on $\ddot q$ as well but still lead to equations of motion which are  only second order in time. We will now   analyse this class which will generalize to the holographic actions in field theory.

To do this, let consider the following question: We want to modify the Lagrangian $L_q$ such that  the same equations of motion are obtained when  --- instead of fixing $q$ at the end points --- we  fix some other (given) function
 $C(q,\dot q)$  at the end points. This is easily done by modifying the Lagrangian by adding a term
$-df(q,\dot q)/dt$ which depends on $\dot q$ as well. (The minus sign is  for future convenience.) 
The new Lagrangian is:
\begin{equation}
L_C(q,\dot q,\ddot q)=L_q(q,\dot q)-\frac{df(q,\dot q)}{dt}
\end{equation}
We want  $L_C$ to lead to the same equations of motion as $L_q$, when some given function
 $C(q,\dot q)$  is held fixed at the end points. We assume $L_q$ and $C$ are given and  that we need to find $f$. The standard variation now gives
 \begin{equation}
\delta A_C =\int_{t_1}^{t_2} dt\left[\left(\frac{\partial L}{\partial q}\right)-\frac{dp}{dt}\right]\delta q
-\int_{t_1}^{t_2} dt \frac{d}{dt}\left[\delta f -p\delta q \right]
\label{elf}
\end{equation} 
 At the end-points, we invert the relation $C=C(q,\dot q)$ to determine $\dot q=\dot q(q,C)$ and express $p(q,\dot q)$ in terms of $(q,C)$ obtaining the function $p=p(q,C)$. We then treat $f$ as a function of $q$ and $C$, so that the variation of the action  becomes:
\begin{eqnarray}
\delta A_C&=&\int_{t_1}^{t_2} dt\left[\left(\frac{\partial L}{\partial q}\right)-\frac{dp}{dt}\right]\delta q+
\left[p(q,C)-\left(\frac{\partial f}{\partial q}\right)_C\right]\delta q\Big|_{t_1}^{t_2}  -\left(\frac{\partial f}{\partial C}\right)_q\delta C\Big|_{t_1}^{t_2} \nonumber\\
&=&\int_{t_1}^{t_2} dt\left[\left(\frac{\partial L}{\partial q}\right)-\frac{dp}{dt}\right]\delta q+
\left[p(q,C)-\left(\frac{\partial f}{\partial q}\right)_C\right]\delta q\Big|_{t_1}^{t_2} 
\label{varaction}
\end{eqnarray}
since $\delta C=0$ at the end points by assumption. To obtain the same Euler-Lagrange equations,
the second term in \eq{varaction} should vanish for all $\delta q$. This fixes the form of $f$ to be:
\begin{equation}
f(q,C)=\int p(q,C)dq +F(C)
\label{f}
\end{equation}
where the integration is with constant $C$ and $F$ is an arbitrary function. 

Thus, given a Lagrangian $L_q(q,\dot q)$ which leads to certain equations of motion when $q$ is held fixed, one can construct a \emph{family} of Lagrangians $L_C(q,\dot q,\ddot q)$ that will lead to the {\it same} equations of motion when an arbitrary function $C(q,\dot q)$ is kept fixed at the end points. \textit{This family is remarkable in the sense that though $L_C$ will now also involve $\ddot q$},  the equations of motion are still of second order in $q$ because of the  structure of the Lagrangian. (The results obtained above can also be interpreted in terms of canonical transformations etc. which we purposely avoid since we want to stay within the Lagrangian framework). So, even though a \textit{general} Lagrangian which depends on $\ddot q$ will lead to equations of higher order, there are \textit{special} Lagrangians  which will not. The analysis extends directly to a multicomponent field $q_A(x^a)$ in a spacetime with coordinates $x^a$ where $A$ collectively denotes the tensor indices \cite{tp-ayan}. 

When one considers the action just as a tool  to obtain the field equations, the above procedure can be used with any $C$. But once the dynamical variables in the theory have been identified, there are two natural boundary conditions which can be imposed on the system. The first one holds $q$ fixed at the boundary and the second one keeps the canonical momenta $p$ fixed at the boundary. In the second case, $C=p$ and Eq.(\ref{f}) gives $f=qp=q(\partial L/\partial \dot q)$ leading to the Lagrangian:
\begin{equation}
L_p = L_q - \frac{d}{ dt} \left( q\frac{\partial L_q }{ \partial\dot q} \right) ~.
\label{lbtp}
\end{equation}
This Lagrangian $L_p$ will lead to the same equations of motion when $p(q,\dot q)=(\partial L/\partial \dot q)$ is held
fixed at the end points.
In  field theory, with multicomponent field $q_A(x^a)$ with canonical momenta $\pi^i_A=[\partial L/\partial( \partial_i q^A)]$, when we fix $C_A^a=\pi^a_A$ at the boundary, the new Lagrangian we need to  use is:
\begin{equation}
L_p(\partial^2 q_A,\partial q_A,q_A)=L_q(q_A,\partial q_A)-\partial_i \left[q_A\left(\frac{\partial L_q}{\partial (\partial_i q^A)}\right)\right]
\equiv L_{\rm bulk} + L_{\rm sur}
\end{equation}
We will now summarise \cite{tp-ayan} the implications of the above results:

\begin{enumerate}
\item These results have a natural interpretation in quantum theory: A path integral defined with $L_p$
will give the transition amplitude in momentum space $G(p_2, t_2;p_1, t_1)$, just as a path integral with $L_q$ leads to the transition amplitude in coordinate space $K \left( q_2,t_2;q_1,t_1 \right)$. 
 
\item When the momenta are held fixed, the surface term has, what we shall call,  the ``$d(qp)$" structure. It is obvious from this relation that the surface and bulk terms in the  action are closely related and the form of the surface term will put some constraints on  $L_{\rm bulk} =L_q$. 

\item It turns out that the Einstein-Hilbert action (and more generally the \LL\ action) has precisely this form, which is  evident when the action is written in the $\Gamma\Gamma$ form. {\it This is the key to the holography in \LL\ action functionals which we now describe.} 
\end{enumerate}

\subsubsection{Holography of gravitational action functionals} \label{sec:2.7.2}

Let us first discuss the Einstein-Hilbert action, rewriting the  well known $\Gamma\Gamma$ form of the action in a manner which generalizes  to the \LL\ actions. The  Einstein-Hilbert action can be written using a tensor $Q_{abcd} \equiv (1/2) \l( g_{ac} g_{bd} - g_{ad} g_{bc} \r)$ as:
 \begin{eqnarray}
\sqrt{-g} L_{\rm EH} 
&=& Q^{abcd} R_{abcd}
\nn \\
&=& 2\partial_c\left[\sqrt{-g}Q_a^{\phantom{a}bcd}\Gamma^a_{bd}\right]
+2\sqrt{-g}Q_a^{\phantom{a}bcd}\Gamma^a_{dk}\Gamma^k_{bc}
\nn \\
&\equiv& L_{\rm sur} + L_{\rm bulk} 
\label{gensq}
 \end{eqnarray} 
with
\begin{equation}
L_{\rm bulk}=2\sqrt{-g}Q_a^{\phantom{a}bcd}\Gamma^a_{dk}\Gamma^k_{bc} \;\; ;\qquad
L_{\rm sur}=2\partial_c\left[\sqrt{-g}Q_a^{\phantom{a}bcd}\Gamma^a_{bd}\right]
\equiv\partial_c\left[\sqrt{-g}V^c\right]
\label{sep}
\end{equation}  
where the last equality defines the $D$-component non-tensor $V^c$. As is well known, one can obtain Einstein's equations varying only $L_{bulk}$  keeping $g_{ab}$ fixed at the boundary.

The non-trivial aspect relevant for holography  is a specific relation \cite{tpreview} between $L_{\rm bulk}$ and $L_{\rm sur}$ allowing $L_{\rm sur}$ to be determined completely by $L_{\rm bulk}$. Using $g_{ab}$ and $\partial_cg_{ab}$ as independent variables in $L_{\rm bulk}$ one can prove that:
 \begin{equation}
L_{\rm sur}=-\frac{1}{[(D/2)-1]} 
\underbrace{
\partial_i \left(g_{ab}\frac{\partial L_{\rm bulk}}{\partial (\partial_ig_{ab})}\right)
}_{``\DM(qp)"}
\label{holorel}
\end{equation}
The connection with the   discussion in the last subsection is obvious: the ``$d(qp)$'' structure of $L_{\rm sur}$  suggests that $L_{\rm EH}$ is obtained from $L_{\rm bulk}$ by a transformation from coordinate space to momentum space. The action based on $L_{bulk}$  is the co-ordinate space action and leads to a well-defined variational principle if the metric is fixed at the boundary. On the other hand, the usual Einstein-Hilbert action  
($R\sqrt{-g}=L_{\rm bulk}+L_{\rm sur}$) is actually a \textit{momentum-space} action and leads to well-defined variational principle if the  momenta are fixed at the boundary. This is very transparent if we use $f^{ab}\equiv\sqrt{-g}g^{ab}$ as the dynamical variables with the momenta being 
\begin{equation}
N^i_{jk}\equiv\frac{\partial (\sqrt{-g}L_{bulk})}{\partial(\partial_if^{jk})}
=-[\Gamma^i_{jk}-\frac{1}{2}(\delta^i_j\Gamma^a_{ka}+ \delta^i_k\Gamma^a_{ja})]. 
\label{defN}
\end{equation}  
The equations of motion will follow from $\delta A_{EH}=0$ if we fix the momenta $N^i_{jk}$ on the boundary. 
In fact, if we decide \textit{not} to add any surface term to Einstein-Hilbert action, then we can still obtain \cite{TPstructuralasp} the field equations if we demand:
\begin{equation}
\delta A_{EH} = -\int_{\partial\mathcal{V}}d^3x\sqrt{h}n_i g^{jk}\delta N^i_{jk}
\end{equation} 
instead of the usual $\delta A_{EH} =0$.

Similar results continue to hold for   \LL\ actions, which we will now describe.
Taking a cue from Einstein's theory, consider a Lagrangian of the form:  
\begin{equation}
\sqrt{-g}L=2\partial_c\left[\sqrt{-g}Q_a^{\phantom{a}bcd}\Gamma^a_{bd}\right]
+2\sqrt{-g}Q_a^{\phantom{a}bcd}\Gamma^a_{dk}\Gamma^k_{bc}\equiv L_{\rm sur} + L_{\rm bulk} 
\label{gensq1}
 \end{equation}
where $Q^{abcd}$ is a tensor having  the symmetries of curvature tensor. By straight forward algebra, one can prove  \cite{tp-ayan} the following identity:
\begin{equation}
\sqrt{-g}L=2\partial_c\left[\sqrt{-g}Q_a^{\phantom{a}bcd}\Gamma^a_{bd}\right]
+2\sqrt{-g}Q_a^{\phantom{a}bcd}\Gamma^a_{dj}\Gamma^j_{bc}
 = \sqrt{-g}Q_a^{\phantom{a}bcd}R^a_{\phantom{a}bcd}+2\sqrt{-g}\Gamma^a_{bd}\nabla_cQ_a^{\phantom{a}bcd}
 \label{simple}
\end{equation} 
The general covariance of $L$ requires the condition $\nabla_cQ_a^{\phantom{a}bcd}=0$. Then we get:
\begin{equation}
\sqrt{-g}L=\sqrt{-g}Q_a^{\phantom{a}bcd}R^a_{\phantom{a}bcd}; \qquad \nabla_cQ_a^{\phantom{a}bcd}=0
\label{basicl}
\end{equation}  
We have, therefore, once again arrived at the \LL\ models in a very different context, both mathematically and conceptually. The final step of establishing {\it holography} of such actions  requires some additional care particularly since, unlike in the case of Einstein theory, $L_{\rm bulk}$ will in general depend on second derivatives of the metric. For \LL\ models, it is possible to show that \cite{tp-ayan} the following generalization of holography exists:
\begin{equation}
[(D/2) - m]L_{sur} =-\partial_i \Biggl\{ 
\left[ g_{ab} \frac{\delta }{\delta (\partial_i g_{ab})}
  +\partial_jg_{ab} \frac{\partial}{\partial (\partial_i \partial_jg_{ab})}
  \right] L_{bulk}
  \Biggl\}
\label{result1} 
\end{equation} 
where the {\it Eulerian derivative} $\delta$ is defined as
\begin{equation}
\frac{\delta K[\phi,\partial_i\phi,...]}{\delta\phi}=
\frac{\partial K[\phi,\partial_i\phi,...]}{\partial\phi}-
\partial_a \left[\frac{\partial K[\phi,\partial_i\phi,...]}{\partial(\partial_a\phi)}\right]+\cdots
\end{equation}
Note that the second term on right hand side of \eq{result1} vanishes for $m=1$ \LL\ term (which is the Einstein theory). The proof of above relation, as well as some other forms of holography for the \LL\ class of actions (which arise by choosing the Christoffel symbols in place of first derivatives of the metric as independent variables), can be found in Ref.~\cite{tp-ayan}.

\subsection{Black hole and cosmological solutions of \LL\ models}

There is a vast literature on known solutions for gravitational theories based on the \LL\ Lagrangians, and even a brief discussion of these would form a topic for a new review. We shall therefore use this section just to redirect the interested reader to relevant literature and reviews where more details about specific solutions, and more importantly, additional references, can be found. 

As far as black hole solutions of the theory are considered, one of the very first solution to have appeared (to the best of our knowledge), is the so called Boulware-Deser solution \cite{boulware-deser} which is essentially a spherically symmetric black hole solution of $5$ dimensional Einstein-Gauss-Bonnet (EGB) action. The metric for this solution is given by
\bea
\DM s^2 &=& -f(r) \DM t^2 + \frac{1}{f(r)} \DM r^2 + r^2 \DM \Omega^2_3
\nn \\
f(r) &=& 1 + \frac{r^2}{4 \alpha} \Biggl[ 1 \pm \sqrt{ 1 + \frac{16 \alpha m}{r^4} + \frac{4 \alpha \Lambda}{3} } \Biggl] 
\eea 
where $\DM \Omega^2_3$ is the metric on a unit $3$-sphere, $\Lambda$ is the cosmological constant, and $\alpha$ is the Gauss-Bonnet coupling constant normalised so that the total lagrangian takes the form: $L \propto -2 \Lambda + R + \alpha \l( R^2 - 4 R_{ab}R^{ab} + R_{abcd}R^{abcd} \r)$. The parameter $m$ is an integration constant which is proportional to the ``mass" of the solution; its exact form with all the constants, can be found, for example, in \cite{boulware-deser}. The $\pm$ is a consequence of the fact that field equations are quadratic in $f(r)$. There exists an extensive literature on properties of this solution, as well as its modification upon addition of the electric charge $Q$. An immediate extension including dilations was also given by Boulware and Deser in \cite{boulware-deser-dilations}. 

There also exists considerable literature on study of Birkhoff's theorem in EGB theory, the most relevant of which are 
listed in \cite{birkhoff-egb}. Although the analysis leading to the Boulware-Deser solution presented 
in \cite{boulware-deser} was motivated by appearance of EGB term in low energy expansion of super-symmetric 
string theory, it was followed by extensive work on finding such solutions for generic \LL\ models treated as classical models of higher dimensional gravity in their own right. A nice review of such solutions, and their various asymptotic properties (which are of some interest also for the AdS-CFT correspondence), can be found in Refs.~\cite{myers98} and \cite{ggbh} (see also \cite{cai-ll-sols} for work generalizing known solutions in Einstein theory to \LL\ ). A systematic study of {\it cosmological solutions} in the \LL\ theory can be found in \cite{deruelle-cosmosols}, which deals in particular with maximally symmetric space-times, Robertson-Walker universes, and product manifolds of symmetric subspaces.

\section{\LL\ models: Horizon thermodynamics}\label{sec:llhortherm}

One fascinating aspect of general relativity, which is still not completely understood conceptually, is the association of thermodynamic variables with horizons. Originally, these results were obtained in the context of black hole physics \cite{Bekenstein:1973ur,Hawking:1974rv} but it is now clear that all null surfaces possess these attributes in an observer-dependent manner \cite{daviesunruh}. It is important to understand whether this is a special feature of Einstein's theory or whether all \LL\ models share this intriguing connection. It turns out, as we shall describe in this and ensuing sections, that the latter is the case. 

As we said before, one 
key  utility of the \LL\ models is in their power to discriminate between widely differing conceptual explanations for known results in Einstein gravity.
So the study of horizon thermodynamics in \LL\ models will allow us to distinguish those aspects which transcend Einstein's theory and will lead to a more promising and general route to understanding these aspects.
In this spirit, there has been an extensive study of thermodynamics of black hole solutions in particular and horizons in general in the \LL\ theory
which we now turn to. Since
 it is impossible to do justice to all the literature  in this review, we will be selective in our discussion. A good  introduction to this topic can be found in the (possibly the very first) paper on the topic by Myers and Simon \cite{myers-simons}. There also has been considerable work on analysing various specific thermodynamic aspects of \LL black hole solutions (of particular interest are solutions with a positive or negative cosmological constant; see, for e.g., \cite{ds-sols01}), all of which reveal special features arising essentially due to the structure of the \LL actions. A Smarr like expression for static, asymptotically AdS black holes in \LL gravity has also been established in \cite{kastor-smarr}. However, in what follows, we shall focus on generic aspects of horizons in \LL theory, rather than properties of specific solutions of these theories.

\subsection{Horizon entropy in \LL\ theory} \label{sec:3b1}

The results from the  study of quantum field theory in a given background spacetime does not depend  on the action governing the gravitational dynamics of the background. In particular,  results \cite{daviesunruh} such as the Davies-Unruh effect etc continue to hold in a given metric, independent of the field equations of the theory. In fact, the notion of horizon temperature only
 involves  the near horizon geometric properties of the spacetime metric; therefore,  the notion of temperature for a horizon remains the same independent of the theory of gravity. 

The same is not true for the horizon entropy, though. Algebraically, this is easy to understand in the context of a simple black hole solution in \LL\ theory characterized by a single parameter corresponding to mass $M$ of the black hole (identified in a suitable way, for example, by looking at the asymptotics of the solution). Then, the entropy can be identified by integrating $dM/T(M)$, for which one would need an explicit form for $T(M)$ This depends on the  equations of motion of the theory. 

There exists a general  way of identifying horizon entropy in  higher derivative theories,  given by Wald \cite{wald} and Visser \cite{Visser:1993qa,Visser:1993nu}, which makes the dependence of entropy on gravitational action very evident. In this approach, horizon entropy is identified as a surface integral over the horizon of a $(D-2)$-form corresponding to the Noether charge arising from diffeomorphism invariance of the theory. The basic idea is to find the Noether current $\bm J$ of diffeomorphism invariance (which is a closed $(D-1)$ form), and use the associated Noether potential $\bm Q$ defined by $\bm J= \DM \bm Q$ to define black hole entropy. 
By considering variations (not necessarily stationary) of a stationary background spacetime possessing a Killing horizon, Wald  argued that the first law of black hole dynamics is consistent with the identification of the following expression as the horizon entropy:
\begin{equation}
S_{\rm Noether} =  \beta \int \DM^{D-1}\Sigma_{a} J^{a}= \frac{1}{2}\beta \int \DM^{D-2}\Sigma_{ab} J^{ab}
\label{noetherint}
\end{equation} 
where $\kappa$ is the surface gravity, $\beta^{-1}=2\pi/\kappa$ is the temperature of the horizon and $J^a,J^{ab}$ are the Noether current and the Noether potential.
 In the final expression in \eq{noetherint} the integral is over any $(D-2)$ dimensional spacelike cross-section of the Killing horizon on which the norm of $\xi^a$ vanishes.
For example, in the  case of Einstein gravity, \eq{noedef} reduces to
\begin{equation}
J^{ab} = \frac{1}{16\pi} \left( \nabla^a \xi^b - \nabla^b \xi^a\right)
\end{equation}
where $\xi^a$ is the Killing vector generating the horizon. Writing the area element as $\DM \Sigma_{ab} = (l_a\xi_b-l_b\xi_a)\sqrt{\sigma}d^{D-2}x$ where $l_a$ is an auxiliary vector field satisfying $l_a\xi^a=-1$, the integral in \eq{noetherint} reduces to
\begin{equation}
 S_{\rm Noether} = -\beta\frac{1}{8\pi}\int \; \DM^{D-2}x \; \sqrt{\sigma}
 (l_a\xi_b)\nabla^b \xi^a
\; = \; \beta \frac{1}{8\pi} \kappa \int \DM^{D-2}x \sqrt{\sigma} 
\; = \; \frac{1}{4}A_H
\end{equation} 
where $A_H$ is the horizon area. (We have used the relations $ \xi^a \nabla_a \xi^b = \kappa \xi^b$ and $\beta \kappa=2\pi$.) This, of course, agrees with the Bekenstein-Hawking entropy.

In what follows, we first give a quick derivation (in component notation) of the arguments of Wald as to why the definition of entropy based on Noether potential makes sense for stationary Killing horizons.   
It turns out that the Noether current $J^a$ and it's associated potential $J^{ab}$ (or more accurately, their {\it off-shell} versions) characterize most of the results related to thermodynamic aspects of gravity, and provide the most natural substitute for the equation of motion tensor $E_{ab}$ in attempts to interpret gravitational dynamics as an emergent phenomenon.
So this description is of importance in our later discussions on emergent paradigm.

Let us now sketch the  argument leading to an expression for black hole entropy in a general gravitational theory with the Lagrangian $L[g_{ab}, R_{abcd}]$. The main input needed is the definition of covariant version of Hamiltonian equations of motion, which can be given in terms of a symplectic $(D-1)$ form $ \omega^a$, which is defined, in component notation, as
\bea
\omega^a \equiv \delta \l( \sqrt{-g} \; \delta_{\bm \xi} v^a \r) - \delta_{\bm \xi} \l( \sqrt{-g} \; \delta v^a \r)
\label{eq:symplectic-form}
\eea
where  $\delta_{\bm \xi}$ is the Lie variation with respect to a vector field $\bm \xi$, and $\delta_{\bm \xi} v^a$ is defined from the variation of the Lagrangian under a diffeomorphism:
\bea
\delta \l( L \sqrt{-g} \r) &=& \sqrt{-g} E_{ab} \delta g^{ab} + \sqrt{-g} \nabla_a \l( \delta_{\bm \xi} v^a \r)
\eea
In terms of the above symplectic form, covariant Hamilton's equations take the form (note that $\omega^a$ is a tensor density of
weight unity)
\bea
\delta H[\bm \xi] = \int \limits_{\mathcal C} d\Sigma_a\ \omega^a / \sqrt{-g}
\label{eq:cov-hamilton-eq}
\eea
where $\mathcal C$ is a Cauchy surface. 

A rigorous justification for this can be found in \cite{lee-wald}, which studies the covariant phase space formulation for field theories. We here take a digression to simply motivate the above expression in analogy with point particle mechanics. In this case, the symplectic form is $\bm \Theta = \bm \DM p \wedge \bm \DM q$, in terms of which Hamilton's equations can be written as
\bea
\bm \DM H &=& \bm \Theta(\ldots,\partial_t) 
\nn \\
&=& \dot{q}(q,p) \; \bm \DM p - \dot{p}(q,p) \; \bm \DM q
\eea 
The Hamiltonian $H$ here generates time translations (hence the vector $\partial_t$ in the 2nd slot of $\bm \Theta$), while the $``\DM"$'s are arbitrary variations whose coefficients characterize the $(q,p)$ dependence of the Hamiltonian. One may re-write the above expression as follows.
\bea
\bm \DM H &=& \dot{q} \; \bm \DM p - \dot{p} \; \bm \DM q
\nn \\
&=& \bm \DM (p \dot{q}) - p \bm \DM \dot{q} - \dot{p} \bm \DM q
\nn \\
&=& \bm \DM (p \DM_t q) - \DM_t \l( p \bm \DM q \r)
\eea
Now, the surface term in point particle action is $v(\DM q)=p \DM q$. Hence we can write 
\bea
\bm \DM H = \bm \DM \l[ v(\dot{q}) \r] - \DM_t \l[ v(\DM q) \r]
\eea
which has the same structure as \eq{eq:cov-hamilton-eq} with $\omega^a$ given in \eq{eq:symplectic-form}. (Similar conclusion would be reached if the surface term is $q \DM p$, with the roles of $q$ and $p$ interchanged.)

Returning now to the main discussion, we see that $\omega^a$ can be written as:
\bea
\omega^a &=& \delta \l( \sqrt{-g} \; \delta_{\bm \xi} v^a \r) - \delta_{\bm \xi} \l( \sqrt{-g} \; \delta v^a \r)
\nn \\
&=& \sqrt{-g} \; \Biggl[ \delta (\delta_{\bm \xi} v^a) -  \mathcal L_{\bm \xi} (\delta v^a) \Biggl] + 
(\delta_{\bm \xi} v^a) (\delta \sqrt{-g}) - (\delta v^a) (\delta_{\bm \xi} \sqrt{-g})
\eea
This allows us to relate $\omega^a$ to the Noether
 current $J^a$,  obtained from diffeomorphism invariance of the Lagrangian, as given by \eq{current} which is conserved off-shell. If we omit from it the term involving $E_{ab}$ we get the expression:
\bea
J^a[\bm \xi] &=& \delta_{\bm \xi} v^a - L \xi^a
\eea
which is conserved \textit{on-shell,} i.e., when the equations of motion are satisfied:
\bea 
\nabla_a J^a &=& \nabla_a \l(\delta_{\bm \xi} v^a \r) - \nabla_a (L \xi^a)
\nn \\
&=& \frac{1}{\sqrt{-g}} \mathcal L_{\bm \xi} \l( L \sqrt{-g} \r) - E_{ab} \delta g^{ab} - \nabla_a (L \xi^a)
\nn \\
&=& - E_{ab} \delta g^{ab} 
\eea
since $\mathcal L_{\bm \xi} \l( L \sqrt{-g} \r) = \sqrt{-g} \nabla_a (L \xi^a)$. Therefore, $\nabla_a J^a=0$ when $E_{ab}=0$. We will use the definition of $J^a$ in the expression for $\omega^a$
and also utilize the facts: (i) $\delta_{\bm \xi} v^a = J^a[\bm \xi] + L \xi^a$, (ii) $\delta (\sqrt{-g} \; L) = \sqrt{-g} \; \nabla_a \delta v^a$ on-shell, (iii) $\delta_{\bm \xi} \sqrt{-g} = \sqrt{-g} \nabla_a \xi^a$, and (iv)  $\delta$ represents arbitrary {\it field} variations, and therefore do not act on $\bm \xi$. Straightforward algebra then leads to the result:
\bea
\omega^a = \delta \l( J^a[\bm \xi] \sqrt{-g} \r) \; + \; 2 \sqrt{-g} \; \nabla_b \l( \xi^{[a} \delta v^{b]} \r)
\eea
which can be substituted in \eq{eq:cov-hamilton-eq} to obtain
\bea
\delta H[\bm \xi] = \delta \int \limits_{\mathcal C} d\Sigma_a\ J^a + \int \limits_{\mathcal C} d\Sigma_a\  2 \nabla_b \l( \xi^{[a} \delta v^{b]} \r) 
\eea
(The steps leading to the above relation are best understood by doing a $3+1$ decomposition w.r.t. normal $n_a=\pm N \partial_a n$ to the hypersurface $\mathcal C$, with $\pm$ corresponding to spacelike/timelike surface respectively, and using $
\sqrt{-g} = N \sqrt{|h|}$ with $h_{\mu \nu}$ being the induced metric on $\mathcal C$.) From the above, one can already note that the flux of Noether current is almost the same as the Hamiltonian, except for a surface term. Further, if the variations $\delta g_{ab}$ satisfy the linearised equations of motion, then the perturbed current is also conserved, and one therefore has 
$\delta [J^a \sqrt{-g}] = \partial_b [ \delta (J^{ab} \sqrt{-g}) ]$. Repeating the steps leading to the preceding expression, this time in terms of $J^{ab}$, leads to
\bea
\delta H[\bm \xi] &=& \delta \int \limits_{\mathcal C} d\Sigma_a\ \nabla_b J^{ab} + 
\int \limits_{\mathcal S_{\infty}} d\Sigma_{ab} \; \xi^{[a} \delta v^{b]}  
\nn \\
&=& - \frac{1}{2} \ \delta \int \limits_{\mathcal S_{\rm bif}} d\Sigma_{ab} \; J^{ab} + 
\frac{1}{2} \ \delta \int \limits_{\mathcal S_{\infty}} d\Sigma_{ab} \; J^{ab} + 
\int \limits_{\mathcal S_{\infty}} d\Sigma_{ab} \; \xi^{[a} \delta v^{b]}  
\eea
where $\mathcal S_{\infty}$ is the $(D-2)$ dimensional boundary of $\mathcal C$ at asymptotic infinity. The inner boundary of $\mathcal C$ is taken to be the bifurcation surface $\mathcal B$, where $\xi^a=0$ (hence this surface does not contribute as far as the $\xi^{[a} \delta v^{b]}$ term is concerned). It is clear from the second expression above that, for a meaningful Hamiltonian $H$ to exist, the last integral must be expressible as a variation of a $(D-1)$ form, say $\bm B$. Assuming this is so, then, as argued in \cite{wald}, the combined contribution to $H$ of this term and the second term involving $J^{ab}$ above, at asymptotic infinity $S_{\infty}$, gives the numerical value of the ADM Hamiltonian when the vector field $\bm \xi$ is chosen appropriately to correspond to asymptotic time translation and rotation(s). Now, when $\bm \xi$ is a Killing field, from its very construction, $\delta H[\bm \xi]=0$, although $H[\bm \xi]$ itself need not vanish. Using all these, one may identify the $\mathcal S_{\infty}$ contributions above as the ``work terms" $\delta \mathcal E - \Omega_H \delta \mathcal J$, with $\mathcal E$ and $\mathcal J$ being the energy and angular momentum respectively (see \cite{wald} for a rigorous justification of this). This in turn allows us to identify the black hole entropy $S_{\rm Noether}$ from the $\mathcal S_{\rm bif}$ contribution via the relation $(\kappa/2\pi) \delta S = \delta \mathcal E - \Omega_H \delta \mathcal J$. Symbolically, 
\bea
0 = \underbrace{- \frac{1}{2} \ \delta \int \limits_{\mathcal S_{\rm bif}} d\Sigma_{ab} \; J^{ab} }_{- (\kappa/2\pi) \delta S} 
\; + \;
\underbrace{
\frac{1}{2} \ \delta \int \limits_{\mathcal S_{\infty}} d\Sigma_{ab} \; J^{ab} 
+ 
\int \limits_{\mathcal S_{\infty}} d\Sigma_{ab} \; \xi^{[a} \delta v^{b]}  
}_{\delta \mathcal E - \Omega_H \delta \mathcal J}
\eea
Using all the above arguments, Wald eventually identified the entropy as 
\begin{equation}
S_{\rm Noether} =  \frac{1}{2}\beta \int J^{ab} \; \DM\Sigma_{ab} 
\label{noetherint1}
\end{equation}

The above expression for entropy can be rewritten in a simplified (and appealing) form for stationary horizons in \LL\ models. However, for sake of simplicity, we sketch how this form is arrived at for static horizons. Using \eq{eq:JabLL} for $J_{ab}$ in \LL\ models, we can rewrite \eq{noetherint1} as
\begin{equation}
S_{\rm Noether} = \frac{1}{2} \beta \int \l( 2 P^{abcd} \; \nabla_{[c} \xi_{d]} \r) \; \l(2 \xi_{[a} l_{b]} \r) \sqrt{\sigma} \DM^{(D-2)}x
\end{equation}
where $l_a$ is an auxiliary null vector satisfying $l^2=0$ and $\bm l \cdot \bm \xi=-1$, in terms of which $\DM\Sigma_{ab} = \l(2 \xi_{[a} l_{b]} \r) \sqrt{\sigma} \DM^{(D-2)}x$ because $\bm \xi$ is the horizon generating Killing vector field. One can now use the near horizon static structure of the metric (discussed in detail in section \sec{sec:2b4}; see \eq{static-metric}) to show that $\nabla_{[c} \xi_{d]} = (-2 \kappa^2 n) \; \delta^n_{[c} \delta^0_{d]}$ and $l^a=(1/N^2, 1/N, 0,0 \ldots)$ near the horizon defined by $n=0$ whence the lapse behaves as $N=\kappa n + O(n^3)$. Using these, it is easily seen that
\begin{equation}
S_{\rm Noether} = \frac{1}{2} \beta \int \l( 8 \kappa P^{0n}_{0n} \r) \sqrt{\sigma} \DM^{(D-2)}x
\end{equation}
From the total anti-symmetry of the alternating determinant tensor, and the fact that ${}^{(D-2)}R^{AB}_{CD}=R^{AB}_{CD} + O(n^2)$, it follows that
\begin{equation}
P^{0n}_{0n} = \l(\frac{m}{2}\r) L^{(D-2)}_{m-1}
\end{equation}
which eventually gives
\begin{equation}
S_{\rm Noether} = 4 \pi m \int L^{(D-2)}_{m-1} \sqrt{\sigma} \DM^{(D-2)}x
\label{Lentropy}
\end{equation}
Thus the entropy of $m$-th order \LL\ theory in $D$-dimensions can be related to the \textit{Lagrangian} of the $(m-1)$-th order \LL\ theory in $(D-2)$-dimensions.
The above expression (which holds for stationary horizons as well) provides a description of horizon entropy in \LL\ models purely in terms of intrinsic geometry of the horizon, without any reference to any additional vector fields etc. For a more general (non \LL\ ) Lagrangian, this, of course, will not be the case. In particular, the fact that entropy is purely a functional of intrinsic horizon geometry can presumably be taken as another indication of `holographic' properties of \LL\ models discussed in \sec{sec:2a4}.

Two other aspects of Noether charge entropy  worth mentioning in this context are the following. First, in Einstein gravity, we know that the York-Gibbons-Hawking boundary term, evaluated on the stretched horizon $\mathcal{H}_{\epsilon}$, also gives the correct horizon entropy in the limit $\epsilon \rightarrow 0$. A natural question, therefore, is whether the \LL\ surface term given in \eq{eq:surface-term-defn}  also yields the same entropy. This definition of entropy was first given by Zanelli and Teitelboim \cite{teitelboim-zanelli}, and an argument by Iyer and Wald in \cite{wald} shows that it matches with the Noether charge entropy. However, an explicit correspondence between these very different forms of the two expressions for entropy is not known in general. Perhaps the structural properties of the surface term described in detail in \sec{sec:4d4} and \sec{sec:4d5}  might be useful to establish a direct correspondence between the two entropy expressions (at least for maximally symmetric horizons). (It is also possible to obtain the entropy of the horizon by a procedure (called `replica procedure') related to the conical singularity and deficit angle in the Euclidean sector; this is discussed, e.g., in Ref. \cite{solo}.)

Second, there has been extensive work, based on complex path formalism \cite{tpsrini}, to interpret the Hawking radiation as a tunnelling phenomenon (see e.g., \cite{shankietal,vanzo} and also \cite{maulik-tunneling}) and the question arises as to the status of this approach in \LL\ models. In this context, S. Sarkar and D. Kothawala were able to prove the following result for spherically symmetric horizons \cite{ssdk-tunneling}: For any theory of gravity in which field equations near the horizon can be written in the form $T \DM S - \DM E = P \DM V$, the tunnelling formalism of \cite{maulik-tunneling} yields an entropy which matches the Wald entropy {\it once a specific choice is made for the Hamiltonian used in the tunnelling formalism}. As we shall show in section \sec{sec:4c1}, the above criterion is satisfied near any static horizon (not necessarily spherically symmetric) in \LL\ models, with $S$ being the Wald entropy (see \sec{sec:4c1} for definition of $E$ etc.), and hence the result of \cite{ssdk-tunneling} implies that the tunnelling mechanism can reproduce the correct expression for entropy in \LL\ models as a simple consequence of the first law of thermodynamics.

We conclude this section with a brief mention of an approach to the entropy of horizons, which does \textit{not} seem to generalise from Einstein's theory to \LL\ models easily. This approach is based on entanglement of quantum fields. The modes of the quantum fields outside the horizon are entangled with modes inside and when the latter are traced out, the physical processes outside the horizon is governed by a density matrix $\rho\propto\exp(-\beta H)$. A natural entropy associated with this density matrix is $S=-Tr(\rho\log\rho)$. This quantity has a divergent behaviour of the form $\lim_{\epsilon\to0}(A/4\epsilon^2)$ where $A$ is the horizon area and $\epsilon$ has dimensions of length. If one cuts-off the limit at Planck length, one can match the expression with the standard result in Einstein gravity with entropy coming out as proportional to area. (This cut-off can be introduced in a covariant manner, using, for e.g., the techniques discussed in \cite{tpentanglement}.) But since the entropy in \LL\ models is not, in general, proportional to area it is not clear whether the idea of entanglement entropy generalizes beyond Einstein's theory. (If it does not, it raises the possibility that it may not be the correct approach even in Einstein's theory.) The fact that it is divergent makes it difficult to draw firm conclusions. It is, for example, possible that the regularization procedure to introduce the cut-off somehow leads to the correct expression \cite{tpllcutoff} but these ideas have not been worked out completely. So it is fair to say that whether entanglement entropy approach  reproduces the correct \LL\ entropy is at present unclear.

\subsection{Entropy of self-gravitating systems and Wald entropy} \label{sec:4d4}

 It is well known that statistical mechanics of systems with long range interactions exhibit various peculiar features (see, for a review, Ref.~\cite{tp-statmech}) which could lead to subtleties in the thermodynamic description of such systems. It is therefore natural to ask whether  effects of self-gravity can alter the conventional (extensive) nature of entropy of a matter configuration that is undergoing gravitational collapse,  when the matter configuration is on the verge of forming a black hole. 

This issue was first raised by Israel et. al. \cite{areascalingisrael} in an attempt to give an operational definition of black hole entropy in Einstein gravity. Their results reproduced the well known area scaling of entropy in the limit of a matter shell collapsing to form a black hole (probably not a surprising result for a single shell), but more importantly, gave the correct numerical coefficient of $1/4$ in the near horizon limit. This then motivated Oppenheim \cite{areascalingoppenheim} to study a configuration of closely packed shells and ask how exactly the conventional volume scaling of entropy transmutes to an area scaling as the horizon is approached. The results obtained in these works (also see \cite{sanved-tp-idealgas}) suggested that entropy scaling approaches the Bekenstein-Hawking area law for self-gravitating systems provided the shells are assumed to be instantaneously in  equilibrium with the local acceleration temperature in a manner made precise in the discussion below. 

This result seems too good to be true at a fundamental level, since it seemingly provides an explanation for origin of black hole entropy without any appeal to quantum gravity! Perhaps this is just what happens, but one must be cautious since there are several ways of obtaining the correct entropy scaling using very different arguments. The contribution of  effects of self-gravity to horizon entropy can  be more firmly established if we could test the above analysis for theories other than Einstein gravity, since that would help isolate dynamics from kinematics in the final result. In this section, we shall describe the  re-analysis of this approach for \LL\ models, and show that several non-trivial differences arise.  The conclusion is that horizon entropy and entropy due to  effects of self-gravity are different  from each other in general, although the difference itself turns out to have an interesting form. This is one example in which the study of \LL\ models allows us to clearly isolate certain numerical coincidences which occur in Einstein's theory and do not carry over to \LL\ models.

To calculate the standard thermodynamic entropy of a self gravitating system, we will basically follow the set up suggested by Oppenheim \cite{areascalingoppenheim}, but in a slightly different manner to bring out the key features more prominently \cite{tp-dk-sk}. We consider a system of $n$ densely packed spherically symmetric shells in $D$ spacetime dimensions, assumed to be in thermal equilibrium and supporting itself against its own gravity. We shall be interested in the entropy of this system when the outermost shell is close to the event horizon of the system. We first describe a general set-up to study the system described above for any spherically symmetric spacetime. This is important for two reasons. First, as we shall see, it will highlight the key {\it mathematical feature} responsible for entropy density $S/A$ of the system being $1/4$ for Einstein's theory in arbitrary dimensions. In \cite{areascalingoppenheim}, this result was obtained for $D=4$ by explicitly using the Schwarzschild form of metric functions and the expression for Hawking temperature. However, we shall not require any explicit expression for metric functions or Hawking temperature to obtain the area scaling. Second, the set-up we describe can be used, without any modification, to obtain entropy density for higher curvature gravity actions.

Let us denote the variables  describing the $i$th shell with a subscript $i$. Since the system  is spherically symmetric, we can write the metric outside the $i$th shell as

\begin{equation}
 ds^2 = -c_i f_i(r)dt^2 + b_i(r)^{-1}dr^2 + r^2 d\Omega^2
\label{shellmetric}
\end{equation}
where $d\Omega^2$ is the metric of a unit $(D-2)$ sphere. At the location of the shells (given by $r_i=$const), the above metric  must satisfy the  Israel junction condition which states that the induced metric on the hypersurface should be continuous. 
This leads to the following conditions on the constants $c_i$:
\begin{equation}
 c_i f_i(r_{i+1}) = c_{i+1} f_{i+1}(r_{i+1})
\label{1junction}
\end{equation}
and we shall choose $c_n =1$ (which fixes the interpretation of $t$ as proper time for asymptotic observers). Then \eq{1junction} can be solved to give
\begin{eqnarray}
 c_k = \frac{f_{k+1}(r_{k+1})}{f_{k}(r_{k+1})}.....\frac{f_{n-1}(r_{n-1})}{f_{n-2}(r_{n-1})}\frac{f_{n}(r_{n})}{f_{n-1}(r_{n})}
\label{ccondition}
\end{eqnarray}
Note that when $f_{n}(r_{n}) = 0$  --- i.e.,  when the outermost shell is exactly on the horizon --- we have $c_n=0$ for all $(i \neq n)$ which implies that 
\begin{equation}
 g_{00} = 0 \; \; \forall \ i
\label{g00vanish}
\end{equation}
Here the condition that the $g_{00}$'s vanish even for the inner-shells indicates that our assumption regarding the staticity of the inner shells is not valid when $f_{n}(r_{n})$ is exactly zero. (This is because  we know that a particle cannot be kept at a fixed position inside a black hole.)  However, for the purpose of our discussion, we only need to consider the limit in which the outer shell is very near to the horizon, that is, $f_{n}(r_{n}) \rightarrow 0$, then the result that the of $c_i$'s (other than $c_n$) and $g_{00}$ vanish is just the leading order  result in this approximation. Henceforth, we shall assume that we are working in this limit and will not state it explicitly unless otherwise needed.

We shall next use the second junction condition which relates jumps in geometric quantities across a shell, to the matter stress tensor of the shell, $t_{(i)\mu \nu}$. Let the normal to a $r=$const. surface (dropping the subscript $i$ for convenience) be $n_{a} = b^{-1/2} \partial_a r$, so that the induced metric on $r=$ constant surface  becomes $h_{\mu \nu} = \l( g_{a b} - n_a n_b \r) \delta^a_{\mu} \delta^b_{\nu}$ in coordinates $(t, \theta_A)$. 
Here, we will use Greek letters to define components in the $r=$ constant surface  with coordinates $(t, \theta^A)$. 
Within the hypersurface, we can further define $u_{\mu} = \sqrt{c f} \; \partial_{\mu} t$, and the induced metric on the level surfaces of $t$, $q_{\mu \nu} = h_{\mu \nu} + u_{\mu} u_{\nu}$. It then follows from spherical symmetry that $t_{\mu \nu}$ for each shell has the general form
\begin{eqnarray}
t_{(i)\mu \nu} = \rho_i u_{(i)\mu} u_{(i)\nu} + P_i q_{(i)\mu \nu}
\label{eq:gensab}
\end{eqnarray}
with $E_i = 4\pi r_i^2 \rho_i$ is the energy of the shell. The physical interpretation of $E_i$ and $P_i$ is that of energy and pressure as measured by a local observer at rest on the shell. Further, the condition for thermal equilibrium in curved spacetime implies \cite{tolman}
\begin{eqnarray}
 T_i \sqrt{-g^i_{00}(r_i)} &=& T_{\infty}
 \nn \\
\mu_i \sqrt{-g^i_{00}(r_i)} &=& \mu_n \sqrt{-g^n_{00}(r_n)}
\label{tolman-chemical}
\end{eqnarray}
where $T_{\infty}$ is the temperature of the system as measured by a static observer at infinity, and $\mu_i$'s denote the chemical potential. We can thus express all the $T_i$'s and $\mu_i$'s in terms of just two unknown parameters $T_{\infty}$ and $\mu_n$ respectively. In thermodynamic equilibrium, each shell satisfies the Gibbs-Duhem relation
\begin{equation}
 E_i = T_i S_i - P_i A_i + \mu_i N_i
\end{equation}
where $N_i$ is the number of particles composing the $i$th shell and $A_i$ is the  area of the $i$th shell. Using Eq.~(\ref{tolman-chemical}) in the above expression, we can write the total entropy $S_{\mathrm{matter}}$ of the system in the form
\begin{eqnarray}
 S_{\mathrm{matter}} &=& \sum_i S_i
 \nn \\
 &=& \sum_i \frac{E_i + P_i A_i}{T_{\infty}}\sqrt{-g^i_{00}(r_i)} - \frac{\mu_n N}{T_{\infty}}\sqrt{-g^n_{00}(r_n)}
 \nn \\
  &=& \sum_i \frac{E_i + P_i A_i}{T_{\infty}}\sqrt{c_i f_i(r_i)} - \frac{\mu_n N}{T_{\infty}}\sqrt{f_n(r_n)}
 \nn \\
\label{eq:Stotshell}
\end{eqnarray}
Now assuming that $\mu_n N$ is a finite quantity, the last term in the above expression vanishes in the near horizon limit, since $g^n_{00}(r_n)$ vanishes (see \eq{g00vanish}). Hence the only non-trivial contribution to $S_{matter}$ can come from the first term, which we wish to evaluate in the limit when outermost shell is close to the horizon of the system, $r_n \rightarrow r_H$ (where $f_n(r_H)=0$); we shall refer to this as the near-horizon limit. We can read-off this contribution as follows. 

First, we note that $(E_i + P_i A_i)/T_{\infty}$ has no dependence on $c_i$, since $t_{(i)\mu \nu}$ does not depend on $c_i$ (see below). Further, in the near horizon limit, $c_i=0 ~ \forall i \neq n, c_n=1$, from which it is easy to see that the only non-zero contribution to $S_{\mathrm{matter}}$ will come from the $i=n$ term, and in particular from those terms in $(E_n + P_n A_n)$ which diverge as $1/\sqrt{f_n(r_n)}$. Therefore, our strategy to calculate the entropy of the system will be to analyze the form of $E_n$ and $P_n$ and look for such divergent factors. As we shall now demonstrate, this strategy provides an extremely quick way to obtain the result in Einstein theory (compared to the explicit computations in \cite{areascalingoppenheim}), and also facilitates easy generalization to higher derivative theories. 

In Einstein theory, the surface stress tensor is given by
\begin{eqnarray}
8 \pi t^{(E)}_{(i)\mu \nu} &=& \Biggl \langle K h_{\mu \nu} - K_{\mu \nu} \Biggl \rangle_i
\label{eq:sabeins}
\end{eqnarray}
where $\langle \ldots \rangle_i = \langle \ldots \rangle_{i+1} - \langle \ldots \rangle_i$ evaluated at $r=r_i$. The extrinsic curvature of $r=$const surface, for the metrics described by Eq.~(\ref{shellmetric})), is given by (after dropping the subscript $i$ for convenience):
\begin{eqnarray}
K_{\mu \nu} &=& - \l( \sqrt{b} f'/2 f \r) u_{\mu} u_{\nu} + \l(\sqrt{b}/r\r) q_{\mu \nu}
\nn \\
\nn \\
K &=& h^{\mu \nu} K_{\mu \nu}
\nn \\
&=& \l( \sqrt{b} f'/2 f \r) + \l(D-2\r) \l(\sqrt{b}/r\r)
\label{eq:extcurv}
\end{eqnarray}
Note that, as stated above,  there is no dependence on $c_i$. Based on our strategy outlined in the previous section, we must now find terms which, for $i=n$, diverge as $1/\sqrt{f_n(r_n)}$ as $r_n \rightarrow r_H$. We now specialize to the case $b_i(r)=f_i(r) ~ \forall i$. (This condition is automatically satisfied, in the near-horizon limit,  for all the cases in which the horizon has no curvature singularity; so this choice entails no loss of generality.) Using Eqs.~(\ref{eq:sabeins}), (\ref{eq:extcurv}) and (\ref{eq:Stotshell}), we can immediately write, for the divergent parts,
\begin{eqnarray}
8 \pi [ \rho_n]_{\mathrm{div}} = 8 \pi \l[ t^{(E)}_{(n)\hat 0 \hat 0} \r]_{\mathrm{div}} &=& 8 \pi \l[ t_{(n)\mu \nu} u^{\mu} u^{\nu} \r]_{\mathrm{div}} = 0 
\nn \\
\nn \\
8 \pi [ P_n]_{\mathrm{div}} \; \delta^A_{B} = 
8 \pi \l[ t^{(E)A}_{(n)B} \r]_{\mathrm{div}} &=& - K^{(n)}_{\hat 0 \hat 0} \delta^A_{B} 
\label{eq:sabeinsdiv}
\end{eqnarray}
in the limit $r_n \rightarrow r_H$, $f_n(r_H) \rightarrow 0$ (here $K^{(n)}_{\hat 0 \hat 0} = K^{(n)}_{\mu \nu} u^{\mu} u^{\nu} $). We have recalled the definition of $\rho_i$'s and $P_i$'s from Eq. (\ref{eq:gensab}). Hence, we see that the main divergence comes from 
\begin{eqnarray}
\l[ P_n \r]_{\mathrm{div}} &=& - \l( \frac{1}{8 \pi} \r) K^{(n)}_{\hat 0 \hat 0} 
\nn \\
&=& \l( \frac{1}{16 \pi} \r) \lim_{r_n \rightarrow r_H^+} \frac{f_n'(r_n)}{\sqrt{f_n(r_n)}}
\nn \\
&=& \frac{T_{\infty}}{4}  \lim_{\epsilon \rightarrow 0^+} \frac{1}{ \sqrt{f_n(r_H + \epsilon)} }
\label{eq:PTeos}
\end{eqnarray}
where we have used the definition of Hawking temperature for a system which has collapsed to a black hole. Using the above result in Eq. (\ref{eq:Stotshell}), we finally get
\begin{eqnarray}
S_{\mathrm{matter}} &\underset{r_n \rightarrow r_H}{\longrightarrow}& S_n
\nn \\
&=& \lim_{\epsilon \rightarrow 0^+} \frac{(1/4) T_{\infty} / \sqrt{f_n(r_H+\epsilon)}}{T_{\infty}} A_H \times \sqrt{f_n(r_H+\epsilon)}
\nn \\
&=& \frac{A_H}{4}
\end{eqnarray}
Hence, we see that the entropy of the configuration, as the outermost shell approaches the event horizon of the system, is dominated by the Bekenstein-Hawing entropy. The crucial factor of $1/4$ comes from the last line of Eq.~(\ref{eq:PTeos}), which can be recast as 
\begin{equation}
 [P_n]_{\mathrm{div}}/T_n = 1/4
\label{eqnofstateeg}
\end{equation}
Interestingly, this same relation arises in the context of a derivation of Navier-Stokes equation and its relation to emergent gravity paradigm \cite{navieractionsanved}.

In what follows, we shall repeat the above analysis for \LL\ gravity \cite{tp-dk-sk}, and show that the correspondence between shell entropy as calculated above, and the black hole entropy as given by Wald formula {\it does not always hold}. We will also point out  that our result, nevertheless, has a curious interpretation in terms of the ``equation of state" of the horizon degrees of freedom, and also has a very direct mathematical connection with the membrane paradigm for black hole horizons.

As discussed in \sec{sec:llhortherm}, the surface stress tensor in the $m^{\mathrm{th}}$ order \LL\ theory is given by 

\begin{eqnarray}
8 \pi \left( t^{\; \; \ \ \nu}_{(m) \mu} \right)_i &=&  \frac{m!}{2^{m+1}}\,\alpha _{m} \Biggl \langle \sum_{s=0}^{m-1}C_{s(m)}\,\left( \pi
_{s(m)}\right) ^{\nu}_{\mu} \Biggl \rangle_i \nonumber \\
\left( \pi _{s(m)}\right) ^{\nu}_{\mu} &=& \delta _{\lbrack
\mu \mu_{1}\cdots \mu_{2m-1}]}^{[\nu \nu_{1}\cdots \nu_{2m-1}]}\,\hat{R}_{\nu_{1}\nu_{2}}^{\mu_{1}\mu_{2}}\cdots \hat{R}_{\nu_{2s-1}\nu_{2s}}^{\mu_{2s-1}\mu_{2s}}%
\,K_{\nu_{2s+1}}^{\mu_{2s+1}}\cdots K_{\nu_{2m-1}}^{\mu_{2m-1}} 
\end{eqnarray}%
where the coefficients $C_{s(m)}$ are given by
\begin{equation}
C_{s(m)}=\sum_{q = s}^{m-1} \frac{4^{m-q} \; ^qC_s (-2)^{q-s}}{q!\left( 2m-2q-1\right) !!}
\end{equation}
Here $\hat R_{\mu \rho \sigma \nu}$ etc. indicates that these quantities are to be evaluated for the induced metric $h_{\mu \nu}$. To calculate the matter entropy of the system of $n$ shells, we proceed in the manner similar as in the case of Einstein's gravity described in the previous section, that is, by finding $\l[ t^{\ \ \nu}_{(n) \mu} \r]_{\mathrm{div}}$. We note that, in the near horizon limit (and assuming that the intrinsic Riemann tensor of horizon surface is finite), the only divergence in the stress tensor will come from the $K^{\hat 0}_{\hat 0}$ component of the extrinsic curvature of the $r= r_n$th shell, which diverges as $(1/\sqrt{f_n})$ as can be seen from \eq{eq:extcurv}. The transverse components $K^A_B$ vanish as $(\sqrt{f_n}/r)$. Further, note that the determinant tensor has the property 
\begin{equation}
\delta^{0 \alpha_1 \alpha_2 \cdots \alpha_n}_{0 \beta_1 \beta_2 \cdots \beta_n} = \delta^{A_1 A_2 \cdots A_n}_{B_1 B_2 \cdots B_n} \times \l( \delta^{\alpha_1}_{A_1} \delta^{B_1}_{\beta_1} \ldots \delta^{\alpha_n}_{An} \delta^{B_n}_{\beta_n} \r)
\label{determinanttensor}
\end{equation}
That is, the presence of $0$ in each row of the determinant tensor forces all the other indices to take the values $2,3, \cdots (D-1)$. Thus, at  most, we can have only one $K^0_0$ present in any term of \eq{junctionconditionLL} which shows that the maximum possible divergence is $O(1/\sqrt{f_n})$ even though the surface stress tensor is a $(2m-1)$th degree polynomial in $K$. From the structure of \eq{junctionconditionLL}, it is easy to see that only the $s=(m-1)$ term gives this divergent contribution while all the other terms being of $O(\sqrt{f})$ or higher vanish while from \eq{determinanttensor}, it is again obvious that only the transverse component of the surface stress tensor will contribute. Thus we get
\begin{eqnarray}
8 \pi [ \rho_n]_{\mathrm{div}} = 8 \pi \l[ t^{}_{(n)\hat 0 \hat 0} \r]_{\mathrm{div}} &=& 8 \pi \l[ t_{(n)\mu \nu} u^{\mu} u^{\nu} \r]_{\mathrm{div}} = 0 \nn \\ \nn \\
8 \pi [ P_n]_{\mathrm{div}} \; \delta^A_B = 
8 \pi \l[ t^{A}_{(n)B} \r]_{\mathrm{div}} 
&=& \frac{m \alpha _{m}}{2^{m-1}} \delta ^{
A A_{1}\cdots A_{2m-2}}_{B B_{1}\cdots B_{2m-2}}\,\hat{\hat{R}}^{B_{1}B_{2}}_{A_{1}A_{2}}\cdots \hat {\hat{R}}^{B_{2m-3}B_{2m-2}}_{A_{2m-3}A_{2m-2}} \; K^{\hat 0}_{\hat 0}
\label{sabdivergenceLL}
\end{eqnarray}
where in the second line we have used $C_{m-1} = 4/(m-1)!$ from its definition in \eq{coefficient}, and replaced $\hat{R}^{A B}_{C D}$ with the intrinsic curvature $\hat{\hat{R}}^{A B}_{C D}$ defined completely in terms of the induced metric $q_{AB}$ of the constant $t$, constant $r$ surface. Such a replacement is valid since the corresponding extrinsic curvature vanishes. We can now obtain the pressure easily using the fact that $q_{AB}$ is maximally symmetric (see \eq{eq:gensab}), and hence
\begin{eqnarray}
 [ P_n]_{\mathrm{div}} = \frac{1}{(D-2)} {q^{AB}\l[ t_{(n)AB} \r]_{\mathrm{div}}} &=& \frac{2 m \alpha_{m}}{(D-2)} \frac{1}{16 \pi} \frac{1}{2^{m-1}} \delta ^{
A A_{1}\cdots A_{2m-2}}_{A B_{1}\cdots B_{2m-2}}\,\hat{\hat{R}}^{B_{1}B_{2}}_{A_{1}A_{2}}\cdots \hat {\hat{R}}^{B_{2m-3}B_{2m-2}}_{A_{2m-3}A_{2m-2}} \; K^{\hat 0}_{\hat 0} \nonumber \\
&=& \l( \frac{D-2m}{D-2} \r) 2 m \alpha_{m} L^{(D-2)}_{m-1} \; K^{\hat 0}_{\hat 0}
\label{pressureLL}
\end{eqnarray}
In arriving at the second equality, we have used the following two relations concerning \LL\ actions of order $m$ in $D$ dimensions derived earlier in Eqs.~(\ref{eom-lovelock}), (\ref{treom})
\begin{eqnarray}
E^i_{j(m)} &=&  - \frac{1}{2} \frac{1}{16 \pi} \frac{1}{2^m} \delta^{i a_1 b_1 \ldots a_m b_m}_{j c_1 d_1 \ldots c_m d_m} R^{c_1 d_1}_{~ a_1 b_1} \cdots R^{c_m d_m}_{~ a_m b_m}
\nn \\
E^i_{i(m)} &=& - \frac{D-2m}{2} L^D_m
\end{eqnarray}
where $E^i_{j(m)}$ is the equation of motion tensor for \LL\ action. Hence, we once again  find that the only non-zero contribution to matter entropy comes from the pressure term as in the case of Einstein gravity (see \eq{eq:sabeinsdiv}). No further computation is needed since we can simply obtain the entropy by comparing expression in Eq.~(\ref{pressureLL}) with the first of Eqs.~(\ref{eq:PTeos}) in Einstein theory. This immediately allows us to read-off the matter entropy as
\begin{eqnarray}
S^{(m)}_{\mathrm{matter}} = \l( \frac{D-2m}{D-2} \r) 4 \pi m \alpha_m L^{(D-2)}_{m-1} A_H \label{matterentropyLL}
\end{eqnarray}
where $A_H$ is the $D-2$ dimensional hypersurface area of the horizon. On the other hand, the Wald entropy for general \LL\ theory obtained earlier in 
\eq{Lentropy}, when evaluated for the spherically symmetric case is given by:
\begin{eqnarray}
S^{(m)}_{\mathrm{Wald}} &=&  4 \pi m \alpha_m L^{(D-2)}_{m-1} A_H
\end{eqnarray}
which leads to the following relation
\begin{eqnarray}
 S^{(m)}_{\mathrm{matter}} = \l( \frac{D-2m}{D-2} \r) S^{(m)}_{\mathrm{Wald}}
\label{entropymatterwald}
\end{eqnarray}
  So the matter entropy in \LL\ theory is proportional to the corresponding Wald entropy of the \bh. We also see that $S_{\mathrm{matter}} \leq S_{\mathrm{Wald}}$ with the equality holding only in Einstein's gravity $m=1$. This inequality strongly suggests that, in general, the \mdof \ responsible for the entropy of the \bh\ are certainly not the extrinsic (matter) degrees of freedom forming the \bh. Further, since a general \LL\ theory will be described by Lagrangian of the form $L=\sum_m \alpha_m L_m$, the total matter entropy in such theories would be 
\begin{eqnarray}
 S_{\mathrm{matter}} = \sum \limits_{m=0}^{[D-1)/2]} \l( \frac{D-2m}{D-2} \r) S^{(m)}_{\mathrm{Wald}}
\end{eqnarray}
which has no simple relation, or proportionality to the Wald entropy 
\begin{eqnarray}
 S_{\mathrm{Wald}} = \sum \limits_{m=0}^{[(D-1)/2]} S^{(m)}_{\mathrm{Wald}}
\end{eqnarray}

We highlight another point which is relevant to the membrane paradigm for \LL\ models, to be discussed in \sec{sec:4d5}. As is evident from above analysis, it is only the pressure term which leads to a non-zero contribution to the matter entropy and hence, in the near horizon limit, we have the result:
\begin{eqnarray}
\frac{S_{\mathrm{matter}}}{A_H} &=& \frac{[P_n]_{\mathrm{div}}}{T_n} \nonumber \\
&\equiv& \frac{P}{T_n}
\end{eqnarray}
where for the sake of brevity, we have replaced $[P_n]_{\mathrm{div}}$ by $P$. Thus the scaling  of entropy with the horizon radius $r_H$, as well as the proportionality constant, is completely determined by the relation between $P$ and $T$, which is to say that the equation of state $P = P(T)$ determines the matter entropy. In the case of \LL\ gravity, we can read off the equation of state $P = P(T) = [t^{\ \ \theta}_{(n) \theta}]_{\mathrm{div}}$ from \eq{pressureLL} by using the relation $K_{\hat 0 \hat 0} = -K^{\hat 0}_{\hat 0} = -f^{\prime}_H/(2\sqrt{f}) = -2\pi T_n$ to get
\begin{eqnarray}
 \frac{P}{T_n} = \l( \frac{D-2m}{D-2} \r) 4 \pi m \alpha_m L^{(D-2)}_{m-1}
\label{generaleqnofstate}
\end{eqnarray}
Using \eq{entropymatterwald}, the above relation can also be written in terms of Wald entropy as:
\begin{eqnarray}
 \frac{P A_H}{T_n} = \l( \frac{D-2m}{D-2} \r) S^{(m)}_{\mathrm{Wald}}
\end{eqnarray}
In fact, the pressure $P$ obtained above is exactly what one would call as surface tension of the stretched horizon in the membrane paradigm. This is mathematically obvious from the way we arrived at the near-horizon for $P$, but it must be emphasized that the equivalence with membrane pressure is strictly valid only in the near-horizon limit.

The main implication of the above result is that matter entropy of a system is not necessarily the only contribution to entropy of a black hole formed from collapse of that system. This  unambiguously establishes the fact that there might be more to horizon entropy than just the effects of self-gravitation, at least to the extent that the approximations made in the above analysis are valid. (We have not studied the system in full generality, since the dynamics of the collapse has not been taken into account. We only considered the shells at various stages of collapse, imposing consistency conditions in the form of junction conditions at each instant of the collapse. While this may be sufficient, it will be interesting to study what happens in a fully dynamical situation.) However,  whatever be the nature of additional corrections introduced by a more general analysis, it is unlikely to establish the complete equivalence of the resultant entropy with the Wald entropy. 

It is also interesting to view the above result in the light of several entropy bounds on matter entropy which exist in the literature. The model calculation above gives precise expressions for matter entropy and its relation to Wald entropy, rather than just an inequality. Finally, the analysis above also highlights a possible connection between horizon entropy and the ``equation-of-state" $P(T)$ satisfied by the degrees of freedom residing on the horizon surface. This would become  more evident when we present a study of the {\it membrane paradigm} for the \LL\ models in the next section, deriving all the transport coefficients to linear order in the perturbation theory.

\subsection{Membrane paradigm in  \LL\ models} \label{sec:4d5}

We now use the mathematical tools of the previous section to derive several properties of the so called stretched horizon in a theory described by the \LL\ action \cite{sk-dk}. That is, we set up the so called {\it membrane paradigm} for the \LL\ models. After a brief introduction to the membrane paradigm in Einstein gravity, we shall discuss the analysis for the \LL\ models, keeping in focus the new features introduced by the non-triviality of the \LL\ action. 

Specifically, we shall present a derivation of the various {\it transport coefficients} associated with the membrane  to first order in perturbation theory, and demonstrate how the structure of these coefficients brings into focus several fundamental features of the membrane paradigm itself, which are otherwise not evident when attention is restricted solely to Einstein gravity. This  reinforces our original point of view about the utility of \LL\ models in gaining insight into  aspects of Einstein gravity itself.

The membrane paradigm of black holes \cite{membranekipthorne} takes the point of view that, as far as the static observers outside the horizon are concerned, the black hole horizon can be replaced by a \textit{stretched} horizon, a membrane, endowed with specific physical properties which encode the presence of the inaccessible black hole region. Although originally developed to facilitate inclusion of the  physics of black hole background in astrophysical applications, work over the last decade or so has indicated that physical properties of the membrane (such as viscosity etc.) might have a deeper relevance,  from the point of view of holographic dualities which map gravitational systems to non-gravitational systems in one lower dimension. This has led to a renewed interest in the study of membrane paradigm.

The membrane paradigm in Einstein gravity has also been analysed from a view point of the emergent gravity paradigm. The Damour-Navier Stokes equation governing the dynamics of the \bh\ membrane could be obtained \cite{navieractionsanved} starting from an action which could be given a thermodynamic interpretation as an entropy production rate when expressed in terms of thermodynamic variables such as temperature, entropy, pressure, etc of the horizon. This comes very close to quantifying a precise connection of properties of the horizon membrane with the thermodynamics of horizons, but as we shall show in this section, a more direct connection between the transport coefficients of the membrane and its thermodynamic properties can be established by studying the \LL\ models. 

In Einstein gravity (as well as with the Gauss-Bonnet term added to it), the \textit{geometric} structure of the fluid transport coefficients of the membrane is not evident from their \textit{algebraic} expressions. For example, in Einstein's gravity, it is known that the shear viscosity $\eta$ is equal to $1/(16\pi)$. But the origin of this factor is not well understood in terms of the \textit{geometric} properties of the horizon such as  the intrinsic curvature of the horizon etc. As we shall show, setting up the membrane paradigm \cite{tp-dk-sk} for the \LL\ models once again helps us in not only understanding such geometric aspects of the transport coefficients in a more general manner, but even explains the origin of factors such as $1/(16 \pi)$ for shear viscosity in Einstein theory!

Let us briefly describe the geometric set-up and the procedure to construct the membrane paradigm. More details can be found in \cite{membranekipthorne}, and we only refresh the essential expressions below to set up the notation etc.
We shall work in a $D$ dimensional spacetime containing an event horizon $H$ generated by the null geodesics $\bm l$. The surface gravity $\kappa$ is then determined through $\nabla_{\bm l} \bm l = \kappa \bm l$. We shall deal with a stationary background spacetime, in which case $\bm l$ corresponds to the timelike Killing vector $\bm \xi$ restricted to $H$ where $\xi^2=0$. For asymptotically flat spacetimes, the norm of $\bm l$ can be fixed at infinity and there is no ambiguity in the definition of $\kappa$. The stretched horizon $\mathcal{H}_s$ is then defined as a timelike surface infinitesimally close to $H$, and separated from $H$ along spacelike geodesics $\bm n$. This surface is spanned by a unit timelike vector field $\bm u$ and $(D-2)$ vectors $\bm e_{(A)}$'s. The correspondence between points on $H$ and $\mathcal{H}_s$ is made via ingoing null geodesics $\bm k$ normalized to have unit Killing energy, $\bm k \bm \cdot \bm l=-1$. The horizon $H$ is then located at $\lambda=0$, where $\lambda$ is the affine parameter along $\bm k$, related to the norm of $\bm \xi$ by $\lambda = - (1/2 \kappa) \xi^2$. We shall define $N = \sqrt{-\xi^2}$ to simplify notation and also to facilitate comparison with known results in the literature. In the limit $\lambda \to 0$, $\mathcal{H}_s \rightarrow H$ and $\bm u \to N^{-1} \bm l$ and $\bm n \to N^{-1} \bm l$.

The induced metric $h_{\mu \nu}$ on $\mathcal{H}_s$ is given by $h_{\mu \nu} = h_{ab} e^a_{(\mu)} e^b_{(\nu)}$, where $h_{ab} = g_{ab} - n_a n_b$ and $\bm e_{(\mu)}$'s are basis vectors spanning $\mathcal{H}_s$ ($\bm e_{(\mu)} \bm \cdot \bm n=0$). Similarly, the induced metric $\gamma_{AB}$ on the $(D\!-\!2)$-dimensional space-like cross-section of $\mathcal{H}_s$ orthogonal to $\bm u$ is given by $\gamma_{AB} = \gamma_{ab} e^a_{(A)} e^b_{(B)}$ where $\gamma_{ab} = h_{ab} + u_a u_b$ and $\bm e_{(A)}$'s are basis vectors spanning $\gamma$ ($\bm e_{(A)} \bm \cdot \bm u=0=\bm e_{(A)} \bm \cdot \bm n$). The extrinsic curvature of $\mathcal{H}_s$ is defined as $ K_{\mu \nu} = e^a_{(\mu)} e^b_{(\nu)} \nabla_a n_b$ and it is easy to verify that in the limit  $N \to 0$, we have: 
\begin{eqnarray}
K_{{\hat 0} {\hat 0}} &=& K_{\mu \nu} u^{\mu} u^{\nu} =- N^{-1} \kappa
\nn \\
K_{{\hat 0} A} &=& K_{\mu \nu} u^{\mu} e^{\nu}_{(A)}=0
\nn \\
K_{A B} &=& K_{\mu \nu} e^{\mu}_{(A)} e^{\nu}_{(B)}
\nn \\ &=& N^{-1} k_{AB}
\label{nulllimit},
 \end{eqnarray}
where $k_{AB}$ is the extrinsic curvature of the $(D-2)$-dimensional space-like cross-section of the true horizon $H$. This can be decomposed with respect to its trace-free part as 
\begin{eqnarray}
k_{AB} = \sigma_{AB} +  \frac{1}{(D-2)} \theta\, \gamma_{AB},
\label{ksplit}
\end{eqnarray}
where $\gamma^{AB} \sigma_{AB}=0$, and $\theta$ and $\sigma_{AB}$ are the expansion scalar and shear respectively of the $\bm l$ congruence generating the horizon. For a Killing horizon, both of these vanish as $N^2$ and therefore $K_{{\hat 0} {\hat 0}}$ remains the only divergent contribution to $K_{\mu \nu}$. However, since we wish to investigate dissipative properties associated with the horizon, we shall perturb the background away from staticity and then obtain the membrane stress tensor to first order in perturbations to determine the transport coefficients. The extrinsic curvature can be written, upto linear order in perturbations, as
\begin{eqnarray}
 K^{\alpha}_{\beta} = \k + \kp 
\end{eqnarray}
where $\k$ is the unperturbed extrinsic curvature. Since we assumed the background geometry to be static, we have
\begin{eqnarray}
\overset{(0)}{K_{{\hat 0} {\hat 0}}} &=&- \kappa / N, \; \; \;
\overset{(0)}{K_{{\hat 0} A}} = 0, \; \;
\overset{(0)}{K_{A B}} = \left( N/r \right) \delta_{AB}
\label{zerothorder}
\end{eqnarray}
To study the additional divergences brought in when the horizon is perturbed, it is convenient to choose perturbations which do not affect $\overset{(0)}{K_{{\hat 0} {\hat 0}}}$.  
(As we shall see, our final relations will only have divergence of $O(N^{-1})$, and perturbing $\kappa$ will not lead to any qualitative difference.) Formally, since we assume the background $\kappa$ to be non-zero, the perturbation in $\kappa$ is not gauge invariant and our choice corresponds \cite{membranekipthorne} to a gauge wherein $\delta \kappa=0$.)
Therefore, we have
\begin{eqnarray}
\overset{(1)}{K_{{\hat 0} {\hat 0}}} &=& 0, \; \; \;
\overset{(1)}{K_{{\hat 0} A}} = 0, \; \;
\overset{(1)}{K_{A B}} = (1/N) k_{AB}
\label{firstorder}
\end{eqnarray}
where we have not bothered to put a ``$(1)$" over $k_{AB}$ since its unperturbed part is strictly zero. The expansion and shear parameters of the perturbed horizon will now introduce additional divergences in the stress tensor via $K_{AB}$.

We are now ready to use  the above facts to analyze the membrane stress tensor in \LL\ theories of gravity. For further details about the membrane paradigm and aspects of the perturbation scheme, we refer the reader to earlier literature on the subject \cite{membranekipthorne, maulik}, and also to the introductory sections in the recent paper \cite{membraneGB} which focuses on Einstein-Gauss-Bonnet theories. Since the derivation for \LL\ involves considerable combinatorics, we present a summary of the final results first and then provide a sketch of the derivation \cite{tp-dk-sk}.
 
\subsubsection{Membrane paradigm: summary of the results} 

We shall be mainly interested in membranes which have isotropic tangential stresses (pressure), which is only possible when  all directions,  everywhere within the horizon surface, are equivalent. (That is, the horizon surface is maximally symmetric). Therefore, the cross-section of the horizon of the background geometry has the curvature 
\begin{equation}
^{(D-2)} \overset{(0)}{R}_{ABCD} = (\mathcal{K}/r^2) \,(\gamma_{AC}\gamma_{BD}-\gamma_{AD}\gamma_{BC})                                                                                                      \end{equation}  
where $\mathcal{K}=\pm 1$ and $r$ is a constant of dimension $length$ which is related to the intrinsic Ricci scalar of the horizon cross-section as $\mathcal{K} (D-3)(D-2) r^{-2} =~ ^{(D-2)}\overset{(0)}{R}$. For simplicity of notation, we will assume $\mathcal{K}=+1$ in the intermediate steps, and only restore it in the final expressions (which is easy to do). Doing so will come in handy while discussing the case of planar horizons, for which we will simply put $\mathcal{K}=0$. 
With this assumption, it can be shown that the horizon membrane has the stress tensor 
\begin{eqnarray}
t_{\alpha \beta}=\rho_s u_\alpha u_\beta + e^{(A)}_\alpha e^{(B)}_\beta \left( p_s
\gamma_{AB} - 2 \eta_s {\sigma_s}_{AB} - \zeta_s \theta_s \gamma_{AB}
\right)
\end{eqnarray}
with $p_s = \sum_m p^{(m)}_s$ etc., and the different variables are given by:
\begin{eqnarray}
\textrm {Pressure : } ~~p^{(m)}_s &=& \l( \frac{D_{2 m}}{D_2} \r) \l( \frac{\kappa}{2 \pi} \r) ~ 4 \pi m \alpha_m L^{D-2}_{m-1} \nonumber \\
\textrm {Energy density : }~ \rho^{(m)}_s &=& - \l( \frac{\theta_s}{\kappa}\r) p^{(m)}_s \nonumber \\
\textrm {Shear Viscosity : }~ \eta^{(m)}_s &=& \l( \frac{m \alpha_m r_H^2}{\mathcal{K} D_2 D_3} \r) \Biggl[ L^{D-2}_m - \l( \frac{2 \kappa}{r_H} \r) 
 (m-1) D_{2m}  L^{D-2}_{m-1} 
\Biggl] \nonumber \\
\textrm {Bulk Viscosity : }~ \zeta^{(m)}_s &=& - \l( \frac{2 D_3}{D_2} \r) \eta^{(m)}_s
\label{eq:memtranscoeff}
\end{eqnarray}
where we have introduced the notation $D_k \equiv (D-k)$ to avoid clutter, and $\alpha_m$ is the coupling constant of the $m^{\rm th}$ order \LL\ Lagrangian
\begin{eqnarray}
L^D_m = \frac{1}{16 \pi} \frac{1}{2^m} \delta^{a_1 b_1 \ldots a_m b_m}_{c_1 d_1 \ldots c_m d_m} R^{c_1 d_1}_{a_1 b_1} \ldots R^{c_m d_m}_{a_m b_m}
\end{eqnarray}

The transport coefficients above can be connected with the thermodynamic properties of the horizon in the \LL\ gravity, by noting that the Wald entropy and quasi-local energy of the horizon in these theories are given by
\begin{eqnarray}
S^{(m)}_{_{\mathrm{Wald}}} &=& 4 \pi m \alpha_m \int_H d\Sigma \; L^{D-2}_{m-1}
\nn \\
E^{(m)}_{\mathrm{Wald}} &=& \alpha_m \int^{\lambda} d \lambda \int_H \DM \Sigma \; L^{D-2}_{m}
\label{waldentropy}
\end{eqnarray}
This immediately implies that
\begin{eqnarray}
\frac{p_s}{T_{\rm{\infty}}} \;\; \underset{N \rightarrow 0}{\equiv} \;\; \sum \limits_{m=0}^{[D-1]/2} \l( \frac{D-2m}{D-2} \r) S^{(m)}_{\rm{Wald}} 
\end{eqnarray}
Similarly, it is obvious from the dependence on $L^{D-2}_{m-1}$ and $L^{D-2}_{m}$ factors appearing in the third and fourth equalities in Eqs.~(\ref{eq:memtranscoeff}) for the shear and bulk viscosities, that both $\eta$ and $\zeta$ are also expressible solely in terms of $S^{(m)}$ and $E^{(m)}$.
 
\subsubsection{Membrane paradigm: details of the analysis}

As discussed in \sec{sec:2a3}, for the $m^{\mathrm{th}}$ order \LL\ Lagrangian $L_m$ in \eq{eq:ll-action}, 
the surface stress tensor is given by (see \eq{junctionconditionLL}):
\begin{eqnarray}
8 \pi t^{\nu}_{(m) \mu} &=&  \frac{\alpha _{m} m!}{2^{m+1}} \, \sum_{s=0}^{m-1} C_s \, \pi
_{(s) \mu}^{\nu} 
\nonumber \\
\pi^{\nu}_{(s) \mu} &=& \delta _{\lbrack
\mu \mu_{1}\cdots \mu_{2m-1}]}^{[\nu \nu_{1}\cdots \nu_{2m-1}]}\,{\hat R}_{\nu_{1}\nu_{2}}^{\mu_{1}\mu_{2}}\cdots {\hat R}_{\nu_{2s-1}\nu_{2s}}^{\mu_{2s-1}\mu_{2s}}%
\,K_{\nu_{2s+1}}^{\mu_{2s+1}}\cdots K_{\nu_{2m-1}}^{\mu_{2m-1}} 
\end{eqnarray}%
with 
\begin{equation}
C_s=\sum_{q = s}^{m-1} \frac{ (-2)^{q-s} 4^{m-q} \; \binom{q}{s} }{q!\left( 2m-2q-1\right) !!}
\end{equation}
[We have replaced $c_m$ in \eq{junctionconditionLL} by $\alpha_m$ for notational clarity.]
One can check that for $m=1$ and $m=2$, the above expression reduces to that of the surface stress tensor in Einstein \cite{mtw} and Gauss-Bonnet \cite{junctionGB} theories respectively. 

As in the case of the membrane paradigm for Einstein gravity, we will interpret the $t_{\alpha \beta}$ as due to a fictitious matter source residing on the stretched horizon $\mathcal{H}_s$. Using the assumptions and expressions given above we can show that the surface stress energy tensor can be written in a form analogous to that of a viscous fluid, namely,
\begin{eqnarray}
t_{\alpha \beta}=\rho_s u_\alpha u_\beta + e^{(A)}_\alpha e^{(B)}_\beta \left( p_s
\gamma_{AB} - 2 \eta_s {\sigma_s}_{AB} - \zeta_s \theta_s \gamma_{AB}
\label{viscousfluidtensor}
\right)
\end{eqnarray}
and thus read off the corresponding energy density $\rho_s$, pressure $p_s$, shear viscosity co-efficient $\eta$ and co-efficient bulk viscosity $\zeta_s$ in terms of the geometrical quantities describing the horizon.

From the fact that ${\pi _{(s)}}^{\nu}_{\mu}$ in \eq{junctionconditionLL} is a polynomial of odd degree $\l[2(m-s)-1\r]$ in $K_{\mu \nu}$, it is easy to see that, to first order in perturbations, the only divergences in $t_{\alpha \beta}$ are of $O(N^{-1})$, and they arise from $s=(m-1)$ and $s=(m-2)$ terms in the series. This follows (upon simple counting) from Eqs.~(\ref{zerothorder}), (\ref{firstorder}), (\ref{junctionconditionLL}) and (\ref{determinanttensor}). All the other terms in the series are of $O(N)$ and hence vanish in the limit $N \rightarrow 0$. We give below an outline of the contribution of divergent terms, stating only the relevant expressions and skipping the  intermediate steps which are algebraically straightforward. We will also define
\begin{equation}
\l( Q^{(D-1)}_m \r)^{\alpha \mu \beta \nu} = (1/m) \; \partial L^{(D-1)}_{m}/\partial \hat R_{\alpha \mu \beta \nu}                                                                                                                   \end{equation} 

\begin{enumerate}

\item[A.] \textit{Divergence due to the zeroth order term}:

Only the $s = (m-1)$ term will contribute to the zeroth perturbative order. The corresponding contribution  \cite{tp-dk-sk} turns out to be
\begin{eqnarray}
\hspace{-1cm} \mathrm{For~} s = (m-1):  \hspace{0.3in} 
\l[  \pi_{(s) {\hat 0} {\hat 0}}  \r]_{\rm div} &=& 0 
\nn \\
\nn \\
\l[  \pi^A_{(s) B}  \r]_{\rm div}&=& N^{-1}  \times 2^m \; 16 \pi \left( Q^{(D-1)}_m \right)^{{\hat 0} A}_{{\hat 0} B} \; \kappa
\end{eqnarray}

\item[B.] \textit{Divergence due to the first order term}:

At first order in the perturbations, we get divergences from $s=(m-1)$ and $s=(m-2)$. The contributions from these can also be found in a straightforward way, and are given by
\begin{eqnarray}
\hspace{0.5in} \mathrm{For~} s = (m-1):  \hspace{0.3in} 
\l[  \pi_{(s) {\hat 0} {\hat 0}}  \r]_{\rm div} &=& N^{-1}  \times 2^m \; 16 \pi \left( Q^{(D-1)}_m \right)^{{\hat 0} C}_{{\hat 0} D} k^D_C
\nn \\
\nn \\
\l[  \pi^A_{(s) B}  \r]_{\rm div} &=& N^{-1}  \times 2^m \; 16 \pi \left( Q^{(D-1)}_m \right)^{A C}_{B D} k^D_C
\nn \\
\nn \\
\hspace{0.5in} \mathrm{For~} s = (m-2):  \hspace{0.3in}
\l[ \pi_{(s) {\hat 0} {\hat 0}}  \r]_{\rm div} &=& 0
\nn \\
\nn \\
\l[  \pi^A_{(s) B}  \r]_{\rm div} &=& 6 \, N^{-1} \times \frac{\kappa}{r} \times 
\delta_{{\hat 0}FBD \; \bullet \bullet \cdots \star \star}^{{\hat 0}FAC \; \bullet \bullet \cdots \star \star} \, 
\underbrace{  \hat{R}_{\bullet \bullet}^{\bullet \bullet} \cdots \hat{R}_{\star \star}^{\star \star}  }_{2(m-2) \rm{factors}}
\; k^D_C
\nn \\
\end{eqnarray}
where the symmetry factor of $6$ in the last equation comes because we need to choose, from $3$ factors of $K_{\alpha \beta}$, one factor of $\overset{(1)}{K}_{AB}$ ($3$ ways) and one factor of $K_{uu}$ from the remaining $2$ $K_{\alpha \beta}$'s ($2$ ways), which can be done in a total of $3\times 2=6$ ways.
\end{enumerate}
The above equations can be further simplified by using the following identities (the proofs for which are given in Appendix \ref{app:membrane-ids})
\begin{eqnarray}
 \left( Q^{(D-1)}_m \right)^{0A}_{0B} &=& \l( \frac{D_{2m}}{2 D_2} \r) L^{D-2}_{m-1} \; \delta^A_B 
 \label{identity1}  \\
\left( Q^{(D-1)}_m \right)^{AC}_{BD} &=& \frac{1}{2} \l( \frac{L^{D-2}_{m}}{^{D-2}R} \r) \; \delta^{AC}_{BD} 
\label{identity2} \\
\delta_{
B{\hat 0}DF \; \bullet \bullet \cdots \star \star}^{A{\hat 0}CE \; \bullet \bullet \cdots \star \star } \, 
\underbrace{  \hat{R}_{\bullet \bullet}^{\bullet \bullet} \cdots \hat{R}_{\star \star}^{\star \star}  }_{2(m-2) \rm{factors}}
&=& \l( \frac{ 2^{m-2} D_{2m}}{D_4} \r) \left( \frac{16\pi L^{D-2}_{m-1}}{^{D-2}R} \right) \; \delta^{ACE}_{BDF}
\label{identity3}
\end{eqnarray}
where we have again used the  notation $D_k=(D-k)$ introduced earlier. We can now collect the $O(N^{-1})$ terms in the total stress tensor $t_{\alpha \beta} = \overset{(0)}{t}_{\alpha \beta} +\overset{(1)}{t}_{\alpha \beta}$ upto first order in perturbation to get
\begin{eqnarray}
 t^{{\hat 0}}_{{\hat 0}} &=& N^{-1} \times 2 m \alpha_m \; \l( \frac{D_{2m}}{D_2} \r) L^{D-2}_{m-1} \; \theta 
\\
t^{A}_{B} &=& N^{-1} \times 2 m \alpha_m 
\Biggl\{  
\l( \frac{D_{2m}}{D_2} \r) L^{D-2}_{m-1} \; \kappa \; \delta^A_B
+ \l( \frac{L^{D-2}_{m}}{^{D-2}R} \r) \; k^A_B
- 2 (m-1) D_{2m} \left( \frac{L^{D-2}_{m}}{^{D-2}R}\right) \frac{\kappa}{r_H} k^A_B
\Biggl\}
\nn \\
\end{eqnarray}

We have thus accomplished our main task of finding the divergent contributions to surface stress tensor $t_{\mu \nu}$ of the membrane to linear order in horizon perturbations. The only divergence is of $O(N^{-1})$, which is precisely the same as in Einstein theory, and can therefore be regulated in the same way as is done in conventional membrane paradigm for Einstein gravity. That is, we simply multiply by $N$ to obtain the finite part of $t_{\mu \nu}$, with the regularization having the interpretation of canceling the infinite redshift at the horizon. Writing $k^A_B$ in terms of the traceless part $\sigma^A_B$ and trace $\theta$ as in \eq{ksplit}, we can  express the stress tensor in the required form
\begin{eqnarray}
t_{\alpha \beta}=\rho^{(m)}_s u_\alpha u_\beta + e^{(A)}_\alpha e^{(B)}_\beta \left( p^{(m)}_s
\gamma_{AB} - 2 \eta^{(m)}_s {\sigma_s}_{AB} - \zeta^{(m)}_s \theta_s \gamma_{AB}
\right)
\end{eqnarray}
and read off the pressure, transport co-efficients, etc as
\begin{eqnarray}
\textrm {Pressure : } ~~p^{(m)}_s &=& \l( \frac{D_{2 m}}{D_2} \r) \l( \frac{\kappa}{2 \pi} \r) ~ 4 \pi m \alpha_m L^{D-2}_{m-1} \nonumber \\
\textrm {Energy density : }~ \rho^{(m)}_s &=& - \l( \frac{\theta_s}{\kappa}\r) p^{(m)}_s \nonumber \\
\textrm {Shear Viscosity : }~ \eta^{(m)}_s &=& \l( \frac{m \alpha_m }{^{^{(D-2)}}R} \r) \Biggl[ L^{D-2}_m - \l( \frac{2 \kappa}{r_H} \r) 
 (m-1) D_{2m}  L^{D-2}_{m-1} 
\Biggl] \nonumber \\
\textrm {Bulk Viscosity : }~ \zeta^{(m)}_s &=& - \l( \frac{2 D_3}{D_2} \r) \eta^{(m)}_s
\label{finalexpressioneta}
\end{eqnarray}
where, again,  $D_k \equiv (D-k)$. One can check that for Einstein's gravity $m=1$ and Gauss-Bonnet gravity $m=2$, the above expressions reduce to the corresponding  expressions in the literature \cite{membranekipthorne, membraneGB}. But of course, as is evident, the final expressions yield much more than just a generalization of conventional results from Einstein theory to \LL\ models. As we already indicated  earlier the above expressions highlight a connection between the fluid transport coefficients and thermal properties of the horizon and calls for a more thorough investigation into possible physical implications of the results. 

In particular, one quantity which has gained significance from the point of view of the AdS-CFT correspondence is the ratio of shear viscosity to entropy density, $\eta/s$. Using the expression of Wald entropy given in \eq{waldentropy} in above expressions, we arrive at
\begin{eqnarray}
 \frac{\eta^{(m)}}{s^{(m)}} &=& \l( \frac{1}{4\pi \; ^{^{D-2}}R } \r) \Biggl[ \frac{L^{D-2}_m}{L^{D-2}_{m-1}} - \l( \frac{2 \kappa}{r_H} \r) 
 (m-1) D_{2m} \Biggl]
 \nn \\
 \nn \\
 &=& \l( \frac{D_{2m}}{4\pi D_2 D_3} \r) \Biggl[ D_{2m+1} - 2 \left( \kappa r_H \right) 
 (m-1) \mathcal{K}^{-1} \Biggl]
 \nn \\
\nn \\ 
&\equiv& \frac{C_m}{4 \pi} 
\end{eqnarray}
where we have used the relation $L^{D-2}_{m} = \l( \mathcal{K}/r^2 \r) D_{2m} D_{2m+1} L^{D-2}_{m-1}$ which holds for a  $(D-2)$ dimensional maximally symmetric spacetime. We can therefore write
\begin{eqnarray}
 \frac{\eta}{s} &=& \frac{1}{4 \pi} \l[ 
 \frac{1+\sum \limits_{m=2}^{m_c} C_m {\bar s}_m}{1+\sum \limits_{m=2}^{m_c} {\bar s}_m \textcolor{white}{C_m} }
\r]
\end{eqnarray}
where $m_c=[(D-1)/2]$ and ${\bar s}_m := s_m/s_1=4 s_m$. 

One can now investigate the ratio $\eta/s$ on a case-by-case basis, considering specific solutions with horizons in \LL\ theory. Specifically, for {\it planar} 
\footnote{A couple of clarifying points while dealing with planar horizons are in order: (i) although ($\kappa r_H$) would in general depend on $\mathcal K$, we expect it to be finite for $\mathcal K=0$, and (ii) the most general form of a planar horizon metric, consistent with the assumption of staticity, is: $\DM s_H^2 = \Omega(r/L_i) \l( \DM x^2 + \DM y^2 + \ldots \r)$, where $\Omega(r/L_i)$ is a dimensionless function constructed out of ratios of the coordinate $r$ and any other length scale(s) $L_i$ appearing in the metric. For known solutions which are asymptotically AdS, $\Omega(r/L_i) \equiv r/\ell$, $\ell$ being the AdS scale.}
horizons, $\mathcal{K}=0$, and we note that in this case, $C_m \rightarrow (\ldots)\mathcal{K}^{-1}$. 
Therefore, since ${\bar s}_m \rightarrow (\ldots)\mathcal{K}^{m-1}$, only the $m=2$ term in the sum in the numerator survives for $\mathcal{K}=0$. One is therefore led to the conclusion that no Lovelock term other than Gauss-Bonnet contributes to the $\eta/s$ ratio for the planar ($\mathcal{K}=0$) case; this contribution is given by
\begin{eqnarray}
\hspace{-1cm} \mathrm{For~\mathcal{K}=0} :  \hspace{0.5in} 
 \frac{\eta}{s} &=& \frac{1 + [C_2 {\bar s}_2]_{\mathcal{K}=0}}{4 \pi} 
\nn \\
&=& \frac{1}{4 \pi} \l[ 1 - 4 \alpha_2 (D-4) \l( \frac{\kappa}{r_H} \r) \r]
\end{eqnarray}
Although only the Gauss-Bonnet contributes for the planar case, it must be noted that $\eta/s$ will, in general, depend on other Lovelock coupling constants as well, through the ratio $\l(\kappa/r_H\r)$ which must be calculated using a specific $\mathcal{K}=0$ solution of the full Lovelock action. 

A similar result has been noted in the literature \cite{shu, edelstein} in the context of AdS-CFT correspondence with the difference being in the boundary used --- which is the AdS infinity whereas we have considered the boundary to be the stretched horizon. However, it has been argued in \cite{rgflow} that the $\eta/s$ ratio for the two cases are related by a renormalization-group flow equation in Einstein's gravity. It would indeed be interesting to check whether such a flow exist in \LL\ theory, since our result for the membrane paradigm, in conjunction with the AdS-CFT results for \LL\ gravity, does indicate such a possible connection in the $\mathcal{K}=0$ case. Further, in the {\it special case} of a black brane solution in Einstein-Gauss-Bonnet theory, we have $\kappa/r_H=(D-1)/2 \ell^2$, in which case the result above matches exactly with the one obtained in the context of AdS-CFT (see \cite{membraneGB} and references within). Using the expression for $\eta/ s$, it would be interesting to check whether it obeys or violates the KSS bound \cite{KSSbound}, $\eta/ s \geq 1/(4\pi)$, by explicitly computing it for black hole solutions known for \LL\ gravity. For $\mathcal{K}=0$ case, the KSS bound is found to be violated in the AdS-CFT context in Einstein-Gauss Bonnet gravity \cite{kssboundGB, membraneGB} and also in a general \LL\ theory of gravity \cite{shu, kssboundLL}.

We can now summarize the broad implications of the results discussed in this (and the previous) section.

\begin{enumerate}
 \item 
The connection between various transport coefficients and the thermodynamic properties of the horizon, such as temperature, entropy and quasi-local energy, becomes immediately obvious in our analysis, and it would be  interesting if such a connection holds for horizons in more general class of gravity theories.
\item
The  calculation  highlights  the connection between Wald entropy $S_{\rm{Wald}}$ of the horizon and the pressure $p_s$ of the horizon membrane. This relation is encoded in the form of an equation of state, 
\begin{eqnarray}
\frac{p_s A}{T_{\rm{loc}}} \;\; \underset{N \rightarrow 0}{\equiv} \;\; \sum \limits_{m=1}^{[D-1]/2} \l( \frac{D-2m}{D-2} \r) S^{(m)}_{_{\rm{Wald}}}
\end{eqnarray}
The algebraic steps leading to the above result are precisely the same as those leading to the evaluation of entropy of a self-gravitating configuration of densely packed shells on the verge of becoming a black hole,  discussed in the last section. 
\item
The relation between bulk and shear viscosities for \LL\ order $m$, $\zeta^{(m)}/\eta^{(m)} = -2 (D-3)/(D-2)$ is also noteworthy. This ratio is {\it independent} of $m$, which immediately implies
 \begin{equation}
  \frac{\zeta}{\eta} = -2\  \frac{D-3}{D-2}
 \end{equation}
for an action which is a \textit{sum} of  \LL\ terms. This corroborates the results already known for Einstein gravity, and recently also established for Einstein-Gauss-Bonnet case \cite{membraneGB}. It would be worth investigating whether there is a deeper reason for the robustness of this ratio. 
\end{enumerate}

\subsection{Second law for black hole entropy in \LL\ models}

Having analyzed the Noether charge entropy in \LL\ models from various different perspectives in the previous sections, we now discuss the issue of whether the change in this entropy can be shown to be positive definite; that is, whether the second law can be shown to hold  during a physical process for black hole entropy in \LL\ theory. 

In general relativity,  the ``area theorem'' implies that area of a black hole cannot decrease in any physical process, so long as the matter stress tensor satisfies the null energy condition \cite{Hawking:1971vc}. However, the equilibrium state version of first law for black holes, which has been established by Wald and collaborators, says nothing in general as to whether the Wald entropy always increases under physical processes in arbitrary higher curvature theories. In fact, black hole entropy is no longer proportional to area in arbitrary higher curvature theories in general, and \LL\ models in particular, and hence the question of whether the second law is valid in such cases is important and not yet solved (except for $f(R)$ models, see \cite{Jacobson:1995uq}). The reference \cite{Jacobson:1995uq} also gives an argument for second law for arbitrary diffeomorphism invariant theories, but the argument relies on (a rather strong) assumption that stationary comparison version of the first law implies the physical process version for quasi-stationary processes. 

We shall now sketch a recent proof of the validity of the second law for quasi-stationary physical processes involving black holes in Lanczos-Lovelock gravity \cite{tp-ss-sk}. Such a result was first proved for the Einstein-Gauss-Bonnet theory in \cite{ac-ss}, but since the result in \cite{ac-ss} is a special case of the one proved in \cite{tp-ss-sk}, we shall essentially outline the steps given in the latter reference. The main result of the analysis is that, Wald entropy of stationary black holes in \LL\ theories increases monotonically for physical processes which are quasi-stationary, at the end of which the black hole settles down to a quasi-stationary state. One of the reasons why this result is important is that it reinforces the physical fact that Wald entropy indeed behaves like ordinary thermodynamic entropy. 

We shall concentrate on stationary, non-extremal, Killing horizons. In $D$ spacetime dimensions, the event horizon is a null hypersurface ${\cal H}$ with tangent vectors $\bm k = \bm \partial_\lambda$, where $\lambda$ is an affine parameter and $\bm k$  satisfies geodesic equation. The constant $\lambda$ slices foliate the horizon and are spacelike. An arbitrary point $p$ on such slices can be assigned coordinates $\{\lambda, x^A\}$  where $x^A, \,(A=2, \cdots ,D)$ are coordinates of the point on $\lambda = 0$ slice connected to $p$ by a horizon generator. To complete the basis, one needs to introduce a second null vector $\bm l$ such that $\bm l \bm \cdot \bm k=-1$, thereby getting $\{k^a, l^a, e^{a}_{A}\}$. Finally, we would require the induced metric on any slice, $\gamma_{ab} = g_{ab} + 2 k_{(a} l_{b)}$, with $k^a \gamma_{ab}  = 0= l^a \gamma_{ab}$. The change in induced metric from one slice to another is given by \cite{Wald:1984rg},
\beq
{\cal L}_{k} \gamma_{ab} = 2 \left( \sigma_{ab} + \frac{\theta}{(D-2)} \gamma_{ab} \right),
\label{eq:metric_evolution}
\eeq
where $\sigma_{ab}$ and $\theta$ are shear and expansion, respectively, of the horizon. We will further assume that the event horizon is also a Killing horizon. This fact is true for general relativity due to strong rigidity theorem \cite{Hawking:1973uf}. However, no such proof seems to exist for the Lanczos-Lovelock models to the best of our knowledge. With this assumption, the horizon is then generated by orbits of a Killing field $\bm \xi = \bm \partial_v$ which becomes null on the horizon, and surface gravity $\kappa$ is defined in the standard manner as $\xi^a \nabla_a \xi^b = \kappa\, \xi^b$. For stationary spacetimes with a Killing horizon, $\sigma_{ab}=0; \theta=0$. Using evolution equation for $\theta$ (Raychaudhuri equation) and shear, we obtain \cite{Wald:1984rg},

\beq
R_{ab} k^a k^b = \xi^a \gamma^{b}_{i}\gamma^{c}_{j} \gamma^{d}_{k} R_{abcd} = k^a k^c \gamma^{b}_{m} \gamma^{d}_{n} R_{abcd} = 0. 
\label{eq:stationary_conditions}
\eeq

Now consider a physical process wherein a weak stress-energy tensor perturbs a stationary black hole for certain duration, and after which  the black hole settles to a stationary state in asymptotic future. This last point implies that $\bm \xi$ is an exact Killing vector in asymptotic future. Further, the flow of stress-energy is assumed to be slow, so that changes in dynamical fields are first order in some suitable (bookeeping) parameter $\epsilon$ and viscous effects can be neglected. To be more specific, $\theta \sim {\cal O}(\epsilon), \sigma_{ab}\sim {\cal O}(\epsilon)$. 

In general relativity, a concrete example of such a physical process is slow accretion of matter by a black hole of mass $M$ for a finite time, after which the black hole settles down to a stationary state. Linearized Raychaudhuri equation then gives,
\beq
\frac{d \theta}{d \lambda} \approx - R_{ab} k^a k^b = -8\, \pi \, T_{ab} k^a k^b, 
\label{eq:physical_law}
\eeq
where Einstein's equation have been used to obtain the second equality. If $ T_{ab}$ satisfies null energy condition $ T_{ab} k^a k^b \geq 0$, the rate of change of $\theta$ is negative on any slice prior to asymptotic future. Since $\theta=0$ in the future, the generators must have positive expansion during the accretion process. This suffices to prove the monotonic increase of area in the physical process. It is evident that the result depends crucially on field equations, and hence monotonicity of horizon area is only valid in case of general relativity. 

For Lanczos-Lovelock gravity, the starting point, of course, is the Wald entropy of a stationary Killing horizon \cite{Visser:1993qa, Jacobson:1993xs, Visser:1993nu},
\beq
S = \frac{1}{4} \int \rho~ \sqrt{\gamma}~dA 
\eeq
where the entropy density (after separating out the $m=1$ term) is given by (see \sec{sec:3b1}):
\beq
\rho = \left( 1 + \sum \limits_{m=2}^{[D-1)/2]} 16 \pi m \alpha_m L^{(D-2)}_{(m-1)} \right).
\eeq
The aim is to prove that this entropy always increases when a black hole is perturbed by a weak matter stress energy tensor of ${\cal O}(\epsilon)$ obeying null energy condition.
To study the entropy change
\begin{eqnarray}
\Delta S &=&\frac{1}{4}\int_{{\cal H}} \left(\frac{d\rho}{d\lambda} +\theta\, \rho \right)\,d\lambda \,\sqrt{\gamma}\, dA.
\label{change_wald_ent}
\end{eqnarray}
it is useful to define a quantity $\Theta$ as,
\beq
\Theta = \left(\frac{d\rho}{d\lambda} +\theta\, \rho \right).
\eeq
In general relativity, $\Theta$ is the same as the expansion parameter $\theta$ of null generators, but for a general \LL\ model, $\Theta$ is the rate of change of the entropy associated with a infinitesimal portion of horizon (\cite{Jacobson:1995uq} discusses similar construction in $f(R)$ gravity). Our aim is to prove that, given our assumptions, $\Theta$ is positive on any slice in a physical process. 

To proceed further, we use the fact already noted in Sec.~\ref{sec:4c1}, that the change of the $(D-2)$-dimensional scalar $L^{(D-2)}_{(m-1)}$ can be thought of due to the change in the intrinsic metric, and this  variation of $L^{(D-2)}_{(m-1)}$ simply gives the corresponding equations of motion of $(m-1)$-th order Lanczos-Lovelock term in $(D-2)$ dimensions. So, for a general Lanczos-Lovelock gravity, we can write
\begin{eqnarray}
\frac{d\rho}{d\lambda} &=& \sum \limits_{m=2}^{[D-1)/2]} 16 \pi m \alpha_m \ k^a \nabla_a ( L^{(D-2)}_{(m-1)})  \nonumber \\
&=& - \sum \limits_{m=2}^{[D-1)/2]} 16 \pi m \alpha_m \ ^{(D-2)}{\cal R}^{ab}_{(m-1)}\, {\cal L}_{k} \gamma_{ab}, \label{variation_section}
\end{eqnarray}
where the surface term has been ignored since horizon sections are assumed to be compact surfaces without boundaries. The tensor $^{(D-2)}{\cal R}_{ab}$, which is the generalization of Ricci tensor for $(m-1)$-Lanczos-Lovelock term, is given by 
\begin{eqnarray}
^{(D-2)} {\cal R}_{b \ (m-1)}^{a} &=&   \frac{1}{16 \pi} \frac{(m-1)}{2^{m}} \delta^{ a_1 b_1 \ldots a_m b_m}_{b c_1 d_1 \ldots c_m d_m} {}^{(D-2)}R_{~ a_1 b_1}^{a d_1} \cdots  
{}^{(D-2)} R_{~ a_m b_m}^{c_m d_m}
\end{eqnarray}
Using \eq{eq:metric_evolution}, one now obtains,
\begin{eqnarray}
\Theta = \theta + 16 \pi \sum \limits_{m=2}^{[D-1)/2]} \alpha_m \biggl[-2\biggl(\frac{^{(D-2)}{\cal R}_{(m-1)} \theta}{(D-2)} 
+ ^{(D-2)}{\cal R}^{ab}_{(m-1)} \sigma_{ab} \biggr) + \theta L^{(D-2)}_{(m-1)} \biggr].
\end{eqnarray}
To study the rate of change of $\Theta$ along the congruence, we use the evolution equations for expansion (the Raychaudhuri equation) and shear \cite{Wald:1984rg}. Since we work to first order in perturbation over a background stationary spacetime, we can always extract the part linear in perturbation from a product of quantities $X Y$ as 
\beq
X Y \approx  X^{(B)} \, Y^{(P)} +  X^{(P)} \,Y ^{(B)},\label{perturbation scheme}
\eeq
where $X^{(B)}$ is the value of $X$ evaluated on the stationary background, and $X^{(P)} $ is the linear part of its perturbation. Note that, for stationary background, Raychaudhuri equation demands $R^{(B)}_{ab} k^a k^b = 0$. Since $T^{(B)}_{ab} k^a k^b = 0$, this implies $E^{(B)}_{(m)ab} k^a k^b = 0$. We shall further simplify the calculation by using diffeomorphism freedom to make the null geodesic generators of the event horizon of the perturbed black hole coincide with the null geodesic generators of the stationary black hole \cite{Gao:2001ut}.

Using the evolution equation of $\theta$ and $\sigma_{ab}$ to linear order: $d\theta/d\lambda \approx - R^{(P)}_{ab} k^a k^b$ and $d\sigma_{ab}/d\lambda \approx  C^{(P)}_{acdb} k^c k^d$, and \eq{eq:stationary_conditions} for the background, the evolution of $\Theta$ to linear order in perturbation can be written as
\beq
\frac{d \Theta}{d\lambda} = - 8\pi\, T_{ab} k^a k^b + {\cal D}_{ab}k^a k^b,
\eeq
where
\begin{eqnarray}
 {\cal D}_{ab}k^a k^b &=& \sum \limits_{m=2}^{[D-1)/2]}16 \pi \alpha_m \biggl[ E^{(P)}_{(m) ab} k^a k^b 
 + 2m \,^{(D-2)}E^{(B)ab}_{(m-1)}\,
R^{(P)}_{acbd} k^{c}k^{d}\biggr].
\label{Ddefn}
\end{eqnarray}
and we have used the expression of perturbed Weyl tensor in terms of curvature and Ricci tensors, along with the relation ${}^{(D-2)}E_{(m)ab} = {}^{(D-2)}{\cal R}_{(m) ab} - (1/2)\gamma_{ab} {}^{(D-2)}L_{(m)}$.

The next, and the most crucial, step is to prove that the first order part of ${\cal D}_{ab}k^a k^b $ vanishes identically. We omit the proof of this, which can be found in \cite{tp-ss-sk}. It is, however, worth mentioning that the proof depends crucially on various algebraic properties of \LL\ models that have been discussed in earlier parts of this review. Whether there exists another way of proving the result which is adaptable to gravity theories more general than \LL\ remains an open issue.

Once the first order part of ${\cal D}_{ab} k^a k^b$ is shown to vanish identically, we finally arrive at,
\beq
\frac{d \Theta}{d\lambda} = - 8\pi\, T_{ab} k^a k^b + {\cal O}(\epsilon^2). \label{final_form}
\eeq
The above equation shows that if null energy condition holds, the rate of change of $\Theta$ is always negative during a slow classical dynamical process which perturbs the black hole to a new stationary state. Since, the final state by assumption is stationary, both $\theta$ and $\sigma$, and as a consequence $\Theta$, vanishes in the asymptotic future. We can use therefore use the same argument as in the case of general relativity to conclude that $\Theta$ must be positive at every slice during the process, thereby proving that the horizon entropy of black holes in \LL\ gravity is indeed a monotonically increasing function during any quasi-stationary physical process, i.e.
\beq
\frac{d S}{d \lambda} \geq 0. \label{final_result}
\eeq
Further comments on the implications of this result, and suggestion for further investigations, can be found in \cite{tp-ss-sk}.

\subsection{Entropy from horizon Virasoro algebra} \label{sec:4d3}

There is now considerable evidence to show that  features like entropy, temperature etc--- which were originally thought of as properties of black holes \cite{Bekenstein:1973ur,Hawking:1974rv}--- arises in a  much more general context. Several investigations  suggest that there exist certain level of universality in the thermodynamic properties of null surfaces. Hence approaches which compute the entropy density of a null surface in a generic, universal, manner might provide us with deeper insight into the quantum structure of spacetime. 
This aspect was emphasized, among others, by Carlip who has developed such an approach to compute the entropy using horizon \va. 

The basic idea is based on an approach first presented by Brown and Henneaux \cite{Brown:1986nw}, which was originally used in the context of  asymptotically AdS space-time  in ($1+2$) dimensional gravity. They found that the Fourier modes $Q_j$ of the charges corresponding to the asymptotic diffeomorphism symmetry generators obey a Virasoro algebra with central extension:
\begin{eqnarray}
i[Q_m,Q_n] = (m-n)Q_{m+n}+\frac{C}{12}m^3\delta_{m+n,0}~,
\label{virasoro}
\end{eqnarray}
where $C$ is known as the central charge. The work by Strominger \cite{Strominger:1997eq} and others showed that
 if one uses the above central charge in the Cardy formula \cite{Cardy:1986ie,Carlip:1998qw},
the resulting entropy comes out to be the Bekenstein-Hawking entropy for the ($1+2$) dimensional black hole. 
This was further developed by Carlip  \cite{Carlip:1998wz,Carlip:1999cy} using the diffeomorphism symmetry generators \textit{near the horizon}
to obtain the black hole entropy. In this approach, one begins with the  diffeomorphism generators $\xi_n^a(x)$ which preserve a set of boundary conditions near the horizon. The Fourier modes of these generators obey one sub-algebra isomorphic to Diff $S^1$ given by
\begin{eqnarray}
i\{\xi_m,\xi_n\}^a = (m-n)\xi^a_{m+n}
\label{Virasoro1}
\end{eqnarray}
where  $\{,\}$ is the Lie bracket. One can then construct the Fourier  modes $Q_n$  of the charges  corresponding to each $\xi^a_n$, either by Hamiltonian \cite{Regge:1974zd} or a covariant Lagrangian formalism \cite{Katz:1996nr,wald,Wald:1993nt,Julia:1998ys,Julia:2000er,Silva:2000ys} and evaluate the Lie brackets among them. A comparison between this algebra and \eq{virasoro} allows us to identify the central charge. Finally, one finds the zero mode eigenvalue $Q_0$ and computes the entropy of the black hole using Cardy formula.

Several related approaches have been developed using these ideas
with the hope that diffeomorphism symmetry generators may shed some light towards the microscopic degrees of freedom responsible for entropy of the horizon \cite{Solodukhin:1998tc,Brustein:2000fw,Lin:1999gf,Das:2000zs,Jing:2000yn,Terashima:2001gn,Koga:2001vq,Park:2001zn,Carlip:2002be,Cvitan:2002cs,Cvitan:2002rh,Park:1999tj,Park:1999hs,Dreyer:2001py,Silva:2002jq}.
All these approaches developed in the literature have the following ingredients: 
\begin{itemize}
\item 
The Noether current used in the approaches are defined on-shell, usually by ignoring a term which vanishes when equations of motion are used.
\item
The calculation of Lie brackets is  complicated and different approaches lead to slightly different results and it is  not clear how to interpret these differences in the calculations. 
\item
To obtain the correct result, one  has to further impose specific boundary conditions on the horizon in order to set certain terms to zero. Again, the physical meaning of these boundary conditions is often not clear.
\item
Most of the analysis (except the one in \cite{Cvitan:2002rh}) is confined to general relativity and it is not clear how to generalize the results for a wider class of theories.
\end{itemize}

We will now describe an approach to horizon entropy based on the existence of a \va\ and its central charge as developed in Ref.~\cite{tp-bibhas1} in which it was shown that there is a relatively simple way of obtaining the central charge and the horizon entropy using the off-shell Noether current in any generally covariant theory of gravity, and in particularly for the \LL\ model. The method does not require any boundary conditions to make unwanted terms to vanish. The essential idea is to use diffeomorphism invariance  of the Lagrangian under  $x^i \to x^i + \xi^i_1$ to define the Noether current $J_a[\xi_1]$  and then use \textit{its} variation $\delta_{\xi_2} J_a[\xi_1]$ under a second diffeomorphism $x^i \to x^i + \xi^i_2$ to define the Lie bracket structure.  This can be done without using the explicit form of Noether current, and the calculation of the resulting Lie bracket is algebraically quite simple  and leads to the standard results. One can then identify the resulting \va, the central charge and the zero-mode eigenvalue; plugging these into the Cardy formula, one obtains entropy density of a null surface which turns out to be the same as Wald entropy for the \LL\ models. 

The key new features of this approach are as follows:

\begin{itemize}

\item The current  is now \textit{defined and conserved} off-shell i.e. we do not use the  equations of motion in its definition and in its conservation. A  derivation of Noether current and its conservation for \textit{on-shell} condition is given in \cite{New1}. In this case a specific boundary condition is required to obtain the necessary results. But the approach in Ref.~\cite{tp-bibhas1} deals  with off-shell situation and does not require any ad-hoc boundary condition. This is in contrast to --- and an improvement on --- the previous approaches which use equations of motion extensively.  

\item The definition of the bracket among the charges is completely general in the sense that  one does not need the explicit expression for the Noether current to obtain its structure.

\item  This approach also provides a  derivation of the central term (and zero mode energy) from the surface term contribution $\delta v^a$ in the Noether current. This is  new and does not seem to have been noticed in the previous works.

\item  In the derivation of bracket we do not use the  equations of motion or specific  boundary conditions (like Dirichlet or the Neumann boundary condition) to make non-covariant terms vanish; therefore  this approach is  general enough to include any covariant theory of gravity. (This feature was again  missing in the earlier literature which used the equations of motion for the evaluation of the relevant brackets as an on-shell construct.) As a direct consequence of the above, the central term obtained is also an off-shell construct.

\end{itemize}

Consider any generally covariant Lagrangian, for which the conserved Noether current $J^a$ (with a corresponding current density $P^a$) that can be expressed as the covariant derivative of an anti-symmetric Noether potential $J^{ab}$. These  satisfy the standard conservation laws which are valid off-shell:
\begin{eqnarray}
J^a \equiv \nabla_bJ^{ab}, \quad P^a \equiv \sqrt{g}J^a; \quad  \quad \nabla_a J^a =0, \quad \partial_a P^a =0~.
\label{1.07}
\end{eqnarray}
 Let us now consider the variation of the current density itself for an arbitrary diffeomorphism $x^a\rightarrow x^a+\xi^a$. and use it to obtain the variation of the Noether charge by straight forward calculation: 
\begin{eqnarray}
\delta_\xi Q \equiv \int d\Sigma_a \delta_{\xi} P^a = \int d\Sigma_{ab}\sqrt{h}\xi^b J^a~,
\label{1.119}
\end{eqnarray}
where $h$ is the determinant of the induced metric of ($d-2$)-dimensional boundary in a $d$-dimensional spacetime.
We next define a (Lie)   bracket structure for the charges  which leads to the  usual Virasoro algebra with the central extension for a general Lagrangian of the kind $L=L(g_{ab},R_{abcd})$.
The relevant bracket among the charges is defined as: 
\begin{equation}
[Q_1,Q_2] \equiv (\delta_{\xi_1} Q[\xi_2] - \delta_{\xi_2} Q[\xi_1])  
\label{q1q2}                                                               
\end{equation} 
Then using Eq.~(\ref{1.119}), we obtain:
\begin{eqnarray}
[Q_1,Q_2]: = \int \sqrt{h}d\Sigma_{ab}\Big[\xi^a_2J^b[\xi_1] - \xi^a_1J^b[\xi_2] \Big]
\equiv\int \sqrt{h}d\Sigma_{ab}\Big[\xi^a_2J^b_1 - \xi^a_1J^b_2 \Big]
\label{1.78}
\end{eqnarray}
where we use the notation $J^b_1=J^b[\xi_1]$ etc.
We note that: 
(a) This definition is quite general and has not used any field equations. 
(b) The derivation of \eq{1.78} is simple and straightforward. 
(c) If we use the form $J^a = \nabla_b(\nabla^a\xi^b - \nabla^b\xi^a)$ for general relativity one can easily obtain the result obtained earlier in the literature, like e.g., in  Ref. \cite{Carlip:1999cy}.
We will hereafter concentrate  on Lanczos-Lovelock gravity and evaluate  Eq. (\ref{1.78}) on the ($d-2$)-dimensional null surface which is a Killing horizon.

To evaluate  Eq. (\ref{1.78}) over the Killing horizon, we will follow the stretched horizon approach of Carlip \cite{Carlip:1999cy}. Let us first mention some of the key results needed for this computation. The location of the horizon is defined by the vanishing of the norm of a timelike (approximate) Killing vector $\chi^a$. Near the horizon, one can define a vector $\rho^a$, orthogonal to the orbits of the Killing vector $\chi^a$, by the following relation
\begin{eqnarray}
\nabla_a\chi^2=-2\kappa\rho_a~,
\label{1.121}
\end{eqnarray}
where $\kappa$ is the surface gravity at the horizon, with $\chi^a\rho_a=0$. Consider a class of diffeomorphism generators given by:
\begin{eqnarray}
\xi^a = T\chi^a + R\rho^a~.
\label{1.122}
\end{eqnarray}
where $T$ and $R$ are scalar functions chosen such that the generators obey the (near-horizon) condition
$[(\chi^a\chi^b)/\chi^2]\delta_{\xi} g_{ab}\rightarrow 0$ which preserves the 
horizon structure.  This condition leads to a relation among $R$ and $T$ given by:
\begin{eqnarray}
R= \left( \frac{\chi^2}{\rho^2 \kappa}\right) \, \chi^a \nabla_a T \equiv\frac{\chi^2}{\kappa\rho^2}DT~,
\label{1.123}
\end{eqnarray}
where $D\equiv \chi^a\nabla_a$. We can now compute the Noether charge:
\begin{eqnarray}
Q[\xi] &=&  -\frac{1}{32\pi G}\int\sqrt{h}d^{d-2}X\mu_{ab}\frac{\rho}{|\chi|}P^{abcd}\mu_{cd}\Big[2\kappa T - \frac{1}{\kappa}D^2T\Big]
\nonumber
\\
&=& -\frac{1}{32\pi G}\int\sqrt{h}d^{d-2}XP^{abcd}\mu_{ab}\mu_{cd}\Big[2\kappa T - \frac{1}{\kappa}D^2T\Big]
\label{1.90}
\end{eqnarray}
and the central term of the \va\ defined by the relation,
\begin{eqnarray}
K[\xi_1,\xi_2] = [Q_1,Q_2] - Q[\{\xi_1,\xi_2\}]~,
\label{1.77}
\end{eqnarray}
where $[Q_1,Q_2]$ is given by \eq{q1q2}.
We find that the central term is given by
\begin{eqnarray}
K[\xi_1,\xi_2] = - \frac{1}{32\pi G}\int\sqrt{h}d^{d-2}XP^{abcd}\mu_{ab}\mu_{cd}\frac{1}{\kappa}\Big[DT_1~D^2T_2 - DT_2~D^2T_1\Big]~.
\label{1.93}
\end{eqnarray}
This was obtained earlier in Ref.\cite{Cvitan:2002rh} by a more complicated procedure and using the on-shell expressions. We have obtained it without using the field equations or any special boundary conditions thereby demonstrating the generality of the result.

Using a suitably defined Fourier decomposition of the $T_1$ and $T_2$ in Eq.~(\ref{1.93})  we can find the central charge and zero mode eigenvalue. We use a Fourier decomposition of $T_1$ and $T_2$ given by:
\begin{eqnarray}
T_1 = \displaystyle\sum_m A_m T_m; \,\,\,\
T_2 = \displaystyle\sum_n B_nT_n~,
\label{Fourier}
\end{eqnarray}
with $A_n^* = A_{-n}$, $B_m^*=B_{-m}$ and
\begin{eqnarray}
T_m = \frac{1}{\alpha}\exp\left[im\left(\alpha t  + g(x) + p.x_{\perp}\right)\right]~
\label{T}
\end{eqnarray}
where $\alpha$ is a constant, $g(x)$ is a  function that is regular  at the Killing horizon, $p$ is an integer and $x_{\perp}$ are the ($d-2$) tangential coordinates. Here the $t-x$ plane defines the null surface. Obviously, the result will depend on the choice made for $\alpha$ and we need to fix this to get a unique value for entropy. A natural choice, arising from the fact that near-horizon Rindler geometry exhibits periodicity in imaginary time with period $2\pi/\kappa$, is
\begin{equation}
\alpha=\kappa
\label{condalpha}
\end{equation} 
Substituting Eq.~(\ref{T}) in  Eq.~(\ref{1.93}) and defining a quantity
\begin{eqnarray}   
\hat{\cal{A}} =  - \frac{1}{2}\int \sqrt{h}d^{d-2}X P^{abcd}\mu_{ab}\mu_{cd}~,
\label{area}
\end{eqnarray}
which is
proportional to the Wald entropy,
we obtain,
\begin{eqnarray}
K[\xi_m,\xi_n]:
= -im^3\frac{\hat{\cal{A}}}{8\pi G} \delta_{n+m,0}~.
\label{2.01}
\end{eqnarray}
Similarly, using the Fourier decomposition, $Q[\xi] = \displaystyle\sum_{m} A_{m} Q[\xi_m]$ in Eq.~(\ref{1.90}), we obtain,
\begin{eqnarray}
Q[\xi_m] = \frac{\hat{\cal{A}}}{8\pi G}\delta_{m,0}~.
\label{Q1}
\end{eqnarray}
and the algebra of charges turns out to be:
\begin{eqnarray}
i[Q_m,Q_n] = (m-n)Q[\xi_{m+n}]  + m^3\frac{\hat{\cal{A}}}{8\pi G} \delta_{n+m,0}~.
\label{Q2}
\end{eqnarray}
This is the 
standard form of the Virasoro algebra in Eq.~(\ref{virasoro}) with $Q[\xi_{m+n}]\equiv Q_{m+n}$. We can identify the central charge and the zero mode eigenvalue  as:
\begin{equation}
\frac{C}{12} = \frac{\hat{\cal{A}}}{8\pi G};
\qquad
Q[\xi_0] =  
 \frac{\hat{\cal{A}}}{8\pi G} ~.
\label{C}
\end{equation}
The standard Cardy formula for the entropy is given by \cite{Cardy:1986ie,Carlip:1998qw},
\begin{eqnarray}
S=2\pi\sqrt{\frac{C\Delta}{6}}~;\qquad \Delta \equiv Q_0-\frac{C}{24}
\label{cardy}
\end{eqnarray}
which leads to
\begin{eqnarray}
S=\frac{\hat{\cal{A}}}{4G}
\label{rindlerentropy}
\end{eqnarray}
This exactly matches with the Wald entropy of the theory.

The conceptual basis of this approach was concretely demonstrated (in the case of general relativity) from another point of view \cite{tp-bibhas2}. One can show that the Noether current that arises from the \textit{surface term} of the action in general relativity, for diffeomorphisms which preserve the horizon structure, also possess a \va\ with similar properties. More precisely ref. \cite{tp-bibhas2} considers a class of diffeomorphisms which leave a particular form of the Rindler metric invariant and introduces the Noether current associated with the \textit{surface term} in the action functional for gravity (in contrast to previous approaches in the literature in which either the bulk term or the  Einstein-Hilbert action was used to define the Noether current). Given the Noether current and diffeomorphism generators,  there is a natural \va\ which can be associated with the horizon. The central extension of this \va\  again leads --- via Cardy formula --- to the correct entropy density of the horizon.
Just to provide a contrast with the previous analysis, we will mention couple of technical points. In  the earlier work, 
to obtain the correct entropy  one had to either shift the zero mode energy (as done in. e.g., \cite{Carlip:1999cy}) or choose a parameter contained in the Fourier modes of $T$ (for instance $\alpha$) as the surface gravity (as done in, e.g.,\cite{Silva:2002jq}) or both \cite{tp-bibhas1} depending on the action of the theory we have started with. \textit{Here, interestingly, one did not require any such ad hoc prescription}. This is because the parameter $\alpha$ in the expression of $T_m$ does \textit{not} appear in the final expression of entropy in this approach.

Using (a) the Noether current corresponding to the York-Gibbons-Hawking surface term and (b) choosing the vector fields by demanding invariance of the Rindler form of the metric, have not been attempted before and are new features of the work in Ref.~\cite{tp-bibhas2}. This is important because it further strengthens the peculiar feature present in all gravitational actions, viz., that the same information is encoded in both bulk and boundary terms of the action.

From a conceptual point of view, the analysis in Ref.~\cite{tp-bibhas2} provides a nice, simple, physical picture of the connection between horizon entropy and the degrees of freedom which contribute to it. In conventional physics, we are accustomed to thinking of degrees of freedom of a system as absolute and independent of observer or the coordinate system used by the observer. In such a picture, the entropy --- which is related to the logarithm of the degrees of freedom --- will also be absolute and independent of the observer. We however know that horizon entropy, horizon temperature etc. must be treated as observer dependent notions because a freely falling observer through a black hole horizon and a static observer outside the black hole will attribute different thermodynamic properties to the horizon. \textit{It follows that any microscopic degrees of freedom which leads to horizon entropy must also be necessarily  observer and coordinate dependent.}
So the question ``what are the degrees of freedom responsible for black hole entropy?'' has no observer independent answer. We find that the notion of residual gauge symmetries which are perceived as real degrees of freedom under a restricted class of diffeomorphism, allows us to quantify this notion. An observer who perceives a horizon works with a theory having less diffeomorphism symmetry if she wants to retain the structure of the metric near the horizon. This necessarily upgrades some of the original gauge degrees of freedom to effectively true degrees of freedom as far as this particular observer is concerned. As a result, this particular class of observers will attribute an entropy to the horizon in an observer dependent manner.

The result of \cite{tp-bibhas2} has already been generalised to $m=2$ and $m=3$ \LL\ terms in \cite{noether-m12}, using the corresponding surface terms for these lagrangians given in \sec{sec:2a3}, and we expect it to hold for all \LL\ terms in a straightforward manner \cite{bdt-underprep}.

\section{\LL models and the  emergent gravity paradigm} \label{sec:4}

Recent results indicate that gravity could be an emergent phenomenon, like fluid mechanics or elasticity, with the field equations governing gravitational dynamics having the same status as, say, the equations of fluid mechanics.
This  alternative perspective for classical gravity appears to be  conceptually more satisfying than the usual perspective in certain aspects. (See for a review, Refs.~\cite{rop,TPreviews},  \cite{holocosmo}.) and several features of \LL\ models strengthen this paradigm. In this section we shall describe these features.

The  back drop in which
such an approach arises is the following: it is reasonable to assume that there are certain pre-geometric variables underlying the spacetime structure and that there exists a microscopic theory describing  their dynamics. If we think of continuum spacetime as  analogous to a continuum description of a fluid, the microscopic ---    quantum gravitational --- description is analogous to statistical mechanics of the molecules of the fluid.
(That is, the pre-geometric variables will act as the ``atoms of spacetime''.)  The exact quantum theory of gravity, when it is available, will allow us to construct, in a coarse-grained long wavelength limit,  a smooth spacetime and the effective degrees of freedom (like the metric tensor) in terms of pre-geometric variables. This procedure is analogous to obtaining variables like pressure, temperature etc. for a fluid in terms of microscopic variables. The dynamical equations of the microscopic theory will also lead to some effective description for the emergent degrees of freedom. 
For example, given the microscopic description of molecules moving randomly inside a container and colliding with the walls, one can obtain the ideal gas law $PV = Nk_BT$ governing the  macroscopic (``emergent'') variables like pressure, temperature etc. Similarly, once we have the correct theory of quantum gravity (in terms of pre-geometric variables), we should be able to obtain the field equations governing the geometry by a suitable approximate description.

Such an approach, however, works only when we know the underlying microscopic theory.  How do we proceed 
when  we do \textit{not} know the underlying theory, which is the current situation as regards the spacetime? In the case of normal matter, this was indeed the situation couple of centuries back and physicists could successfully describe the behaviour of matter using thermodynamic concepts, even before we understood the molecular structure of matter. The emergent paradigm for gravity can be viewed as a similar attempt to  describe gravity in a thermodynamic language, given the fact that we do not know the exact, microscopic description of spacetime. 

In the case of normal matter, the evidence for the existence of microscopic degrees of freedom is provided by the elementary observation that matter could be heated up. Boltzmann  emphasized the fact that matter could store and transfer energy in the form of heat \textit{only because} it had microscopic degrees of freedom.  If a fluid admits continuum description all the way and has no internal degrees of freedom, it cannot store heat or have a  notion of temperature. Thus the existence of thermal phenomena provides strong internal evidence that continuum fluid mechanics misses some essential aspect of physics. This is clearly seen, in the case of an ideal gas, if we write 
\begin{equation}
 Nk_B = \frac{PV}{T}
\label{idealgasn}
\end{equation} 
The quantities on the right hand side of \eq{idealgasn} are well defined and meaningful in thermodynamics but the $N$ in the left hand side has no physical meaning in thermodynamics. It is proportional to  the number of microscopic degrees of freedom of the gas and is irrelevant in the strictly continuum limit. With hindsight, we now know that the existence of the gas constant $R=Nk_B $ for  a mole of gas has a microscopic interpretation related to the internal degrees of freedom. This is a situation in which we could have guessed (as Boltzmann did) the existence of microscopic degrees of freedom  --- and even counted it, as 
Avogadro number for a mole of gas ---
from the existence of thermal phenomena in normal matter.

In a similar vein, we can ask whether the continuum description of spacetime dynamics \textit{exhibits evidence} for a more fundamental level of microscopic description in terms of ``atoms of spacetime''. It would,  of course,  be impossible to describe the  statistical mechanics of such microscopic degrees of freedom knowing only the continuum description of spacetime, just as one could not understand the constituents of matter in the days of Boltzmann. But it must be possible to provide internal evidence strongly suggesting the \textit{existence of} such atoms of spacetime just by  probing the continuum theories of gravity. 

The first task of the emergent paradigm will be to come up with such `internal evidence' by closely examining classical theories of gravity.
We will discuss  several peculiar features in the structure of classical gravitational theories which have no deeper explanation  within the standard framework and have to be  accepted  as just algebraic accidents.
 These peculiar features 
provide us with hints about  the underlying microscopic theory. In that sense, they 
are conceptually similar to the equality of inertial and gravitational mass (which could have been thought of as an algebraic accident in Newtonian gravity  but does find a deeper explanation when gravity is described as spacetime geometry) or the fact that matter can be heated up (which defied a fundamental explanation until Boltzmann postulated the existence of microscopic degrees of freedom).
These features strongly suggest interpreting classical gravity as an emergent phenomenon
with its field equations having the same status as equations of, say, fluid mechanics. Careful analysis of classical gravity  (with a single quantum mechanical input, viz., the Davies-Unruh temperature \cite{daviesunruh}
of local Rindler horizon) leads us to this conclusion.

 In the case  of normal matter,
a more formal  link between microscopic and macroscopic description is established  by specifying certain thermodynamic potentials like entropy, free-energy etc. as suitable functions of emergent variables like pressure, temperature etc. 
The functional form of these
 potentials can, in principle, be derived from the microscopic theory;  but,
 when we do not know the microscopic theory, they are
  postulated phenomenologically from the known behaviour of the macroscopic systems. Similarly, we would expect such a thermodynamic approach to work in this case of spacetime as well. 
 It should be possible to express, say, an entropy or free energy density for spacetime in terms of suitable variables, the extremum of which should lead to a consistency condition on the background spacetime --- which will act as the equation of motion. Such a consistency condition arises even in normal thermodynamics, though it is not often described as one such.
For a gas of $N$ molecules in a volume $V$, we can express the kinetic energy and momentum transfer through collisions to the walls of the container per unit area and time, entirely in terms of the microscopic variables. 
Coarse graining these quantities we obtain the macroscopic variables $T$ and $P$. The equations of microscopic physics now demand the \textit{consistency condition}
\begin{equation}
 \frac{\begin{pmatrix}
         \displaystyle{\text{mean momentum transfer to walls}} \\ \displaystyle{\text{ per unit time per unit area}}
        \end{pmatrix}
      }
       {\left(\text{mean kinetic energy}\right)}
 = \frac{N}{V}\, =\frac{P}{T}
\end{equation}
 between the  coarse grained variables. We will show in \sec{sec:4c2} that such an approach is indeed possible to obtain the field equations of gravity.

It is also true that, in the case of normal matter, the laws of thermodynamics hold universally and do not depend on the kind of matter (ideal gas, liquid crystal, metal, ... ) one is studying. The information about the \textit{specific} kind of matter which one is studying is provided by the \textit{specific} functional form of the thermodynamic potential lik e.g., the free energy $F = F(T, V)$. Similarly, the thermodynamic framework  can describe  a wide class of possible gravitational field equations for the effective degrees of freedom. Which of these  actually describe nature depends on the specific functional form of the thermodynamic potential, say, the entropy density of spacetime. It turns out that, for the \LL\ models which we will concentrate on,  this information is encoded in the  tensor $P^{ab}_{cd}$ (which we will  call the \textit{entropy tensor}) and the nature of the resulting theory depends on the dimension of spacetime. In particular, if $D=4$, the thermodynamic paradigm selects Einstein's theory uniquely.  

This approach can be thought of as a ``top-down'' view (in real space) from classical gravity to quantum gravity. Specific quantum gravitational models which approach the problem `bottom-up' has to be consistent with the features of classical gravity described in the sequel.
In particular, this approach makes precise the task of the microscopic quantum gravity model viz., that it should lead to a specific functional form for the entropy or free energy density of spacetime, just as microscopic statistical mechanics will lead to a specific entropy (or free energy) functional for a material system. 
In this sense, the emergent approach complements microscopic approaches based on  models of quantum gravity. 
Such a perspective also implies that quantizing any classical gravitational field will be similar to quantizing, say the  equations of fluid dynamics. Gravitons will have the same status as   phonons in a solid. Neither will give us insights into the deeper microstructure --- spacetime or atoms.  
(The alternative paradigm also  has important implications for cosmology and, in fact, suggests linking cosmology with the emergence of space itself in a special manner. The details can be found in ref. \cite{holocosmo}. )

\subsection{Emergent aspects of the \LL\ equations of motion} \label{sec:4c}

The first `internal evidence' suggesting that \LL\ models describe an emergent structure of spacetime arises from the remarkable fact that \LL\ field equations reduce to a thermodynamics identity $TdS=dE+PdV$ on the horizon in any static spacetime. We will begin by describing this result.
Obviously, it will depend  crucially on the near horizon properties of certain tensors associated with  \LL\ field equations which we will describe in \sec{sec:2b4}. These will be shown to lead to the  thermodynamics identity $TdS=dE+PdV$ on the horizon, in  \sec{sec:4c1}.

\subsubsection{Near-horizon symmetries of equation of motion} \label{sec:2b4}

To investigate the near-horizon symmetries, it is convenient to use a setup which is best suited to describe geometry in the neighbourhood of a static horizon. We will describe the static case (viz., the existence of a timelike Killing vector at least in the neighbourhood of the horizon); the more general case of stationary horizons will only be mentioned when we require it.

A coordinate system which 
facilitates taking the near horizon limit is given by the Gaussian normal coordinates, which, for static spacetimes, simplifies further and can be written 
as \cite{visser1}
\begin{eqnarray}
{\DM}s^2 = - N^2 {\DM}t^2 + {\DM}n^2 + \sigma_{A B} {\DM}y^A {\DM}y^B
\label{static-metric}
\end{eqnarray}
where $\sigma_{A B}(n, y^A)$ is the transverse metric, and the Killing horizon, generated by the timelike Killing vector 
field $\bm{\xi} = \bm{\partial}_t$, is approached as $N^2 \rightarrow 0, n \rightarrow 0$. We give below some geometric quantities associated with the above metric, which will be put to good use in
subsequent analysis.

\begin{enumerate}

\item \textit{Taylor expansion of metric components}: Demanding the finiteness of curvature invariants on the horizon leads to the following Taylor 
series expansion for the lapse $N(n,y)$ and the transverse metric $\sigma_{A B}$ 
\cite{visser1}:

\begin{eqnarray}
N(n,y) &=& \kappa n \l[ 1 - \frac{1}{2} R_{\perp}(y;n=0) n^2 + O(n^3) \r] 
\nn \\
\sigma_{A B} &=& \left[ \sigma_H(y) \right]_{A B} + \frac{1}{2} \left[ \sigma_2(y) \right]_{A B} n^2 + O(n^3)
\label{eq:taylor-exp-static}
\end{eqnarray}
where we have collectively called the transverse coordinates as $y$, and $R_{\perp}$ is the Ricci scalar corresponding to $y=$ constant surface. 

\item \textit{Domain of validity}: We shall be interested in a small region in the neighborhood of the spacelike $(D-2)$-surface; more precisely, we shall assume $n \ll R_{\perp}^{-1/2}$, and would be interested in the $\kappa \rightarrow \infty$. That is, we shall 
require the length scale set by $\kappa$ to be the smallest of all length scales in the problem. Note that the coordinate $n$ gives the normal distance from the horizon. 

\item \textit{The null geodesic congruences}: Consider the null vectors 
given by:
\begin{eqnarray}
l_{i} &=& \left( -1, + N^{-1}, 0, 0 \right)
\nonumber \\
k_{i} &=& \left( -1, - N^{-1}, 0, 0 \right)
\end{eqnarray}
Explicit calculation shows that $\nabla_{\bm{l}} \bm{l}$ has only $``y"$-components, given by:
\begin{eqnarray}
\nabla_{\bm{l}} \bm{l} = - \l(2 \kappa^{-2} \r) \l[ \sigma^{AB} \; \partial_B R_{\perp} \r]_{n=0} \; {\bm \partial}_{A} + O(n^2)
\end{eqnarray}
which go to zero in the limit $\partial R_{\perp}/\kappa^3 \rightarrow 0$; therefore, in the limit we are interested 
in, it is easy to check that these vectors satisfy the geodesic equation in affinely parametrized form, that 
is; $\nabla_{\bm{l}} \bm{l} = 0 = \nabla_{\bm{k}} \bm{k}$. 

The affine parameter $\lambda$, defined by $\bm{l} \cdot \nabla \lambda = 1$ can be found by using the above form of $N(n,y)$; to the leading order, we find that, $\lambda \sim \lambda_H + (1/2) \kappa n^2$ where $\lambda = \lambda_H$ is the location of the horizon. Note that, $N^2 \bm{l} \rightarrow \bm{\xi}|_{H}$, which implies, $2 \kappa \l( \lambda - \lambda_H \r) \bm{l} \rightarrow \bm{\xi}|_{H}$. In some part of the subsequent analysis, the differentials of various geometric quantities (such as entropy) defined on the horizon, which are directly involved in the statement of the first law of thermodynamics, would  be interpreted as variations with respect to the affine parameter along the outgoing null geodesics, i.e., $\lambda$. All of above comments hold (except for few sign changes) also for the ingoing null vector $\bm k$.

\item \textit{Comoving observers and their acceleration}: Let us next look at the comoving observers in 
this spacetime, $n=$const, $y=$const, with 4-velocity $\bm u = N^{-1} \bm \partial_{t}$. It is easy to check that their
acceleration $\bm a = \bm \nabla_{\bm u} \bm u$ is given by:
\begin{eqnarray}
\bm a = \frac{1}{N^2} g^{\mu \nu} \partial_{\nu} N \; \bm \partial_{\mu}
\label{eq:comoving-acc-static}
\end{eqnarray}
The components can be worked out explicitly by using Taylor series expansions in Eqs.~(\ref{eq:taylor-exp-static}); 
these are given by [with $C=-(1/2) R_{\perp}(y;n=0)$]
\begin{eqnarray*}
a_n &=& a^n = n^{-1} + 2 C n  + O(n^2)
\\
a_{\perp} &=& ( \partial_{\perp} C ) n^2 + O(n^4)
\\
a^2 &=& a_n^2 + \sigma^{AB} a_A a_B
\\
&=& n^{-2} + 4 C + O(n^2) + O(n^4)
\end{eqnarray*}
so that
\begin{eqnarray}
\frac{a^n}{a} = 1 + O(n^3) \;\; ; \;\; \frac{a_{\perp}}{a} = O(n^3)
\end{eqnarray}
Therefore, we see that, 
\begin{eqnarray}
\bm{\hat a} = [1+O(n^3)] \; \bm \partial_n + O(n^3) \bm \partial_{A}
\end{eqnarray}
This result indicates how the curvature $R_{\perp}$ of the $t-n$ part of spacetime, which is a plane spanned by $\bm u$ and $\bm n$, will in general tend to produce an acceleration in the ``$y$" directions. We see that even in a curved spacetime, the unit vector along $\bm a$ points along $\bm n$ to an excellent approximation [$O(n^3)$]. 
\end{enumerate}

Getting back to the symmetries of equations of motion, we can now use the above near-horizon form of the metric to show that the \LL\ equations of motion satisfy the following near-horizon symmetries, generalizing the result first derived in \cite{visser1}:
\begin{eqnarray}
E^{\hat a}_{\hat b}|_H = \left[ \begin{array}{cc|c}
E_{\perp} &0 &0\\
0& E_{\perp} & 0\\
\hline
0&  0 & {E_{\parallel }}^{{\hat A} }_{{\hat B}} \vphantom{\bigg{|}} 
\end{array} 
\right] \;
\label{eom-matrix}
\end{eqnarray}
where $E_{\perp}$ is given by
\begin{eqnarray}
E_{\perp} &=& \frac{1}{16 \pi} \frac{m}{2^m} \left[ \sigma_2 \right]_{C B}  \sigma^{C A} \mathcal{E}^B_A - \frac{1}{2} L^{(D-2)}_m \nonumber 
\\ \label{eq:ll-tensor-near-hor}
\\
\mathcal{E}^B_A &=& \delta^{B A_1 \ldots B_{m-1}}_{A C_1 \ldots D_{m-1}} ~ ^{(D-2)}R^{C_1 D_1}_{A_1 B_1} \cdots ~ ^{(D-2)}R^{C_{m-1} D_{m-1}}_{A_{m-1} B_{m-1}}  \nonumber \\ 
\end{eqnarray}
In Sec.~\ref{sec:4c1}, we will use the above form when proving that \LL\ field equations reduce to $TdS = dE + P\, dV$. Here, we give a brief sketch of how the above, highly symmetric, form of \LL\ equations of motion arise near a static horizon as a result of the geometric structure of the \LL\ Lagrangian. More details can be found in the appendices of Ref. \cite{Kothawala:2009kc}.

Let us first briefly outline the proof for Einstein tensor. We begin with the following expressions for the decomposition of the Riemann tensor for the metric  in Eq.~(\ref{static-metric}) (see, for example, Section 21.5 and Exercise 21.9 of \cite{mtw}). 
\begin{eqnarray}
R^{\tslt}_{~ \mu \nu \rho} &=& 0 \nonumber \\
R_{\mu \tslt \nu \tslt} &=& N N_{| \mu \nu} \nonumber \\
R^{\mu}_{~ \nu \rho \sigma} &=& ~^{(3)} R^{\mu}_{~ \nu \rho \sigma} 
\label{rd11}
\end{eqnarray}
and
\begin{eqnarray}
~^{(3)} R_{{A B C D}} &=& ~^{(2)} R_{{A B C D}} - \l( K_{{A C}} K_{{B D}} - K_{{A D}} K_{{B C}} \r) \nonumber \\
~^{(3)} R_{\tn {B C D}} &=& K_{{A C} :{B}} - K_{{A B} :{C}} 
\label{rd1}
\end{eqnarray}
Here, $|$ and $:$ are the covariant derivatives compatible with the induced metric on \{$t=$ constant\} and \{$t=$ constant, $n=$ constant\} surfaces respectively, and $K_{{A B}} = -(1/2) \partial_n \sigma_{{A B}}$ is the extrinsic curvature of the \{$t=$ constant, $n=$ constant\} 2-surface as embedded in the \{$t=$ constant\} surface. These expressions lead to  
\begin{eqnarray}
\rnna &=& \sigma^{{A} { C}} \left[ \partial_n K_{{C} {B}} + \left( K^2 \right)_{{B} { C}} \right]
\nonumber \\
\rtta &=& \sigma^{{ A} { C}} \left[ - \frac{N_{:{ C} { B}} - K_{{ C} { B}} \partial_n N}{N} \right]
\label{rd2} 
\end{eqnarray}
Using the relevant Taylor series expansions (see Eqs. (\ref{eq:taylor-exp-static})), we obtain
\begin{eqnarray}
\partial_n \mathrm{tr}K |_{n=0} = - \frac{1}{2} \mathrm{tr}[\sigma_2]
\end{eqnarray}
which gives, correct to $O(n^2)$,
\begin{eqnarray}
\rnn = - \frac{1}{2} \mathrm{tr}[\sigma_2] = \rtt  
\label{rd3} 
\end{eqnarray}
and
\begin{eqnarray}
R^{{ A B}}_{{ C D}} = {}^{(2)}R^{{ A B}}_{{ C D}} + O(n^2)
\nonumber
\end{eqnarray}
which, in turn, implies
\begin{eqnarray}
R^{{A B}}_{{A B}} = \; R_{\parallel} + O(n^2)
\label{rd4}   
\end{eqnarray}

Finally, we use the following general expression for the Einstein tensor (see, for example, \cite{mtw}, section 14.2, pp. 344):
\begin{eqnarray}
G^{\tslt}_{\tslt} &=& - \left( \rnn + \frac{1}{2} R^{{ A B}}_{{ A B}} \right) \nonumber
\\ 
G^{\tn}_{\tn} &=& - \left( \rtt + \frac{1}{2} R^{{ A B}}_{{ A B}} \right)
\label{mtw-einstein-tensor}
\end{eqnarray}
Plugging in the above expressions for the Riemann tensor, and finally taking the limit $n \rightarrow 0$, we immediately obtain part of  Eq.~(\ref{eom-matrix}). (Note that, with the up-down components, $G^{\hatxi}_{\hatxi} = G^{\tslt}_{\tslt}$.) One can go further and analyze, in the same way, the remaining components of the Einstein tensor; the final result is \cite{visser1}
\begin{eqnarray}
E^{\hat a}_{\hat b}|_H = \left[ \begin{array}{cc|c}
E_{\perp} &0 &0\\
0& E_{\perp} & 0\\
\hline
0&  0 & {E_{\parallel }}^{{\hat A}}_{{\hat B}} \vphantom{\bigg{|}} \end{array} 
\right] \;
\label{eom-matrix1}
\end{eqnarray}
where $E^{\hat a}_{\hat b}|_H = \l( 16 \pi \r)^{-1} G^{\hat a}_{\hat b}|_H$. 

One can now proceed to do a similar analysis and obtain the near-horizon symmetries for a generic \LL\ Lagrangian. The full result, turns out to be the same as Eq.~(\ref{eom-matrix}). We skip the proof which involves a bit of combinatorics, but otherwise straightforward to establish. Here, we shall only give a proof of the identity $E^{\hat \xi}_{\hat \xi} = E^{\hatn}_{\hatn} ~ |_H$, which would be directly relevant when we discuss thermodynamic structure of the field equations.

Using the first equality in Eq.~(\ref{eom-lovelock}), it is easy to deduce the form of $E^{\textsl k}_{\textsl k}$ (no summation over $k$). The alternating determinant simplifies upon using $\delta^{\textsl k}_{\textsl k} = 1$, and the fact that none of the other indices can be $k$ due to total anti-symmetry. Therefore, we are left with
\begin{eqnarray}
E^{\textsl k}_{\textsl k} = - \frac{1}{2} L_m \l\{ \overline{k} \r\}
\end{eqnarray}
where $L_m \l\{ \overline{k} \r\}$ denotes terms which do not contain the index $k$ at all. We now specialize to the case when the index $k$ corresponds to $n$. The RHS can be further split depending on the number of occurrences of the index $\tslt$, as
\begin{eqnarray}
E^\tn_\tn = - \frac{1}{2} \l[ L_m \l\{ \overline{n}, t \r\} + L_m \l\{ \overline{n}, 2 t \r\} + L_m \l\{ \overline{n}, \overline{t} \r\} \r]
\end{eqnarray}
The first set on the RHS contains terms like $R^{\tslt { D}}_{~ A B}$ which are identically zero (see first of Eqs.~(\ref{rd11})), while the last set is the same as appears in $E^\tslt_\tslt$. So we only have to prove that $L_m \l\{ \overline{n}, 2 t \r\} = L_m \l\{ \overline{t}, 2 n \r\} $. The set $L_m \l\{ \overline{n}, 2 t \r\}$ will have two $t$'s appearing either on \textit{different} factors of $R^{a b}_{~c d}$, which would again vanish identically for the same reason as the first set, or it can have the two $t$'s appearing on the \textit{same} factor, which would contribute
\begin{eqnarray}
L_m \l\{ \overline{n}, 2 t \r\} &=& 4m \times \frac{1}{16 \pi} \frac{1}{2^m} \delta^{\tslt { B}_1 \ldots { A}_m { B}_m}_{\tslt { D}_1 \ldots { C}_m { D}_m} R^{\tslt { D}_1}_{~ \tslt { B}_1} \cdots R^{{ C}_m { D}_m}_{{ A}_m { B}_m}
\nonumber \\
&=& 4m \times \frac{1}{16 \pi} \frac{1}{2^m} \delta^{{ B}_1 \ldots { A}_m { B}_m}_{{ D}_1 \ldots { C}_m { D}_m} R^{\tslt { D}_1}_{~ \tslt { B}_1} \cdots R^{{ C}_m { D}_m}_{{ A}_m { B}_m}
\nonumber \\
\end{eqnarray}
Using Eqs.~(\ref{rd2}), we see that $L_m \l\{ \overline{n}, 2 t \r\} = L_m \l\{ \overline{t}, 2 n \r\} + O(n^2)$. Therefore, $E^\tn_\tn = E^{\tslt}_{\tslt} + O(n^2)$, and the equality holds on the horizon. The form for ${E_{\parallel }}^{{\hat A}}_{{\hat B}}$ maybe deduced in the same way, but since it is not of much use in study of horizon thermodynamics, we do not give that analysis here.

\subsubsection{Thermodynamic structure of \LL\ field equations} \label{sec:4c1}

To demonstrate the equivalence of field equations with the identity $TdS = dE + P\, dV$, we shall  use the results derived above in \sec{sec:2b4}. We begin with the $t-t$ component of the field equations, and, following an analysis similar to the one leading to Eq.~(30) of \cite{aseem-lovelock}, we arrive at,
\begin{eqnarray}
E^{\hatxi}_{\hatxi} &=& E^\tslt_\tslt = \frac{1}{16 \pi} \frac{m}{2^m} \left[ \sigma_2 \right]_{C B}  \sigma^{C A} \mathcal{E}^B_A - \frac{1}{2} L^{(D-2)}_m + O(n) 
\nonumber \\
\end{eqnarray}
where
\begin{eqnarray}
\mathcal{E}^B_A &=& \delta^{B A_1 \ldots B_{m-1}}_{A C_1 \ldots D_{m-1}} ~ ^{(D-2)}R^{C_1 D_1}_{A_1 B_1} \cdots ~ ^{(D-2)}R^{C_{m-1} D_{m-1}}_{A_{m-1} B_{m-1}}  \nonumber \\ 
\end{eqnarray}
In \sec{sec:2b4}, we proved the equality $E^{\hat \xi}_{\hat \xi} = E^{\hatn}_{\hatn}$ on the horizon. 
Therefore, we have
\begin{eqnarray}
E^{\hatxi}_{\hatxi} \;|_H = E^{\hatn}_{\hatn} \;|_H = \frac{1}{16 \pi} \frac{m}{2^m} \left[ \sigma_2 \right]_{C B}  \sigma^{C A} \mathcal{E}^B_A - \frac{1}{2} L^{(D-2)}_m \nonumber 
\end{eqnarray}
Using again the Taylor series from \sec{sec:2b4}, and the affine parameter $\lambda$ introduced in that section, we also have
\begin{eqnarray}
\delta_{\lambda} \sigma_{A B} = \frac{\delta \lambda}{\kappa} \left[ \sigma_2 \right]_{A B} + O[ (\lambda - \lambda_H)^{1/2} \; \delta \lambda ]
\end{eqnarray}
Using this, we arrive at the expression:
\begin{eqnarray}
2 E^{\hatxi}_{\hatxi} \; \sqrt{\sigma} \; \delta \lambda \;\; &\overset{H}{=}& T \underbrace{\l( \frac{1}{8} \frac{m}{2^{m-1}} \r) \mathcal{E}^{B C} \; \delta_{\lambda} \sigma_{B C} \; \sqrt{\sigma}}_{\delta_{\lambda} S} 
- \underbrace{\frac{}{} L^{(D-2)}_m \sqrt{\sigma} \; \delta \lambda}_{\delta_{\lambda} E}  
\label{eom-tds-1}
\end{eqnarray}
where we have introduced the Hawking temperature, $T=\kappa/(2 \pi)$. As the final step, we will now show that the factor multiplying $T$, as well as the second factor on the right hand side, can both be expressed as variations of quantities defined in terms of intrinsic horizon geometry, as is anticipated by the notations $``\delta_{\lambda} S"$ and $``\delta_{\lambda} E"$ we have used for them. {\it This is a remarkable fact, and is most probably true only for the \LL\ models.} Moreover, the quantities $S$ and $E$ will turn out to be the entropy and energy associated with the horizon in \LL\ theory.

To begin with, consider the quantity
\begin{eqnarray}
S = 4 \pi m \int \DM \Sigma \; L^{(D-2)}_{m-1} \label{wald-entropy-2} 
\end{eqnarray}
and note that, purely algebraically, the variation of the above expression must give equations of motion for the $(m-1)^{\mathrm{th}}$ order \LL\ term in $(D-2)$ dimensions. (The variation would also produce surface terms, which would not contribute when evaluated at $\lambda=\lambda_H$ because we will assume that the horizon is a compact surface with no boundary.) We have:
\begin{eqnarray}
\delta_{\lambda} S = - 4 \pi m \int_H \DM \Sigma \; \mathcal{E'}^{B C} ~ \delta_{\lambda} \sigma_{B C} 
\end{eqnarray}
where we have evaluated the variation on $\lambda=\lambda_H$. Noting that the Lagrangian is
\begin{eqnarray*}
L^{(D-2)}_{m-1} = \frac{1}{16 \pi} \frac{1}{2^{(m-1)}} \delta^{A_1 B_1 \ldots B_{m-1}}_{C_1 D_1 \ldots D_{m-1}} ~ \cdots ~ ^{(D-2)}R^{C_{m-1} D_{m-1}}_{A_{m-1} B_{m-1}} 
\end{eqnarray*}
and using the first of Eqs.~(\ref{eom-lovelock}), we see that
\begin{eqnarray}
\mathcal{E'}^B_C = - \frac{1}{2} \frac{1}{16 \pi} \frac{1}{2^{(m-1)}} \mathcal{E}^B_C
\end{eqnarray}
Therefore, we obtain
\begin{eqnarray}
\delta_{\lambda} S = \frac{1}{8} \frac{m}{2^{(m-1)}} \int_H \DM \Sigma \;\; \mathcal{E}^{B C} ~ \delta_{\lambda} \sigma_{B C}
\end{eqnarray}
\textit{which is precisely the integral of the factor multiplying $T$ in Eq.\;(\ref{eom-tds-1})}. As mentioned above, $S$ defined in Eq.\;(\ref{wald-entropy-2}) is a function of $\lambda$, and its derivative with respect to $\lambda$ is well defined and finite on the horizon. The expression for $S$, evaluated at $\lambda=\lambda_H$, 
\begin{eqnarray}
S\l( \lambda=\lambda_H \r) = 4 \pi m \int_H \DM \Sigma \;\; L^{(D-2)}_{m-1} \label{wald-entropy-1} 
\end{eqnarray}
is what we shall interpret as the entropy of the horizon. As mentioned before, $S$ has no physical significance away from the horizon. (As an aside, we would also like to point out the following fact; when $D=2m$, that is, when the corresponding \LL\ action is the Euler characteristic of the full $D$-dimensional manifold, then $D-2=2(m-1)$, and we see from Eq.~(\ref{wald-entropy-1}) that $S$ is proportional to the Euler characteristic of the $(D-2)$-dimensional horizon surface, determined by $L^{(D-2)}_{m-1}$; therefore, $S$ is just a constant number. For example, the Gauss-Bonnet term ($m=2$) in $D=4$ will contribute a constant additive term to the standard Bekenstein-Hawking entropy.)  

Multiplying Eq.~(\ref{eom-tds-1}) by ${\DM}^{(D-2)} y$, integrating over the horizon surface, and taking the $n \rightarrow 0$ limit, we now see that it can be written as
\begin{eqnarray}
T {\delta}_{\lambda} S - \int_H \DM \Sigma \;\; L^{(D-2)}_{m} ~ {\delta} \lambda  &=& \int_H \DM \Sigma \;\; T^{\hatxi}_{\hatxi} ~ {\delta} \lambda
\nonumber \\
&=& \int_H \DM \Sigma \;\; T^{\hatn}_{\hatn} ~ {\delta} \lambda
\nonumber \\
&=& \int_H \DM \Sigma \;\; P_{\perp} ~ {\delta} \lambda
\end{eqnarray}
where we have used the field equations $E^{\hatxi}_{\hatxi} = (1/2) T^{\hatxi}_{\hatxi}$ in the first equality, and the relation $E^{\hatxi}_{\hatxi} \;|_H = E^{\hatn}_{\hatn} \;|_H$ in the second equality. We have also used the interpretation of $T^{\hatn}_{\hatn}$ as normal pressure, $P_{\perp}$, \textit{on the horizon}. The above equation now has the desired form of the first law of thermodynamics, provided: (i) We identify the quantity $S$, defined by Eq.~(\ref{wald-entropy-1}) as the entropy of horizons in \LL\ gravity; indeed, \textit{exactly} the same expression for entropy has been obtained in the literature using independent methods, see e.g, Ref.\cite{Jacobson:1993xs}.
(That the same expression is obtained using the Noether charge approach of Wald, was demonstrated in Ref.~\cite{entropyLL}.)
 (ii) We also identify the second term on left-hand-side as $\delta_{\lambda} E$; this leads to the definition of $E$ to be
\begin{eqnarray}
E = \int^{\lambda} \delta \lambda \int_H \DM \Sigma \;\; L^{(D-2)}_{m}
\label{energy-ll}
\end{eqnarray}
where the $\lambda \rightarrow \lambda_H$ limit must be taken \textit{after} the integral is done (therefore, we would require the detailed form of $L^{(D-2)}_{m}$ as a function of $\lambda$ to calculate this explicitly). For $D = 2 (m+1)$, the integral over $H$ above is related to the Euler characteristic of the horizon, in which case $E \propto \lambda_H$ (upon setting the arbitrary integration constant to zero). For $m=2, D=4$, this reduces, of course, to the expression obtained earlier in the case of Einstein gravity. 

Let us briefly comment on the general form of $E$ for \textit{spherically symmetric} spacetimes for general \LL\ Lagrangians, with horizon at $r=r_H$, and $\lambda=r$. In this case, $L^{(D-2)}_{m} \sim (1/{\lambda}^2)^{m}$ and also $\sqrt{\sigma} \sim {\lambda}^{D-2}$. The integrand in the energy expression therefore scales as ${\lambda}^{(D-2) - 2 m}$. Integrating the RHS of Eq.~(\ref{energy-ll}), and taking the $\lambda \rightarrow \lambda_H$ limit after doing the integration, we see that $E \sim \lambda_H^{(D-2) - 2 m + 1}$. As mentioned above, for $D = 2 (m+1)$, $E \sim \lambda_H$. In fact, in the case of spherically symmetric spacetimes in \LL\ theory, the above expression can be formally shown to be \textit{exactly equivalent} to the one already known in the literature (see \cite{aseem-lovelock}, and also Ref.~[16] therein). As far as we are aware, no general expression for energy in \LL\ theory exists in the literature, and the one we have arrived at could be thought of as one definition which appears to be reasonable from physical point of view. In any case, the expression for energy clearly deserves further investigation.

Putting everything together, we see that, for \textit{generic static spacetimes} in \LL\ gravity, the field equations can be written as a thermodynamic identity:
\begin{eqnarray}
T \delta_{\lambda} S - \delta_{\lambda} E = \overline{F} \delta \lambda  
\end{eqnarray} 
where we have found it appropriate to interpret
\begin{equation}
\overline{F} = \int_H P_{\perp} \sqrt{\sigma} \; {\DM}^2 y 
\end{equation} 
as the the average normal force over the horizon ``surface" (somewhat in line with the spirit of the so called {\it membrane paradigm}, which we discuss in further detail in \sec{sec:4d5}). The term $\overline{F} ~ {\delta} \lambda$ has the interpretation of virtual work done in displacing the horizon by an affine distance ${\delta} \lambda$, in perfect analogy with the work term in the standard first law of thermodynamics. 

We have therefore established the equivalence of \LL\ field equations with the thermodyanamic identity, generalizing the result already known in Einstein gravity. In fact, as will be a recurring point in this review, we have achieved more than just a generalization of result from Einstein to all of \LL\ -- the analysis here provides a more rigorous and robust geometric basis for the thermodynamic interpretation of field equations in Einstein theory itself. To put the result in appropriate context, note that we have shown that the following relations
\begin{eqnarray*}
\underbrace{{\bm E}( \bm{\hatn}, \bm{\hatn} ) \; = - {\bm E}( \bm{u}, \bm{u} )}_{\rm near-horizon~symmetry} \hspace{0.25cm} \& \hspace{0.25cm} 
\underbrace{{\bm E}( \bm{u}, \bm{u} ) = ({1}/{2}) \; {\bm T}( \bm{u}, \bm{u} )}_{\rm field~equations} 
\end{eqnarray*}
have a thermodynamic structure. Democracy of all observers then implies that ${\bm E}=(1/2){\bm T}$ also has a  thermodynamic structure:  
\begin{eqnarray*}
{\bm E}( \bm{u}, \bm{u} ) = \frac{1}{2} \; {\bm T}( \bm{u}, \bm{u} ) \;\; \underset{\rm all~\bm{u}}{\longmapsto} \;\; {\bm E} = \frac{1}{2} \; {\bm T}
\end{eqnarray*}
Further,  in the limit $N \rightarrow 0$, we have $N^2 \bm k \rightarrow \bm \xi$. Using this and $\bm u = \bm \xi / N$, we see that ${\bm T}( \bm{u}, \bm{u} ) \rightarrow {\bm T}( \bm{\xi}, \bm{k} )$, which is the flow of energy for a virtual displacement of the horizon along $\bm k$. Another way to interpret this result is as follows: the object $g^{\perp}_{a b} = 2 \xi_{(a} k_{b)}$ is the transverse part of the induced metric, and hence $T_{a b} \xi^a l^b = (1/2) \; \mathrm{tr}_2 [T_{ab}]$, where $\mathrm{tr}_2$ is the trace with respect to the metric $2 \xi_{(a} k_{b)}$. For spherically symmetric spacetimes, since $T^0_0=T^r_r=P$, we have $(1/2) \; \mathrm{tr}_2 [T_{ab}] = P$, which is the result derived in Ref.~\cite{tpreview,gravtherm}.

\subsubsection{Field equations from  the Clausius relation} \label{sec:4c3}

The above result suggests a strong relation between gravitational dynamics and horizon thermodynamics. This idea can be further strengthened by another result, viz. that gravitational field equations can be interpreted as the entropy balance condition on local Rindler horizons. 
In the most straight forward manner, this result can be obtained very simply from the following argument \cite{tp-phyint}.

Consider an infinitesimal displacement of a local patch of the stretched (local Rindler) horizon $\mathcal{H}$ along the direction of its normal $r_a$, by an infinitesimal proper distance $\epsilon$, which will change the proper volume by $dV_{prop}=\epsilon\sqrt{\sigma}d^{D-2}x$ where $\sigma_{ab}$ is the metric in the transverse space.
 The flux of matter energy through the surface will be  $T^a_b \xi^b r_a$ (where $\xi^a$ be the approximate Killing vector corresponding to translation in the local Rindler time) and the corresponding  entropy flux
 can be obtained by multiplying the energy flux by $\beta_{\rm loc}=N\beta$.  Hence
 the `loss' of matter entropy perceived by the outside observer, because the virtual displacement of the horizon has engulfed some matter, is 
\begin{equation}
\delta S_m=\beta_{\rm loc}\delta E=\beta_{\rm loc} T^{aj}\xi_a r_j dV_{prop}. 
\end{equation} 
Interpreting $\beta_{loc}J^a$,
where $J^a$ is the Noether current corresponding
to the local Killing vector $\xi^a$ given by $J^a=2\mathcal{G}^a_b\xi^b+L\xi^a$,
 as the  gravitational entropy current, the
change in the gravitational entropy is given by 
\begin{equation}
\delta S_{\rm grav} \equiv  \beta_{loc} r_a J^a dV_{prop}                                                         \end{equation} 
(Note the occurence of the local, redshifted, temperature through $\beta_{\rm loc}=N\beta$ in  these expressions.)
As the stretched horizon approaches the true horizon,   $N r^a \to \xi^a$
 and $\beta \xi^a \xi_a L \to 0$. Hence we get, in this limit:
$
\delta S_{\rm grav} \equiv  \beta \xi_a J^a dV_{prop} = 2 \beta \mathcal{G}^{aj}\xi_a \xi_j dV_{prop}.
$
Comparing $\delta S_{\rm grav}$ and $\delta S_m$ we see that the field equations $2\mathcal{G}^a_b=T^a_b$ can be interpreted as the entropy balance condition $\delta S_{grav}=\delta S_{matt}$. This provides a direct thermodynamic interpretation of the field equations as local entropy balance in local Rindler frame.  

Though we work with entropy density, the factor $\beta=2\pi/\kappa$ cancels out  in this analysis --- as it should, since the local Rindler observer with a specific $\kappa$ was introduced only for interpretational convenience --- and  the relation $T\delta S_m=T\delta S_{\rm grav}$ would have  the same interpretation. The expression in the right hand side is  the change in the horizon (`heat') energy $H_{sur}=TS$   due to injection of matter energy. (For a more detailed discussion of this expression, see Ref. \cite{TPhorenergy}).
The context we consider corresponds to treating the local Rindler horizon  as  a  physical system (like a hot metal plate) at a given temperature and possessing certain intrinsic degrees of freedom. Then  integration  $\delta S= \delta E/T$ at constant $T$ to relate change in horizon energy to injected matter energy. Any energy injected onto a null surface hovers just outside the horizon for a very long time as far as the local Rindler observer is concerned and thermalizes at the temperature of the horizon if it is assumed to have been held fixed. This is a local version of the well known phenomenon that, the energy dropped into a Schwarzschild black hole horizon appears to hover just outside $R=2M$ as far as an outside observer is concerned. In the case of a local Rindler frame, similar effects will occur as long as the  acceleration is sufficiently high; that is, if $\dot\kappa/\kappa^2 \ll 1$.

While the above results hold quite rigorously, they only show that the field equations lead to an entropy balance condition \textit{while what we are interested in is the converse}. We want to prove that the entropy balance condition implies the field equations. There has been attempts in the literature to 'reverse-engineer' the above argument 
\cite{pmb-gen} but these ideas face a major stumbling block. As pointed out by Padmanabhan \cite{tp-re}  the reverse engineering requires interpreting $J^a$ as an entropy current which in turn is justified only if we assume the field equations (e.g, Wald needed the field equations to show that Noether charge is physical entropy.)

Clearly the idea works only if we can have an expression for entropy \textit{which can be independently justified}. In Einstein's
theory, entropy is proportional to area of the horizon which allowed Jacobson \cite{jacobson-EOS} to use the entropy balance condition to obtain the field equation in general relativity from Clausius relation. The result relied on three crucial inputs: (a) The entropy of the horizon is proportional to the area. (b) Raichaudhury equation  relates the change in the area of cross-section of null congruence to the matter stress tensor. (c) Certain technical assumptions regarding the shear and expansion of the congruence. Given these, it is straight forward to obtain Einstein's equations from the Clausius relation.

Various attempts have been made towards generalization of the original result of Jacobson to higher curvature gravity theories. Since entropy is no longer proportional to the area in a general \LL\ model, the Raichaudhury equation is pretty much useless as a tool and there is no way of directly generalizing the original idea of Jacobson. As a consequence, all the 
 ``generalizations''  are either quite different from the original idea at a conceptual level (in that the Clausius relation is not employed in the way it was originally done), or else marred by technical errors and/or ad hoc assumptions tailor-made to get the desired result. The most immediate generalization given by Jacobson and collaborators themselves \cite{jacobson-EOS-fR}, for the $f(R)$ theories, is  distinct in concept in that additional dissipative terms had to be added to the heat flux to get the ``correct" field equations. (Obviously, \textit{any} field equation can be written in the form $TdS-dE=$ (something) with the right hand side interpreted as giving some 'dissipation'', so this result is conceptually vacuous.)  More recent attempts towards generalization are, on the other hand, based on a judicious use of the expression for the Noether potential $J_{ab}$; essentially, since diffeomorphism invariance of the action is a non-dynamical symmetry, the corresponding current $J^a$ is conserved off-shell, and using its expression for a Killing field, one can hope to derive the equations of motion from a relation like $T \DM S = \delta E$. In these works, while the definition of matter flux is similar to Jacobson's, the Raychaudhuri equation is never invoked, thereby avoiding any need for some assumptions such as vanishing expansion etc. present in original Jacobson work. 

The situation, however, is far from clear since these works do not all agree with each other. For e.g., Parikh and Sarkar have pointed out issues with the Brustein and Hadad \cite{pmb-gen} paper whereas \cite{tp-re} has highlighted certain conceptual issues regarding {\it interpretation} of the results. Specifically, \cite{tp-re} has stressed some of the subtleties (alluded to above) in interpreting these results as a \textit{derivation} of field equations from thermodynamics; these results, as Ref.~\cite{tp-re}  argues, are better viewed as \textit{interpretation} of field equations as a thermodynamic relation, rather than a derivation of the former from the latter. More importantly, Brustein and Hadad derive their result using different definitions for various quantities as compared to the other two papers; this calls for a more detailed and critical look at the analysis therein. 

The most recent work in this direction is that in ref. \cite{guedens-higher-curv} which we shall briefly describe.
First, one starts with the following (very general) ansatz for horizon entropy
\beq \label{entropyform}
S =  \int_{\Sigma}  s^{ab} \epsilon_{ab} dA,
\eeq  
(where the entropy density $s^{ab}$  is an antisymmetric tensor, $\epsilon^{ab}$ is the binormal to the slice and $dA$ is the area element on the slice), and then calculate the change of this entropy between two horizon slices $\Sigma$ and $\Sigma_0$ using Stokes' theorem
\begin{eqnarray} 
\delta S &=& \oint_{\Sigma \cup {\Sigma}_0}  s^{ab} \epsilon_{ab} dA 
= - 2 \int_{H} \nabla_b s^{ab}\,  k_a dV dA
\label{deltaS} 
\end{eqnarray}
where $V$ is a parameter along $\bm k$ (which is tangent to horizon generators). Various assumptions about the slices and enclosed horizon region $H$ can be found in \cite{guedens-higher-curv}.
Second, the Clausius relation, $dS=\delta Q/T$, is  imposed with the heat flux given by
\beq\label{dQ}
\delta Q = \int_H (-T_{ab})\xi^b dH_a
\eeq
Matching the order on two sides of the above relation appropriately using behaviour of the local Killing vector (LKV) $\xi^j$ gives
\beq\label{ClausiusDivs}
- (\hbar/\pi)\nabla_b s^{ab}\,  k_a = T^{ab} \xi_b\, k_a + O(x^2).
\eeq
where $O(x^2)$ refers to higher order terms in expansion of the LKV. Note that, at this stage, the relation derived above is nothing but the local entropy balance condition which was obtained earlier in \cite{tp-phyint}, except that the form of $s_{ab}$ has not yet been fixed. This third step is attempted as follows. 
 One motivates a form for $s_{ab}$ in terms of the LKV and its covariant derivatives, and imposes the above equation to determine conditions on (and inter-relationships between) the various (tensorial) coefficients in this expression. This leads to the final form for $s_{ab}$ in terms of a $4$ rank tensor with symmetries of Riemann, with certain integrability conditions imposed, which eventually identify this tensor with $P_{abcd}$.  
This basically summarizes the key steps in \cite{guedens-higher-curv} which attempts to generalize the Clausius relation maintaining as close a connection as possible with the original result of Jacobson. There are, however, key (and inevitable) differences from earlier work along these lines, and several other subtleties which are discussed in detail in \cite{guedens-higher-curv}.

Finally, we end this section by clarifying an important point concerning distinction between \sec{sec:4c3}, where we discussed the equivalence of field equations with the first law of thermodynamics, and the discussion in the present section where one attempts to derive the field equations from the Clausius relation; that is, we wish to clarify the similarities and differences between $\DM E + P \DM V$ and $\delta Q$, both of which are eventually equated to $T \DM S$. This was done in \cite{dk-thermod-struc}, where  it was pointed out that:

\begin{enumerate}

\item There is a precise sense in which the ``heat flux" term of Jacobson, and the ``$P \DM V$" term of Padmanabhan, are different; this difference is connected with the difference in null vectors which govern the horizon deformation in the two cases.

\item The change in area of a horizon cross-section is determined by the expansion (and not its first derivative) of the ingoing null congruence $\bm k$ normalized to have unit Killing energy. 
An explicit expression for the expansion $\theta$ of this congruence can be given in terms of combination of curvature tensor components as
\begin{eqnarray}
\kappa \frac{\DM ~}{\DM \lambda} \ln \sqrt{\sigma} = R_{ab}\xi^a k^b - R_{abcd} u^a n^b u^c n^d 
\label{eq:riemm-area-change}
\end{eqnarray}
(where $\lambda$ is the afffine parameter along $\bm k$, and $\bm u$ and $\bm n$ are unit timelike and spacelike vectors spanning the horizon, see \cite{dk-thermod-struc}), which clearly shows that {\it the area change involves not just Ricci, but also the Riemann tensor} -- a point which is of relevance in the context of deriving field equations from thermodynamics.

\item From explicit $(D-2)+1+1$ decomposition of the Einstein tensor, one can show that there is also a term corresponding 
to quasi-local energy of the horizon, which must be separately accounted for when considering energy flow across the horizon.

\item The Raychaudhuri equation with the prescribed null congruence has the $O(\lambda)$ term which does not vindicate or necessitate setting the expansion to zero. The most curious aspect of the analysis, however, is the fact this term nevertheless yields the null-null component of Einstein equations, although the algebraic reason is completely different! In fact, in their recent generalization \cite{guedens-higher-curv}, Jacobson et. al. have indeed found that no conditions are imposed on their null congruence by the Clausius relation although this was invoked in the original paper of \cite{jacobson-EOS} and \cite{jacobson-EOS-fR}.

\end{enumerate}

All these facts, along with the previous discussion in this section, shows that while \textit{interpreting} the field equations as thermodynamic relations is mathematically rigorous, the \textit{derivation} of field equations from themodynamic considerations has unresolved conceptual and algebraic issues when it is attempted in the above manner. We shall see later (see \sec{sec:5}) that this is best done from a completely different approach, viz. from a thermodynamic variational principle.

\subsection{Equipartition law for microscopic degrees of freedom} \label{sec:4c2}

Given the fact that spacetime appears to be hot, just like  a body of gas, we can apply the Boltzmann paradigm (``If you can heat it, it has microstructure'') and study  the nature of the \mdof\ of the spacetime --- exactly the way people studied gas dynamics \textit{before} the atomic structure of matter was understood.  
One key relation in such an approach is the  equipartition law $\Delta E = (1/2) k_BT \Delta N$ relating the number density $\Delta N$ of \mdof\ we need to store an energy $\Delta E$ at temperature $T$. (This  number is closely related to the Avogadro number of a gas, which was known even before one figured out what it was counting!).  

If gravity is the thermodynamic limit of the underlying statistical mechanics, describing the `atoms of spacetime', we should be able to relate  $E$ and $T$ of a given spacetime and  determine the number density of \mdof\ of the spacetime when everything is static.
 Remarkably enough, we can do this directly from the gravitational field equations \cite{cqg04,tpsurface1,tpsurface2}. 
The establishment of such a relation would obviously be important from the perspective of the emergent gravity paradigm as well.

Since the basic idea here is quite simple (it is explained, for e.g., \cite{cqg04,tpsurface1,tpsurface2}), it is convenient to first describe the result for Einstein theory (which is essentially some well known relations interpreted differently), and then describe the steps  needed to derive it more formally for the entire class of \LL\ models using the Noether current and potential. We shall work with  a static spacetime with metric components $g_{00}=-N^2, g_{0\alpha}=0, g_{\alpha\beta}=h_{\alpha\beta}$. An observer at rest in this spacetime with four-velocity $u^a=\delta^a_0/N$ will have an acceleration $a^j=(0,a^\mu)$ where $a_\mu=(\partial_\mu N/N)$. We consider a spatial 3-volume $\mathcal{V}$ in this spacetime, having a boundary $\partial \mathcal{V}$.

\subsubsection{Equipartition law in Einstein theory}

To fix the ideas, we will begin with the simplest context in which one can determine the density of \mdof, viz., a static spacetime in Einstein's general relativity. 
In a static spacetime, it is easy to show that  
\begin{equation}
R_{ab} u^a u^b = \nabla_ia^i = N^{-1} D_\mu(Na^\mu)
\label{einsteinrab}
\end{equation}
where $D_\mu$ is the covariant derivative operator corresponding to the 3-space metric $h_{\alpha\beta}$.  
Using the Einstein field equations in \eq{einsteinrab}, we can relate the divergence of the acceleration to the  source $\bar T_{ab}\equiv(T_{ab}-(1/2)g_{ab}T)$ by:
\begin{equation}
D_\mu(Na^\mu)\equiv 8\pi N\bar{T}_{ab}u^au^b
\equiv 4\pi \rho_{\rm Komar}
\label{arhokomar}
\end{equation}
where $\rho_{\rm Komar}$ is the (so called) Komar mass-energy density.
Integrating both sides of \eq{arhokomar} over  $\mathcal{V}$ and using Gauss theorem gives:
\begin{equation}
E\equiv \int_{\mathcal{V}} d^3x\sqrt{h}\rho_{\rm Komar}
=\frac{1}{2}
 \int_{\partial\cal V}\frac{\sqrt{\sigma}\, d^2x}{L_P^2}\left\{\frac{N a^\mu n_\mu}{2\pi}
\right\}
\label{komarintegral}
\end{equation} 
where $\sigma$ is the determinant of the induced metric on $\partial \mathcal{V}$ and $n_\mu$ is the spatial normal to $\partial \mathcal{V}$. (We have temporarily restored $G=L_P^2\neq 1$ keeping $\hbar=c=k_B=1$.) 

\textit{This result has a remarkable interpretation directly related to the equipartition law of conventional thermodynamics}\cite{cqg04,tpsurface1,tpsurface2}. 
To see this, choose $\partial \mathcal{V}$ to be a $N=$ constant surface so that the normal $n_\mu$ is in the direction of the acceleration and $a^\mu n_\mu= |\mbt{a}|$ is the magnitude of the acceleration. The local, redshifted, Davies-Unruh temperature\cite{daviesunruh} is then given by: $T =NT_{\rm loc}=(N a^\mu n_\mu /2\pi)=(N |\mbt{a}| /2\pi)$. Therefore \eq{komarintegral} is exactly in the form of the equipartition law (with $k_B=1$) where
\begin{equation}
E=\frac{1}{2}
 \int_{\partial\cal V}dn T \;\; ; \quad \Delta n \equiv  \frac{\sqrt{\sigma}\,d^2 x}{ L_P^2}
 \label{idn}
\end{equation} 
That is, demanding the validity of  Einstein equations, and interpreting \eq{idn} as a  law of equipartition, we can \textit{determine} the number of \mdof\ in an element of area $\Delta A = \sqrt{\sigma} d^2 x$ on $\partial \mathcal{V}$ to be 
$ \sqrt{\sigma} d^2 x/ L_P^2$. If these microscopic degrees of freedom are in equilibrium at the temperature $T$, then the total equipartition energy contributed by all these degrees of freedom  on $\partial \mathcal{V}$ is equal to the total energy contained in the bulk volume enclosed by the surface --- which could be thought of as another occurence  of the holographic principle. 

\subsubsection{Equipartition law in \LL\ models}
 
Remarkably enough, the equipartition law exists for \textit{any} diffeomorphism invariant theory of gravity and allows us to identify the corresponding surface density $\Delta n/\sqrt{\sigma}d^2x$ of microscopic states in a given theory. However, we restrict our attention to \LL\ models since the general case requires an additional assumption involving bifurcate Killing horizon which makes it's status a bit unclear.

To demonstrate how the result generalizes, we follow precisely the same steps \cite{tpsurface1,tpsurface2} as in the Einstein case, but now giving a generalized version of the equations encountered before. For \LL\ actions, the object which replaces Ricci tensor of Einstein theory is defined by $E_{ab} + (1/2) L g_{ab}$, which, from \eq{eab}, gives
 \begin{equation}
\mathcal{R}_{ab} = P_a^{\phantom{a} cde} R_{bcde} 
\label{genEab}
\end{equation}
(Note that $\mathcal{R}_{ab} \Rightarrow (16 \pi)^{-1}R_{ab}$ in the Einstein case.) We shall also require the expression  for Noether current $J^a$ associated with the timelike Killing vector $\xi^a$, which is quite simple and is given by $J^a=2 \mathcal{R}^a_b \xi^b$. One is now set to rerun the steps outlined for Einstein theory above, as follows.

Using the definition of $J^{ab}$ (or expression \eq{noedef} directly for a Killing vector), we get:
\begin{equation}
 2 \mathcal{R}_{ab} u^a u^b = \nabla_a (J^{ba} u_b N^{-1})
 \label{ruutotdiv}
\end{equation}  
Further, in a static spacetime, $\nabla_i Q^i = N^{-1} D_\alpha (N Q^\alpha) $ for any static vector (with $\partial_0 Q^i=0$) and hence we can  write this  as:
\begin{equation}
2 N \mathcal{R}_{ab} u^a u^b  = D_\alpha (J^{b\alpha} u_b)
\end{equation} 
which generalizes of \eq{einsteinrab} to an arbitrary theory of gravity. 
The source for gravity in a general theory (analogous to Komar energy density) is {\it defined} through 
\begin{eqnarray}
\rho \equiv 4N \mathcal{R}_{ab} u^a u^b
\end{eqnarray}
For \LL\ models, $\rho$ can be written as a function the matter stress-energy tensor $T_{ab}$ (generalizing the $\bar{T}_{ab}$ used to define Komar energy in Einstein case above) using the field equations. This is possible for \LL\ models since trace of the equations of motion is proportional to the Lagrangian. However, the explicit form will not be needed here. 
Integrating both sides over a region $\mathcal{V}$ bounded by $\partial\mathcal{V}$, we get
\begin{eqnarray}
E &\equiv& \int_\mathcal{V} \sqrt{h} d^{D-1}x \ \rho 
\nn \\
&=& \int_\mathcal{V} 2 N\mathcal{R}_{ab} u^a u^b \sqrt{h}\, d^{D-1}x 
\nn \\
&=& \int_{\partial\mathcal{V}}d^{D-2}x  \sqrt{\sigma}(n_iu_bJ^{bi}) 
= \int_{\partial\mathcal{V}}d^{D-2}x  \sqrt{\sigma} (N n_\alpha J^{\alpha 0})
\label{genequi}
\end{eqnarray}
which is the analogue of \eq{komarintegral}. In Einstein's theory, $J^{\mu0}=a^\mu/8\pi$ which will reduce \eq{genequi} to \eq{komarintegral}. In a general theory, the expression for $\Delta n$ is not just proportional to the area and  $J^{\alpha0}$  quantifies this difference. 

To obtain the equipartition law in a more explicit form, we note that (from \eq{noedef}), we get $J^{ab}=2P^{abcd}\nabla_c\xi_d$ and hence $J^{\alpha 0}=4|\mbt{a}|P^{0\alpha}_{0\beta}n^\beta$. So
\begin{equation}
 E=\int_{\partial\mathcal{V}}d^{D-2}x  \sqrt{\sigma} (16\pi P^{0\alpha}_{0\beta}n^\beta n_\alpha)
 \left(\frac{N|\mbt{a}|}{2\pi}\right)\equiv\frac{1}{2}\int dn T
 \label{equipartlaw}
\end{equation}
where $T=N|\mbt{a}|/2\pi$ is the Davies-Unruh temperature as before. The surface density of microscopic degrees of freedom, when the field equations are satisfied (step 2 above), can be now read-off as:
\begin{equation}
 \frac{dn}{\sqrt{\sigma}d^{D-2}x}=32\pi P^{ab}_{cd}\epsilon_{ab}\epsilon^{cd}
 \label{diffeoeqn}
\end{equation} 
where $\epsilon_{ab}\equiv (1/2)(u_an_b-u_bn_a)$ is the binormal to $\partial\mathcal{V}$. The connection with Wald entropy (see section (\ref{sec:3b1})) is evident. In fact, 
\begin{equation}
 S=(1/2) \beta E                  
\end{equation} 
provides the relationship between entropy, energy and temperature in all \LL\ models including general relativity as a special case. (In Ref. \cite{cqg04}, the implications of this result for general relativity was discussed extensively.) It also follows that, in all these models, we have $4 \Delta S = \Delta n$.

To sum up, the equipartition law allows one to identify the number density of microscopic degrees of freedom on a constant redshift surface. Using this one can define the entropy of the horizon in a general theory of gravity, which, as shown above, agrees with Wald entropy. These ideas will play a key role in our discussion in \sec{sec:5}. 
We now discuss one specific context in which the equipartition idea presented here might have some fundamental implication. However, what follows is mathematically independent of the results derived above.

\subsubsection{Holography and quantization of gravitational entropy}

Most of the arguments in the emergent gravity paradigm are  free of unnecessary speculations since the approach taken is ``top-down'' and relies on well-established principles.  For example, the key result of the previous section expressed in the form of \eq{diffeoeqn}, was derived from a classical theory with the only quantum mechanical input being the formula for Davies-Unruh temperature. The $\hbar$ enters the expression only through the formula $k_bT=(\hbar/c)(a/2\pi)$ which, along with $G$ in \textit{classical} gravitational field equations  leads to $G\hbar/c^3$, one of the central constants in \textit{quantum} gravity. So one does not have to make any unwarranted speculations  from the domain of quantum gravity to obtain \eq{diffeoeqn}.

However, it is worth exploring where a result like \eq{diffeoeqn} might emerge from a microscopic candidate theory of quantum gravity like for example, string theory or loop quantum gravity. To begin with, it is clear that \textit{our result in \eq{diffeoeqn} should not depend on the details of the  correct theory of quantum gravity}. Any model of quantum gravity which has (i) the correct classical limit and (ii) is consistent with 
Davies-Unruh  temperature, will lead to our \eq{diffeoeqn}. Since one would expect any quantum gravity model to satisfy these two conditions (i) and (ii) above our ``top-down" approach will meet the ``bottom-up'' approach of such a theory in the overlap domain. So if string theory or loop-quantum-gravity or dynamical triangulations or any other candidate model for quantum gravity satisfies (i) and (ii), it will lead to the equipartition. (On the other hand, if a quantum gravity model does not satisfy (i) and (ii), such a  model is probably wrong in any case.) Even though the thermodynamic limit contains far less information about the system than a microscopic description,  one could still wonder about the possible route or mechanism by which a \textit{wide class} of candidate models in quantum gravity (all satisfying (i) and (ii) above) can lead to a relation like \eq{diffeoeqn} in the appropriate limit. In this section, we will indicate one route through which such results can emerge. (Although the motivation for our discussion comes  from ideas in the emergent gravity paradigm, the results themselves are of independent interest since they generalize certain ideas about quantum mechanical spectrum of a black hole  and its connection with the highly damped quasi-normal modes.)

Let us first consider Einstein's theory in which the equipartition law assigns $A/L_P^2$ degrees of freedom to a proper area element $A$ so that $A\approx n L_P^2$ for $n\gg 1$. (This corresponds to the semiclassical limit in which one can meaningfully talk about proper area elements etc. in  the semiclassical metric.) This is, of course, nothing but area quantization in the asymptotic limit. Though LQG probably leads to such a result, in the  limit we are interested in, it can be obtained from the Bohr-Sommerfeld quantization condition applied to horizons. In fact, Bekenstein conjectured \cite{Bekenstein1} years  back that, in a quantum theory, the black hole {\it area} would be represented by a quantum operator with a discrete spectrum of eigenvalues.
 (Bekenstein showed that the area of a classical black hole behaves like an adiabatic invariant, and hence, according to Ehrenfest's theorem, the corresponding quantum operator must have a discrete spectrum.) Extending these ideas to local Rindler horizons treated as the limit of a stretched horizon, one can understand how a result like quantization of any spatial area element can arise in a microscopic theory. 

The situation becomes more interesting when we go from the Einstein gravity to \LL\ models. In Einstein gravity, entropy of the horizon is proportional to its area and hence one could equivalently claim that it is the \textit{entropy} which has an equidistant spectrum. But, in  the \LL\ models, this proportionality between area and  entropy breaks down. (Generalization of LQG  to \LL\ models does not exist as yet.) The question then arises as to whether it is the quantum of area or quantum of entropy (if  either) which arises  in these models. This question was addressed in ref. \cite{Squant} where it was shown that in the \LL\ models, it is indeed \textit{the entropy that is quantized} with an equidistant spectrum. This matches nicely with the fact that it is the quantity in the right hand side of \eq{diffeoeqn} that takes integral values in the equipartition law. Thus one can alternatively interpret our result as  entropy quantization. 

We  now indicate a sufficiently general  `mechanism' by which any microscopic model of quantum gravity can  lead to such a quantization condition, in terms of  two features. First one is the peculiar `holographic' structure of the action functionals in \LL\ models \cite{holo,tp-ayan,tp-sk-holo}. 
As noted in \sec{sec:2.7.2} , the action functionals in \textit{all} these theories can be separated into a bulk and surface term and the surface term has the structure of an integral over $d(pq)$. It can also be shown \cite{rop,holo,tp-ayan} that the same `$d(pq)$' structure emerges for the on-shell action in the \LL\ models; that is,
\begin{equation}
A|_{\rm on-shell}\propto \int_{{V}}d^Dx \partial_i( \Pi^{ijk}g_{jk})=
\int_{\partial \mathcal{V}}d\Sigma_i\; \Pi^{ijk}g_{jk}
\label{onshellpdq}
\end{equation} 
where $\Pi^{ijk}$ is the suitably defined canonical momenta corresponding to $g_{jk}$.
The second ingredient is  that, in all these theories, the above expression for the action leads to the Wald entropy of the horizon. (For example, in Einstein gravity, the surface term in action, evaluated on a horizon will give one quarter of  area \cite{holo,tp-ayan}.) In the semiclassical limit, Bohr-Sommerfeld quantization condition demands that the integral of $d(pq)$ should be equal to $2\pi n$. Since the action in \eq{onshellpdq} has this `$d(pq)$' structure,  the Bohr-Sommerfeld condition reduces to  $A|_{\rm on-shell}=2\pi n$. Since $A|_{\rm on-shell}$  is also equal to Wald entropy, we obtain
\begin{equation}
S_{\mathrm{Wald}}=A|_{\rm on-shell}=2\pi n
\label{bohr}
\end{equation}
The Bohr-Sommerfeld quantization condition, of course, was the same used originally by Bekenstein and others to argue for the \textit{area} quantization of the  horizon but in the more general context of \LL\ models it appears as entropy quantization \cite{Squant}.
(It is also possible to argue that in the
 \textit{semiclassical} limit, the
 on-shell value of the action will be related to the phase of the semiclassical wave function
$
\Psi \propto \exp \left(iA|_{\rm on-shell}\right)
$. If the semiclassical wave function describing the quantum geometry, relevant for a local Rindler observer, is obtained by integrating out the  inaccessible degrees of freedom beyond the horizon  then one can argue that \cite{Squant}, this phase should be $2\pi n$ in the asymptotic limit leading to  $A|_{\rm on-shell}=2\pi n$. Given the conceptual ambiguities related to interpretation of `wave function of geometry', it is probably better to invoke Bohr-Sommerfeld condition.)
To summarize, the three features: (i) the {\it holographic} structure of the gravitational action
functional (ii) the equality of  on-shell gravitational action functional and Wald entropy
and (iii) the Bohr-Sommerfeld quantization condition, combine together to provide a context in which \eq{diffeoeqn} can arise in any microscopic theory.

While this gives a general result that in \LL\ theories it is the entropy of the horizon that is quantized, it would be useful to  reinforce it by an explicit calculation in the standard context. Fortunately, this can be done for the $m=2$ \LL\ term, that is, for the GB model, using arguments first suggested by Hod \cite{hod} based on quasi-normal modes of black hole oscillations. We give here  a qualitative sketch of the arguments, and refer to 
\cite{Squant} for technical details. 

We first briefly summarize Hod's argument. He started from Bekenstein's arguments regarding quantum area spectrum of a non extremal Kerr-Newman black hole, and showed that the spacing of area eigenvalues can be fixed by associating the classical limit of the quasi-normal mode frequencies, $\omega_c$, with the large $n$ limit of the quantum area spectrum, in the spirit of Bohr's correspondence principle ($n$ being the quantum number).  For a Schwarzschild black hole of mass $M$ in $D=4$ dimensions, the absorption of a quantum of energy $\omega_c$,  would lead to change in the black hole area  eigenvalues as,
$
\Delta{\mathcal A}\equiv {\mathcal A}_{n+1} - {\mathcal A}_n = (\partial {\mathcal
A}/\partial M)   \omega_c
$
and for entropy $\Delta S=(\partial S/\partial M)   \omega_c$. For a Schwarzschild black hole
the level spacing of both area and entropy eigenvalues were indeed found to be equidistant, allowing one to associate the notion of a minimum unit, {\it a quantum}, of area \textit{and} entropy. 

To apply this idea to \LL\ models, we need an expression for the quasi-normal mode frequencies associated with black hole solutions in these theories. These are not known in general, but are known in the case of $5D$ Einstein-Gauss-Bonnet (GB) black holes. The form of the highly damped quasi-normal modes of these black holes suggest that it is the entropy which has a equally spaced spectrum \cite{Squant}. The result essentially depends only on the fact that ${\rm Re} ~ \omega_c \propto T$ leading to $(\partial S/\partial M)   \omega_c\ \propto T(\partial S/\partial M)$ which is a constant. On the other hand, since in this case area is a non-linear function of entropy, ${\mathcal A}=F(S)$, the area spectrum is \textit{not} equidistant. Qualitative comments concerning the precise value of the quantum of entropy (in particular, the condition under which it is $2 \pi$) can be found in \cite{Squant}. 

The case we have made for the quantization of  entropy can be analyzed in several different ways. If one applies our argument based on quasi-normal frequencies with the {\it imaginary} rather than real part of the frequencies (a strong case for doing so was suggested by Maggiore \cite{magg08}), then the result for constant spacing of entropy eigenvalues follows simply from the fact that this imaginary part, being derivable from a scattering matrix approach \cite{tp-qnm, visser-qnm}, is  independent of the specific model of gravity and depends only on the existence of a horizon.

The physical implications of the statement: {\it gravitational entropy is quantized}, are unclear at present. As should have been evident from our discussion in earlier sections, there are various proposed explanations of horizon entropy in the literature, with widely differing interpretations. For \LL\ models, most of these reproduce Wald entropy or yield expressions closely resembling it (as in \sec{sec:4d4}), and therefore the statement that Wald entropy is quantized with equidistant spacing will also generically have implications for the physical principles on which a particular approach to horizon entropy is based. Since at least some of these approaches remain close to well known physics (for instance, the approach described in \sec{sec:4d4}), the statement of quantization of entropy might be of relevance at more fundamental level than has yet been appreciated.

\subsection{Emergent aspects of the \LL\ action} \label{sec:4b}

We now turn to the thermodynamic relevance of results discussed in \sec{sec:2a4} about the peculiar structure of the action functional and related issues. We saw that the splitting of the total action into two terms led to a holographic relationship between the two. It turns out that,  there exists \textit{another} splitting of \LL\ actions which \textit{also} displays similar holographic structure. This separation is  a generalization of the  split of the Einstein-Hilbert Lagrangian as $R=2(G^0_0 - R^0_0)$. For static spacetime with a Killing horizon, the $0$-th component is uniquely determined by the Killing time and hence unambiguous. In the case of Einstein's theory, the first term ($G^0_0$) in the above split is related to energy $E$ while the second one ($R^0_0$) to entropy $S$, of a (static) horizon. More precisely, the first term is associated with the ADM Hamiltonian while the second one to projection of the Noether charge corresponding to diffeomorphism invariance of the theory, which gives the entropy of horizon in such theories. This gives the total action a thermodynamic interpretation as the free energy (at least in spacetimes with horizon). 
Further, it can be shown that the new pair above also {\it obeys the same holographic relation} \eq{result1}; the surface term in the pair arises from the $R^0_0$ part of the split - a result which holds for static spacetimes \cite{tp-sk-holo}. Since the bulk and surface terms in this case are related to energy and entropy, the holographic relation between them may be thought of as analogous to inverting the expression for entropy $S = S(E, \ldots)$ to obtain the energy $E = E(S, \ldots)$ in terms of entropy in normal thermodynamic systems. 

It can be shown that the very same holographic relationship between the bulk and the surface term exists for an {\it arbitrary} splitting of the Lagrangian into $L_{\rm sur}$ and $L_{\rm bulk}$, provided the trace of the equations of motion is proportional to the Lagrangian: $g^{ab} E_{ab} \propto L$. This is certainly true for \LL\ Lagrangians, although there is no rigorous proof as to whether there are any other Lagrangians satisfying it. (The proofs for the above statements can be found in \cite{tp-sk-holo}.)
We will now discuss these features.

\subsubsection{\LL\ action as free energy of spacetime} \label{sec:4b3}

We now present one more aspect of the gravitational action functional which adds strength to the thermodynamic interpretation. One can show that, in any static spacetime with a bifurcation horizon (so that the metric is periodic with a period $\beta = 2\pi/\kappa$ in the Euclidean sector), the {\it action functional for gravity can be interpreted as the free energy of spacetime} for  all \LL\ models. The Noether current for  \LL\ theory 
\begin{equation}
 J^a \equiv \left(2E^{ab} \xi_b+L\xi^a \right). 
\end{equation}
will play a prominent role in our discussion, where
we have assumed staticity (so that $\delta v^a=0$), and will work in the natural coordinate system in which the timelike Killing vector field $\xi^a = (1,\mathbf{0})$ explicitly exhibits the static nature. 

We start with the Noether current $J^a$  and integrate it on-shell (i.e., with $2 E_{ab}=T_{ab}$), over a constant-$t$ hypersurface with the measure $d\Sigma_a = \delta_a^0 N\sqrt{h}\, d^{D-1} x$
where $g^{E}_{00} = N^2$ and $h$ is the determinant of the spatial metric. (We  will  work in the Euclidean sector.)  Multiplying by
the period $\beta$ of the imaginary time, we get
\begin{eqnarray}
\beta \int J^a d \Sigma_a 
&=& \beta \int T^a_b \xi^b d\Sigma_a  + \beta \int L \xi^a d \Sigma_a\nonumber\\
&=& \int  (\beta  N) T^a_b u_au^b \sqrt{h}\, d^{D-1}x  + \int_0^\beta dt_E \int L \sqrt{g}\, d^{D-1}x
\label{useja}
\end{eqnarray} 
where $u^a = \xi^a/N = N^{-1}\delta^a_0$ is the $4$-velocity of observers moving along the orbits of $\xi^a$ and $\DM \Sigma_a = u_a\sqrt{h}\, d^{D-1} x$.  The term involving the Lagrangian gives the Euclidean action
for the theory, while in the term involving $T_{ab}$, we note that $\beta N \equiv \beta_{\rm loc}$ is the redshifted local temperature. If we define the (thermally averaged) energy $E$ as
\begin{equation}
\int  (\beta  N) T^a_b u_au^b \sqrt{h}\, d^{D-1}x =
\int  \beta_{\rm loc}   T^a_b u_au^b \sqrt{h}\, d^{D-1}x \equiv 
\beta E
\label{defE}
\end{equation} 
we obtain 
\begin{equation}
A= \beta \int J^a d \Sigma_a -\beta E
\end{equation} 
Upon using Stokes theorem, we see that the first term (with  the Noether charge) gives horizon entropy. Therefore, we find that 
\begin{equation}
A= S - \beta E = -\beta F 
\end{equation} 
where $F$ is the free energy.  (Usually, one defines the Euclidean action with an extra minus sign as $A=-A_E$  so that the Euclidean action can be interpreted directly as the free energy.) 
One can also obtain from \eq{useja} the  relation:
\begin{equation}
S = \beta \int J^a d \Sigma_a =  \int \sqrt{-g}\, d^Dx (\rho+L ) 
\end{equation} 
where $\rho= T_{ab} u^a u^b$.  This  gives the entropy in terms of matter energy
density and the Euclidean action. 
Alternatively, if we assume that the Euclidean action can be interpreted as 
the free energy, then these relations  provide an alternative argument
for interpreting the Noether charge as the entropy.
(Similar results have been obtained in 
Ref. \cite{Visser:1993qa,Visser:1993nu} by a more complicated procedure.)

There is another way of expressing the above result by introducing a vector $l_k\equiv \nabla_k t=\partial_kt=(1,\mathbf{0})$ which is (un-normalized) normal to $t=$ constant hypersurfaces. The unit normal is $\hat l_k=Nl_k=-u_k$ in the Lorentzian sector.  Using  $J^a=\nabla_b J^{ab}$, we can write
\begin{equation}
2\ E^k_b\xi^b + L \xi^k = J^k = \nabla_a J^{ka}
\end{equation} 
and since 
$
l_k\nabla_aJ^{ka}=\nabla_a(l_kJ^{ka})
$, we get
\begin{equation}
L=\nabla_a(l_kJ^{ka})-2 E^k_b\xi^bl_k=\frac{1}{\sqrt{-g}} \partial_\alpha \left( \sqrt{-g}\, l_kJ^{k\alpha}\right)-2 E^k_b\xi^bl_k
\end{equation} 
which is equivalent to the previous expression with $J^a$, but now expressed in a more formal way and using Noether potential (rather than current).

The free-energy structure of the \LL\ action can be very explicitly demonstrated for black hole solutions which are {\it spherically symmetric}, in which case, the the action can be written as a total derivative. The details can be found in section 4 of Ref.\cite{tp-dk-sk}, and follows from an algebraic identity which holds for \LL\ Lagrangians of a given order $m$ evaluated for spherically symmetric spacetimes with $-g_{00}=f(r)=g^{rr}$ (with horizon at $f(r=a)=0$):
\begin{eqnarray}
r^{D-2} L_m &=& \frac{d}{dr} \left[ \frac{(D-2)!}{(D-2m-1)!}(1-f)^m r^{D-2m-1} \right] 
 - \frac{d}{dr}\left[ \frac{(D-2)!}{(D-2m)!}mf^{\prime}(1-f)^{m-1} r^{D-2m} \right]
 \nn \\
 \label{eq:LLtotderiv}
\end{eqnarray}
The Euclidean action, given by $\int_0^\beta  d\tau \int d^3x \sqrt{g_E}L_m[f(r)]$, when integrated over a spacetime region bounded by the horizon at $r=a$ and an arbitrary radius $r=b(>a)$, becomes
\begin{eqnarray}
- A_{{\mathrm e} \; m} &=& \frac{\beta \Omega}{16\pi} \int_a^b dr \Biggl\{ \left[\frac{(D-2)!}{(D-2m-1)!}(1-f)^m r^{D-2m-1} \right]^{\prime} 
- \left[ \frac{(D-2)!}{(D-2m)!}mf^{\prime}(1-f)^{m-1} r^{D-2m} \right]^{\prime}   \Biggl \} 
\nonumber \\
&=& \frac{\beta \Omega}{16\pi} \left[
\underbrace{\frac{(D-2)!}{(D-2m-1)!} a^{D-2m-1}}_{ \l(16 \pi / \Omega \r) E_m } 
- 
(4 \pi T) \; 
\underbrace{\frac{m (D-2)!}{(D-2m)!} a^{D-2m}}_{4 S_m/\Omega}  
\right] + Q[f(b),f'(b)]
\nn \\
&=& \beta E_m - S_m + Q[f(b),f'(b)]
\label{zres}
\end{eqnarray}
where $\Omega$ is the area of a unit $(D-2)$-sphere, and we have used known expressions for horizon energy and entropy for spherically symmetric black holes in \LL\ theory. (The functional $Q$ depends on behaviour of the metric near $r=b$, and is irrelevant  for our argument.) 
The free energy structure is self-evident in the contribution to the Euclidean action on the horizon, thereby supporting the general arguments given earlier in this section.

\section{Entropy density of spacetime and \LL\ models} \label{sec:5}

As already described in the previous section, one of the most important basis for treating gravity as an emergent phenomenon is the possibility of deducing field equations of gravity from conventional laws of thermodynamics applied to spacetime horizons perceived by relevant class of observers. Since horizons play a central role in such discussions, it is natural to consider an approach in which null surfaces provide the fundamental variables from which gravitational dynamics of the background spacetime emerges. Indeed, since null surfaces act as one-way membranes and can block information for a specific family of observers (time-like curves), it is natural to attribute an entropy to such surfaces.

If we take this point of view seriously, then the deformations of spacetime $(\bar x^i - x^i) \equiv q^i$
associated with a vector field $q^i$ becomes analogous to deformations of a solid in the study of elasticity. By and large, such a spacetime deformation is not of much consequence except  for the  null surfaces.  As we have described earlier,  null surfaces can be thought of as acting as a local Rindler horizon to a suitable set of observers. The deformation of a local patch of a null surface  will change the amount of information accessible to the local Rindler observer. Equivalently, such an observer will associate certain amount of entropy density with the deformation of a null patch with normal  $n^a$. This suggests  that  extremizing  the sum of gravitational and matter entropy associated with \textit{all} null vector fields \textit{simultaneously},  could  lead to the equations obeyed by the background metric. 

Conceptually, this idea is  similar to the manner in which we determine the influence of gravity on other matter fields. If we fill the spacetime with freely falling observers and insist that normal laws of special relativity should hold for all inertial observers simultaneously, we can arrive at the generally covariant versions of equation obeyed by matter in an arbitrary metric. This, in turn, allows us to determine the influence of gravity on matter fields and fixes the \textit{kinematics} of gravity. To determine the \textit{dynamics}, we do the same  but now by filling the spacetime with local Rindler observers. Insisting that the local thermodynamics should lead to extremum of an entropy functional associated with every null vector in the spacetime, we  should be able to obtain the equations which  determine the background spacetime.

\textit{There is no a priori assurance that such a program will succeed} and hence it is a surprise that one can actually achieve this. Let us associate 
with every null vector field  $n^a(x)$ in the spacetime a thermodynamic potential $\Im(n^a)$ (say, entropy) which is quadratic in $n^a$ and given by:
\begin{equation}
\Im[n^a]= \Im_{grav}[n^a]+\Im_{matt}[n^a] \equiv- \left(4P_{ab}^{cd} \D_cn^a\D_dn^b -  T_{ab}n^an^b\right) \,,
\label{ent-func-2}
\end{equation}
where  $P_{ab}^{cd}$ 
and $T_{ab}$ are two tensors that play a role analogous to elastic constants in the theory of elastic deformations. If we extremize this expression with respect to $n^a$, one will  get a differential equation for $n^a$ involving its second derivatives. We, however, want to demand that the extremum holds for all $n^a$, thereby constraining the \textit{background} geometry. Further, the insistence on strictly local description of null-surface thermodynamics translates into the demand that the Euler derivative of the functional $\Im(n^a)$ should not contain any derivatives of $n^a$.  

It is indeed possible to satisfy  these conditions by the following choice: We take $P_{ab}^{cd}$ to be 
 a tensor having the symmetries of curvature tensor and  divergence-free in all its indices and  take 
$T_{ab}$ to be a divergence-free symmetric tensor. 
(The conditions $\nabla_a P^{ab}_{cd}=0, \, \nabla_a T^a_b =0$ can  be thought of as  a generalization of the notion
of ``constancy'' of elastic constants of spacetime \cite{elastic}.)
Once we get the field equations we can read off $T_{ab}$ as the matter energy-momentum tensor; our notation anticipates this result. We also know that the $P^{abcd}$ with the assigned properties   can be expressed as  $P_{ab}^{cd}=\partial L/\partial R^{ab}_{cd}$ where $L$ is the \LL\ Lagrangian and $R_{abcd}$ is the curvature tensor \cite{rop,TPreviews}. (Again, the notation  already anticipates the fact that it will turn out to be the same object as we have introduced in earlier parts of this review. However, since we are now hoping to arrive at the \LL\ actions in the context of emergent gravity paradigm, we hope to present a new perspective on the physical relevance of this tensor.)
This choice in \eq{ent-func-2} will also ensure that the  equations resulting from the entropy extremisation do not contain  derivatives of the metric  of higher order than second.

We now demand that $\delta \Im/\delta n^a=0$ for the variation of all null vectors $n^a$ with the  condition $n_an^a=0$ imposed by  a  Lagrange multiplier  $\lambda(x)g_{ab}n^an^b$ added to $\Im[n^a]$. An elementary calculation and use of
generalized Bianchi identity and the condition $\nabla_aT^a_b=0$ leads us to \cite{rop,TPreviews,aseemtp,tpgrg08} the following construct on background geometry:
\begin{equation}
\mathcal{G}^a_b =  \mathcal{R}^a_b-\frac{1}{2}\delta^a_b L = \frac{1}{2}T{}_b^a +\Lambda\delta^a_b   
\label{ent-func-71}
\end{equation}
where $\Lambda$ is an integration constant.  These are  precisely the field equations for  gravity  in a theory with \LL\ Lagrangian $L$ with an undetermined cosmological constant $\Lambda $  arising as an integration constant. 

The thermodynamical potential 
corresponding to the density $\Im$
can be obtained by integrating the density $\Im[n^a]$ over a region of space or a surface etc. based on the context. The matter part of the $\Im$,  proportional to $T_{ab}n^an^b$,  will pick out the contribution $(\rho+p)$ for an ideal fluid, which is the enthalpy density. Multiplied by $\beta=1/T$, this reduces to the entropy density because of Gibbs-Duhem relation. When the multiplication by $\beta$ can be reinterpreted in terms of integration over $(0,\beta)$ of the time coordinate (in the Euclidean version of the local Rindler frame), the $\Im$ can be interpreted as entropy and the integral over space coordinates alone can be interpreted as rate of generation of entropy.
 (This was the interpretation provided in Refs.~\cite{rop,TPreviews,aseemtp,tpgrg08} but the final result is independent of this interpretation as long as suitable boundary conditions can be imposed). One can also think of $\Im[n^a]$ as an effective Lagrangian for the collective variables $n^a$ describing the deformations of null surfaces.

The    
gravitational entropy density in \eq{ent-func-2} can be expressed in terms of the Killing 
$(\pounds_q g_{ab}= \nabla_a q_b + \nabla_b q_a)$ 
and isoentropic
$(J_{ab} = \nabla_a q_b - \nabla_b q_a)$ parts of the 
 deformations introduced in Ref. \cite{TPstructuralasp} as:
\begin{equation}
 -4P^{ab}_{cd}\nabla_aq^c \nabla_bq^d  
= P^{bijd} (\pounds_q g_{ij}) (\pounds_q g_{bd}) - (1/2) P^{abcd} J_{ab} J_{cd}
\label{entropyden}
\end{equation} 
So the gravitational entropy density has two parts: one contributed by the square of the Noether potential (which vanish for isoentropic deformations with $q_a = \nabla_a \phi$) and another which depends on the change in the metric under the deformation (which will vanish for the Killing deformations). 
  When $q_j$ is a pure gradient, $J_{ab}$  will vanish and one can identify the first term  in \eq{entropyden} with a structure like $Tr(K^2)- (Tr K)^2$. On the other hand,
when $q_a$ is a local Killing vector, the contribution from $S_{ij}$ to the entropy density vanishes and  the entropy density is just the square of the Noether  potential $J_{ab}$. For a general null vector, both the terms contribute to the entropy density. Variation of entropy density with respect to either of the two contributions (after adding suitable Lagrange multiplier to ensure vanishing of the other term) will also lead to the gravitational field equations.

In this approach, there arise several  new features which are worth mentioning.

 First,
the extremum value of the thermodynamic potential, when computed on-shell for a solution with static horizon, leads to the Wald entropy\cite{aseemtp} . This is a non-trivial consistency check on the approach because it was not designed to reproduce this result.  When the field equations hold, the total entropy of a region $\mathcal{V}$ resides on its boundary $\partial\mathcal{V}$ which is yet another illustration of the holographic nature of gravity.

Second, 
the  entropy functional in \eq{ent-func-2}  is invariant under the shift $T_{ab} \to T_{ab} + \rho_0 \gl ab$ which shifts the zero of the energy density. This symmetry allows any low energy cosmological constant, appearing as a parameter in the variational principle, to be removed thereby alleviating the cosmological constant problem to a great extent \cite{tpgrg08,de}. As far as we know, \textit{this is the only way in which  gravity can be made immune to the zero  point level of energy density}. It is again interesting that our approach leads to this result even though it is not designed for this purpose. This works because the cosmological constant, treated as an ideal fluid, has zero entropy because $\rho + p =0$ and thus cannot affect the dynamics in this perspective, in which gravity responds  to  the entropy density rather than energy density.

Third, the \textit{algebraic} reason for the  idea to work is the  identity:
\begin{equation}
 4P_{ab}^{cd} \D_cn^a\D_dn^b=2\mathcal{R}_{ab}n^an^b+\D_c[4P_{ab}^{cd} n^a\D_dn^b]
 \label{magic}
\end{equation} 
which shows that, except for a boundary term, we are extremising the integral of $(2 \mathcal{R}_{ab}-T_{ab})n^an^b$ with respect $n^a$ subject to the constraint $n_a n^a=0$. The algebra is trivial but not the underlying physical concept. 

Fourth, the  gravitational entropy density  --- which is the term in the integrand $\Im_{grav}\propto ( -P_{ab}^{cd} \D_cn^a\D_dn^b)$ in \eq{ent-func-2} --- also obeys the relation:
\begin{equation}
 \frac{\partial \Im_{\rm grav}}{\partial ( \nabla_c n^a)}= - 8 (- P^{cd}_{ab} \nabla_d n^b) =\frac{1}{4\pi} (\nabla_a n^c  - \delta^c_a \nabla_i n^i)
\label{sgravder}
\end{equation} 
where the second relation is for Einstein's theory. This term is analogous to  $t^c_a = K^c_a - \delta^c_a K$ (where $K_{ab}$ is the extrinsic curvature) that arises in the (1+3) separation of Einstein's equations. (More precisely, the  projection to 3-space leads to $t^c_a$.) 
This term  has the interpretation as the  momentum conjugate to the spatial metric in (1+3) decomposition and \eq{sgravder} shows that the entropy density leads to a similar structure. 
That is, the canonical momentum conjugate  to metric in the conventional approach and the momentum conjugate to $n^a$ in $\Im_{\rm grav}$ are similar.

One final question which must be asked is what justifies the starting expression for entropy functional as the ``right" one. This has no answer in absence of a complete quantum gravity model. The next best thing one can do, and as we have done repeatedly in this review, is to compare the on-shell value of the expression near a Rindler horizon with Wald entropy. This can be easily done, and the entropy functional described here can be shown to reproduce correctly the known expression for Wald entropy on-shell. The proof of this is straightforward and can be found in \cite{aseemtp}.

\section{Conclusions}

The \LL\ models provide a simple and elegant generalization of Einstein's theory of gravity to higher dimensions. If it turns out that we actually live in a spacetime with more than 4-dimensions, then \LL\ models probably will prove to be of practical utility in the study of nature
\footnote{There actually have been investigations of \LL\ models in many contexts other than the ones we have discussed here (see, for e.g., \cite{strauss-bimetric} in the context of bimetric theories) which are interesting in their own right, and for which the structural aspects discussed here might be of relevance.}.
But even if this is not the case, \LL\ models are of great interest in providing a natural backdrop to test different conceptual and mathematical aspects of Einstein's theory. This has been the point of view taken up and illustrated throughout the review. On several occasions we have pointed out how looking at a generalized result in \LL\ theory allows us to obtain a better insight into the phenomena in Einstein's theory. Such a study also allows us to delineate special features and numerical coincidences which occur in the $D=4$ Einstein gravity from results of much greater  generality. 

One of the surprising developments in recent years is the effortless generalizations of many concepts related to thermodynamics of horizons from Einstein gravity to these models.  Based on this one could also argue that the entire emergent paradigm of gravity possesses a much broader basis than one would have first guessed from the study of Einstein's theory. Because of its importance, we have used these topics as the focal point for this review. We hope it helps to attract more attention from the community to this fascinating subject.


\section*{Acknowledgements}

The review draws upon the work done by the authors in collaboration with 
 Sanved Kolekar, Bibhas Majhi, Ayan Mukhopadhyay, Aseem Paranjape, Sudipta Sarkar, L. Sriramkumar  and Alexander Yale.
 One of the authors (TP) thanks Sunu Engineer for discussions. The work of TP is partially supported by J.C. Bose research grant.

\section*{Appendices}
\appendix

\numberwithin{equation}{section}

\section{Algebraic aspects of certain tensors in higher curvature theories of gravity} \label{app:aspects-of-eom}

We here give a detailed analysis of certain tensors defined in \sec{sec:2a1} which appear naturally while varying a general Lagrangian which is a function of (and {\it only} of) metric and Riemann tensor and clarify many of the issues listed in \sec{sec:2a1}. We will prove some useful identities which  are non-trivial in the sense that they do not emerge from the \textit{algebraic} symmetry properties of tensors such as $P_{abcd}$ defined in \sec{sec:2a1}; instead they arise due to the  {\it diffeomorphism covariance} of the Lagrangian. This leads to the fact that, when scalar quantities are constructed from  tensorial objects (like in the case of $L$ built from $R_{abcd}$ and $g^{ij}$), the derivatives of the scalar quantity with respect to these tensors must obey some identities. While a few of these results can be proved by other methods in special cases, the procedure outlined here, based on reference \cite{tp-aspects}, is very powerful and readily adaptable to more general cases.

While varying $L=L[g_{ab}, R_{abcd}]$ to obtain the equations of motion, we come across two tensors in the following   partial derivatives
\begin{equation}
 P^{abcd} = \pdlr{L}{R_{abcd}}_{g_{ij}}; \qquad P^{ab}=\pdlr{L}{g_{ab}}_{R_{ijkl}}
 \label{one}
\end{equation} 
The subscripts specify the variables being held fixed, which, of course is, what gives  meaning to the partial derivatives. The $P^{abcd}$ is what we  called the \textit{entropy tensor} of the theory  \cite{wald,rop}. The derivatives appearing in \eq{one} are defined within the subspace of infinitesimal deformations which preserve the algebraic symmetries of the independent variable. Therefore,  $P^{ab}$ is symmetric while $P^{abcd}$  inherits the algebraic symmetries of the curvature tensor: 
\begin{equation}
 P^{abcd} = -P^{bacd} = -P^{abdc};\quad  P^{abcd} = P^{cdab}; \quad  P^{a[bcd]} =0
 \label{symm}
\end{equation} 
One can construct several such tensors --- which have the properties  in \eq{symm} --- from the curvature tensor and the metric. But these properties alone do \textit{not} guarantee that such a tensor can be expressed as the derivative of some scalar  with respect to the curvature tensor. The fact that $P^{abcd}$ has the form in \eq{one} leads to several \textit{further}  properties and inter-relationships which form the main thrust of this appendix. In particular, we will prove the following results:

(1) Let us construct the tensor 
\begin{equation}
\mathcal{R}^{ab} \equiv P^{aijk} \rc bijk
\label{eq:gen-ricci-cal}                                         
\end{equation} 
which can be thought of generalized Ricci tensor in these theories. We will prove that $P^{ab}=-2 \mr^{ab}$. That is, the derivatives of an \textit{arbitrary} scalar function with respect to the metric and the curvature are \textit{not} independent but are connected by the  relation: 
\begin{equation}
 \pdlr{L}{g_{ab}}_{R_{ijkl}} = - 2 \rc bijk \pdlr{L}{R_{aijk}}_{g_{ab}}
 \label{key11}
\end{equation} 
 It  follows from \eq{key11} that the tensor $\mr^{ab}$ is symmetric. (This was noted earlier in a different context in \cite{koga}). This result does not follow merely from the algebraic symmetries
of $P^{abcd}$  in \eq{symm} above.  The nature of the proof which we give below shows that it holds \textit{only because} $P^{abcd}$ is expressible as a derivative of a scalar with respect to the curvature tensor. 

(2) From \eq{key11} we can  obtain several other relations connecting the partial derivatives when different independent variables like 
 $(g^{ab}, R_{abcd})$ or $(g^{ab}, \rc abcd)$ or $(g^{ab},R^{ab}_{cd})$ are used to describe the system. 
 First, if we use the pair $(g^{ab}, R_{abcd})$ as independent variables, we get:
\begin{equation}
 \pdlr{L}{g^{im}}_{R_{abcd}} = - g_{ai} g_{bm} \pdlr{L}{g_{ab}}_{R_{abcd}} = 2\mr_{im}
 \label{derdown1}
\end{equation} 
Instead, if we use the pair $(g^{ab}, \rc abcd)$ as independent variables, one can also show 
\begin{equation}
 \pdlr{L}{g^{im}}_{\rc abcd} = \mr_{im}
 \label{deronethree1}
\end{equation} 
and, more interestingly, if we use the pair $(g^{ab},R^{ab}_{cd})$ as independent variables, we obtain
\begin{equation}
 \pdlr{L}{g^{im}}_{R^{ab}_{cd}} = 0
 \label{dertwotwo1}
\end{equation} 

(3) Using the fact that $P^{bcid}$ is anti-symmetric in $b$ and $c$, one can show 
\begin{equation}
 \nabla_b \nabla_c P^{bcid} 
 = \frac{1}{2} \left(\mr^{id} - \mr^{di}\right) = 0
 \label{eqnseven}
\end{equation} 
where the last equality follows from the symmetry of $\mr^{ab}$. Thus we get another surprising result that, for any scalar $L$ built from curvature tensor and the metric, we have the identity
\begin{equation}
 \nabla_a \nabla_b  \pdlr{L}{R_{abcd}}_{g_{ij}}=0
\end{equation} 
We shall now sketch a proof of these results. 
\\
\\
{\it Proof of the identities:}
\\
\\
As we mentioned in section (\ref{sec:2a1}), all the identities we want to prove follow from diffeomorphism cinvariance of the Lagrangian. So let us begin with an infinitesimal diffeomorphism $x^a\to x^a + \xi^a(x)$ which changes $L$, $g_{ab}$ and $R_{ijkl}$ by infinitesimal amounts. The essential trick is to express the Lie derivative of $L$ in two different  ways and equate the results. First, because $L$ is a scalar which depends on $x^i$ only through $g_{ab}(x)$ and $R_{ijkl}(x)$, we have:
\begin{equation}
\pounds_\xi L = \xi^m\nabla_m L = \xi^m P^{ab}\nabla_m g_{ab}+\xi^m P^{abcd} \nabla_m R_{abcd}
=P^{abcd} \xi^m\nabla_m R_{abcd}
\label{lie1}
\end{equation} 
because  $\nabla_m g_{ab}=0$.
On the other hand, 
if we think of the change $\delta L$ in $L$, due to small changes in $\delta g_{ab} \equiv \pounds_\xi g_{ab} $ and $\delta R_{ijkl} \equiv \pounds_\xi R_{ijkl}$,
we also have:
\begin{equation}
 \pounds_\xi L = P^{ab}\pounds_\xi g_{ab} + P^{ijkl} \pounds_\xi R_{ijkl}
 \label{lie2}
\end{equation} 
We will now expand the right hand side of \eq{lie2} bringing it into a form similar to the right hand side of \eq{lie1} and equate the two expressions. That will lead to  \eq{key11}. The first term on the RHS of \eq{lie2} is easily dealt with; using $ \pounds_\xi g_{ab}=\nabla_a\xi_b + \nabla_b \xi_a$ and the symmetry of $P^{ab} $ we obtain
\begin{equation}
 P^{ab}\pounds_\xi g_{ab} = 2 P^{ab} \nabla_a \xi_b
 \label{plg}
\end{equation} 
The term involving Lie derivative of curvature tensor gives (on contracting with $P^{ijkl}$ and using Eq.~(\ref{symm}))
\begin{equation}
 P^{ijkl}\pounds_\xi R_{ijkl} = P^{ijkl}\xi^m \nabla_m R_{ijkl} + 2P^{ijkl}\left[ (\nabla_i \xi^m) R_{mjkl} + (\nabla_k \xi^m) R_{ijml}\right]
 \label{lie3}
\end{equation} 
With some further index manipulations, this can be written as
\begin{equation}
 P^{ijkl}\pounds_\xi R_{ijkl} = P^{ijkl}\xi^m \nabla_m R_{ijkl}+ 4  (\nabla_i \xi_m)\mr^{im}
 \label{plr}
\end{equation} 
where we have used the definition $\mr^{im} = P^{ijkl} \rc mjkl$.
Using \eq{plr} and \eq{plg} in \eq{lie2} we get 
\begin{eqnarray}
 \pounds_\xi L &=& P^{ijkl} \xi^m \nabla_m R_{ijkl} + 2 (\nabla_i \xi_m) [P^{im} + 2 \mr^{im}]\nonumber\\
 &=& \pounds_\xi L + 2 (\nabla_i \xi_m) [P^{im} + 2 \mr^{im}]
\end{eqnarray} 
Since $\xi_m$ is arbitrary, it follows that the term within square brackets in the last expression has to vanish, leading to the condition in \eq{key11}:
\begin{equation}
 P^{im}= \pdlr{L}{g_{im}}_{R_{abcd}} = -2 P^{ijkl} \rc mjkl = - 2 \mr^{im}
 \label{key3}
\end{equation} 
Since the left-hand-side is symmetric, it  follows that $\mr^{im}$ must be symmetric. {\it This symmetry is certainly not obvious from the definition of $\mr^{im}$ itself.} 

Equation (\ref{eqnseven}) can be  be obtained immediately from the above result, as follows:
\begin{eqnarray}
2 \nabla_b\nabla_c P^{bcid} &=& \big[ \nabla_b,\nabla_c\big] P^{bcid}
 \nn \\
 &=& \rc bkbc P^{kcid} + \rc ckbc P^{bkid} + \rc ikbc P^{bckd} + \rc dkbc P^{bcik}
 \nn \\
 &=& - \mr^{di} + \mr^{id} = 0
 \label{twenty}
\end{eqnarray} 
where the anti-symmetry and pair-exchange symmetries of $P^{bcid}$ have been used. 
\\
\\
{\it Important Corollaries:}
\\
\\
One can derive a few corollaries from \eq{key3} which are useful in expressing the {\it field equations} obtained from our Lagrangian. We first note that, since $\delta g_{ab} = - g_{ai} g_{bj} \delta g^{ij}$, we have: 
\begin{equation}
  \pdlr{L}{g^{im}}_{R_{abcd}} =- \pdlr{L}{g_{im}}_{R_{abcd}} =2 P_i^{\phantom{i}jkl}R_{mjkl}= 2\mr_{im}
\end{equation} 
Let us  next consider how our results change if we use the pair ($g^{im}, \rc abcd$) or the pair $(g^{ab}, R^{ij}_{kl})$ as the independent  variables. Obviously that $\pdsl{L}{g^{im}} $ with $R_{abcd} $ held fixed is \textit{not} the same as 
$\pdsl{L}{g^{im}} $ with $\rc abcd$ held fixed. To find their relationship, we note that 
\begin{eqnarray}
 dL &=& \pdlr{L}{g^{im}}_{R_{abcd}} dg^{im} + P^{abcd}  dR_{abcd} = \pdlr{L}{g^{im}}_{\rc abcd} dg^{im} + P_a^{\phantom{a}bcd} d(g^{ak} R_{kbcd})\nonumber\\
 &=& P^{mbcd} dR_{mbcd} + dg^{im} \left\{ \pdlr{L}{g^{im}}_{\rc abcd} + \mr_{im} \right\}
 \label{interrelation}
\end{eqnarray}
from which it follows that
\begin{eqnarray}
  \pdlr{L}{g^{im}}_{\rc abcd}  &=& \pdlr{L}{g^{im}}_{R_{abcd}} - \mr_{im} \nonumber\\
  &=& 2 \mr_{im} - \mr_{im} = \mr_{im} = P_i^{\phantom{i}bcd}R_{mbcd} 
  \label{five}
\end{eqnarray}
More interestingly, if we work with $(g^{ab}, R^{ij}_{kl})$ as the independent  variables we get in place of \eq{interrelation} the reult:
\begin{equation}
 dL = \pdlr{L}{g^{im}}_{R_{abcd}} dg^{im} + P^{abcd} dR_{abcd} = P^{abcd} dR_{abcd} + dg^{im} \left\{\pdlr{L}{g^{im}}_{R^{ab}_{cd}} + 2 \mr_{im}\right\}
\end{equation} 
leading to
\begin{equation}
 \pdlr{L}{g^{im}}_{R^{ab}_{cd}} = \pdlr{L}{g^{im}}_{R_{abcd}} - 2 \mr_{im}=0
\end{equation} 
This result shows that scalars built from $g^{ij}, R^{ab}_{cd}$ are actually independent of the metric tensor when $R^{ab}_{cd}$ is held fixed.

We conclude by proving another important corollary: {\it The tensor $\nabla_m \nabla_n P^{amnb}$ is symmetric in indices $a$ and $b$.}
This result is important to establish directly the symmetry of $E_{ab}$ in \eq{eab}. We have already proved that $\mr_{ab}$ is symmetric, which means the last term $\nabla_m \nabla_n P^{amnb}$ in \eq{eab} must also be symmetric in $a$ and $b$. To show this explicitly, consider the anti-symmetric part of this tensor
\begin{equation}
 \nabla_m \nabla_n \left( P^{amnb} - P^{bmna}\right) =\nabla_m \nabla_n \left( P^{amnb} + P^{anbm}\right) 
 = -\nabla_m \nabla_n P^{abmn}
  \label{eight}
\end{equation} 
To obtain the first equality we have used the pair exchange symmetry and the anti-symmetry of the first two indices $P^{ijkl}$; in arriving at the second equality we have used the cyclic relation 
$P^{i[jkl]} =0$. It follows that the last term in \eq{eab} is symmetric only if the right hand side of \eq{eight} vanishes, which is assured by \eq{twenty} and the symmetry of $P^{abcd}$ under pair exchange. 

Therefore, we see that each of the terms in \eq{eab} are individually symmetric. The results of this section are based on arguments which would apply to Lagrangians more general than the ones we have considered (and hence more general than the \LL\ models), presumably giving more identities etc when, say, covariant derivatives of Riemann are also involved. In some sense, the main basis for deriving such identities lies in the very first equation \eq{lie1} which actually encapsulates the scalar nature of the Lagrangian $L$. 

A more detailed discussion of the results proved here, and some further implications of the same, can be found in \cite{tp-aspects}.

\section{Construction of  the Gauss-Bonnet lagrangian}\label{app:gblag}

Let us consider building a Lagrangian $L_2$ which is a linear combination of quadratic terms in curvature
 $R^2$, $R_{ab}R^{ab}$ and $R_{abcd} R^{abcd}$,  The general form for $L_2$ can be written as

\begin{eqnarray}
L_2 = \Biggl[ \alpha R^2 + \beta R_{a b} R^{a b} + \gamma R_{a b c d} R^{a b c d} \Biggl]
\label{gbl}
\end{eqnarray}
and its variation will have the form
\begin{eqnarray}
\delta \l(\sqrt{-g}~ L_2\r) = \sqrt{-g}~ E_2^{a b} ~ \delta g_{a b} + \sqrt{-g} ~ \nabla_k \delta v^k
\end{eqnarray}
A long but straightforward calculation gives
\begin{eqnarray}
E_2^{a b} = \frac{1}{2} g^{a b} L_2 &-& \Biggl[ 2 \alpha R R^{a b} + \l(2 \beta + 4 \gamma \r) R_{k c} R^{k a c b} - 4 \gamma R^{a c} R^{b}_{c} + 2 \gamma R^{a}_{k c d} R^{b k c d} \Biggl] \\
&+& \Biggl[ \l(2 \alpha + \beta + 2 \gamma \r) \nabla^a \nabla^b R - \left(2 \alpha + \frac{\beta}{2}\right)  g^{a b} \Box R - \left( \beta + 4 \gamma \right) \Box R^{a b} \Biggl]
\end{eqnarray}
and
\begin{eqnarray}
\delta v^k  &=& M^{a c k b} \nabla_{c} \delta g_{a b} \\
2 M^{c b a k} &=& \left( 4 \alpha R g^{a [c} g^{b] k} + \beta \left( 2 g^{b k} R^{a c} - g^{k c} R^{b a} - g^{b a} R^{k c} \right) + 4 \gamma R^{b c k a}  \right) + \left( a \leftrightarrow b \right)
\label{eq:x1-gb-bdyterm}
\end{eqnarray}
For $E_{ab}$ not to contain higher than second derivatives of the metric, one must demand
\begin{eqnarray}
2 \alpha + \beta + 2 \gamma = 0 \\
4 \alpha + \beta = 0 \\
\beta + 4 \gamma = 0
\end{eqnarray}
The solution is $\alpha = c_2 = \gamma,~\beta = - 4 c_2$, where $c_2$ is some arbitrary constant. We therefore arrive at the following final form for the full second order Lagrangian
\begin{eqnarray}
L = (16 \pi G)^{-1} \l[ R + (32 \pi^2)^{-1} \alpha L_2 \r] + {\rm boundary~terms}
\end{eqnarray}
where $[\alpha]=({\rm length})^2$ and we have set $c_2=(32 \pi^2)^{-1}$ (this convention follows from the fact that Gauss-Bonnet term will then be the Euler character in $4D$) with
\begin{equation}
L_2\propto \left[R^{abcd}R_{abcd}-4R^{ab}R_{ab}+R^2\right]                                               
\end{equation} 

The Einstein-Gauss-Bonnet (EGB) system, with lagrangian given by a linear combination of Einstein-Hilbert lagrangian and $L_2$ above is the simplest truncation of Lovelock model which has been well studied in the literature. For example, some consequences of the quasi-linearity of Einstein-Gauss-Bonnet (EGB) equations on the Cauchy problem as well as on ``junction conditions" are discussed in \cite{deruelle-GB-1}, while \cite{deruelle-GB-2} focusses on discussion of the conserved charges in EGB.

\section{Proof of identities in Eqs.~(\ref{identity1}) - (\ref{identity3}) } \label{app:membrane-ids}

To prove the required identities, we will use the following two relations concerning \LL\ actions of order $m$ in $D$ dimensions (see Eqs. (\ref{eom-lovelock}))
\begin{eqnarray}
E^{i (D)}_{j(m)} &=&  - \frac{1}{2} \frac{1}{16 \pi} \frac{1}{2^m} \delta^{i a_1 b_1 \ldots a_m b_m}_{j c_1 d_1 \ldots c_m d_m} R^{c_1 d_1}_{~ a_1 b_1} \cdots R^{c_m d_m}_{~ a_m b_m}
\nn \\
E^{i (D)}_{i(m)} &=& - \frac{D-2m}{2} L^D_m \label{eomproperty}
\end{eqnarray}
where $E^{i (D)}_{j(m)}$ is the equation of motion tensor for \LL\ action; for e.g., for $m=1$, $E^{i (D)}_{j(1)} = (16 \pi)^{-1} G^i_j$ where $G^i_j$ is the Einstein tensor. Note that for maximally symmetric spacetime, which is the case we are considering for the $(D-2)$ sub-manifold corresponding to the $(D-2)$ dimensional horizon, we have $G^2_2 = G^3_3 = \cdots = G^{D-1}_{D-1}$ due to isotropy. Hence we can write the second relation in \eq{eomproperty} for equation of motion tensor for the $(D-2)$ dimensional maximally symmetric subspace as 
\begin{eqnarray}
E^{A (D-2)}_{B(m)} &=& - \frac{\delta^A_B}{2} \left[ \frac{(D-2)-2m}{D-2} \right] L^{D-2}_m \label{eommaxsymm}
\end{eqnarray}
Using the above identity it is easy to prove \eq{identity1} as
\begin{eqnarray}
\left( Q^{0B}_{0A} \right)^{(D-1)}_{m} &=& \frac{1}{16 \pi} \frac{1}{2^m} \delta _{\lbrack
0A \mu_{1}\cdots \mu_{2m-2}]}^{[0B \nu_{1}\cdots \nu_{2m-2}]}\,\hat{R}_{\nu_{1}\nu_{2}}^{\mu_{1}\mu_{2}}\cdots \hat{R}_{\nu_{2m-3}\nu_{2m-2}}^{\mu_{2m-3}\mu_{2m-2}} 
\nonumber \\
&=& \frac{1}{16 \pi} \frac{1}{2^m} \delta _{\lbrack
A A_{1}\cdots A_{2m-2}]}^{[B B_{1}\cdots B_{2m-2}]}\,\hat{\hat{R}}_{B_{1}B_{2}}^{A_{1}A_{2}}\cdots \hat{\hat{R}}_{B_{2m-3}B_{2m-2}}^{A_{2m-3}A_{2m-2}} 
\nonumber \\
&=& - E^{B (D-2)}_{A(m-1)} = \frac{1}{2} \left( \frac{D-2m}{D-2} \right) L^{D-2}_{m-1} \; \delta^A_B
\end{eqnarray}
where to obtain the second equality we have used \eq{determinanttensor} and replaced $\hat{R}^{A B}_{C D}$ with the intrinsic curvature $\hat{\hat{R}}^{A B}_{C D}$ defined completely in terms of the induced metric $\gamma_{AB}$ of the horizon. Such a replacement is valid since the corresponding extrinsic curvature vanishes due to our assumption of staticity. The last equality then follows from using \eq{eommaxsymm}.
To prove the second identity of \eq{identity2} note that
\begin{eqnarray}
\left( Q^{BD}_{AC} \right)^{(D-1)}_{m} &=& \frac{1}{16 \pi} \frac{1}{2^m} \delta _{\lbrack
AC \mu_{1}\cdots \mu_{2m-2}]}^{[BD \nu_{1}\cdots \nu_{2m-2}]}\,\hat{R}_{\nu_{1}\nu_{2}}^{\mu_{1}\mu_{2}}\cdots \hat{R}_{\nu_{2m-3}\nu_{2m-2}}^{\mu_{2m-3}\mu_{2m-2}} \nonumber \\
&=& \frac{1}{16 \pi} \frac{1}{2^m} \delta _{\lbrack
AC A_{1}\cdots A_{2m-2}]}^{[BD B_{1}\cdots B_{2m-2}]}\,\hat{\hat{R}}_{B_{1}B_{2}}^{A_{1}A_{2}}\cdots \hat{\hat{R}}_{B_{2m-3}B_{2m-2}}^{A_{2m-3}A_{2m-2}} \nonumber \\
&=& \left( Q^{BD}_{AC} \right)^{(D-2)}_{m} = M^{(D-2)}_{m} \delta^{BD}_{AC}
\end{eqnarray}
where to obtain the second equality we have used $\hat{R}^{\hat{{\hat 0}} \mu}_{\rho \nu} = 0$ which is true due to (i) static nature of the background geometry and (ii) maximal symmetry of the $(D-2)$ dimensional subspace and  we have again replaced $\hat{R}^{A B}_{C D}$ with the intrinsic curvature $\hat{\hat{R}}^{A B}_{C D}$. Further, in the last equality we have again exploited maximal symmetry to get a $\delta^{BD}_{AC}$ proportionality with a proportionality constant $M^{(D-2)}_{m} $ which we can determine by contracting the above equation with $\hat{\hat{R}}^{A B}_{C D}$ to get
\begin{eqnarray}
L^{(D-2)}_{m} = \left( Q^{BD}_{AC} \right)^{(D-2)}_{m} \hat{\hat{R}}^{A C}_{B D} &=&  M^{(D-2)}_{m} \delta^{BD}_{AC} \hat{\hat{R}}^{A C}_{B D} = 2 M^{(D-2)}_{m} \; ^{D-2}R
\end{eqnarray}
This completes the proof of identity in \eq{identity2}. To prove the third identity of \eq{identity3} we again use the same procedure and write 
\begin{eqnarray}
 \delta _{\lbrack
A0CE \mu_{1}\cdots \mu_{2m-4}]}^{[B0DF \nu_{1}\cdots \nu_{2m-4}]}\,\hat{R}_{\nu_{1}\nu_{2}}^{\mu_{1}\mu_{2}}\cdots \hat{R}_{\nu_{2m-5}\nu_{2m-4}}^{\mu_{2m-5}\mu_{2m-4}} &=&  \delta _{\lbrack
ACE A_{1}\cdots A_{2m-4}]}^{[BDF B_{1}\cdots B_{2m-4}]}\,\hat{\hat{R}}_{B_{1}B_{2}}^{A_{1}A_{2}}\cdots \hat{\hat{R}}_{B_{2m-5}B_{2m-4}}^{A_{2m-5}A_{2m-4}} \nonumber \\
&=& N^{(D-2)}_{m-1} \delta^{BDF}_{ACE}
\end{eqnarray}
with the proportionality constant $N^{(D-2)}_{m-1} $ determined by contracting both sides of the above expression by  $\hat{\hat{R}}^{A B}_{C D}$ and then taking a trace to get
\begin{eqnarray}
(16\pi)2^{m-1} \frac{D-2m}{D-2} L^{D-2}_{m-1} = N^{(D-2)}_{m-1} (16\pi)2^{1} \frac{D-4}{D-2} L^{D-2}_{1} 
\end{eqnarray}
This completes the proof of identity in \eq{identity3}.

\end{document}